\documentclass[%
  reprint,
  superscriptaddress, % Grouped authors cleanly by institution superscripts
  amsmath,amssymb,
  aps,
  pra,
  floatfix,
  longbibliography,
]{revtex4-2}

\usepackage{graphicx}% Include figure files
\usepackage{dcolumn}% Align table columns on decimal point
\usepackage{bm}% bold math

\usepackage{amsthm, mathtools, amsfonts}
\usepackage{braket}
\usepackage{quantikz}
\usepackage{appendix}

\usepackage{booktabs}
\usepackage{multirow}
\usepackage{array}
\usepackage{tabularx}
\usepackage{enumitem}
\usepackage{tcolorbox}

\usepackage[utf8]{inputenc}
\usepackage[T1]{fontenc}
\usepackage{nicefrac}
\usepackage{url}

\IfFileExists{microtype.sty}{\usepackage[expansion=false,protrusion=true]{microtype}}{}

\usepackage{xcolor}
\usepackage{silence}
\usepackage[colorlinks=true, linkcolor=blue, citecolor=blue, urlcolor=blue]{hyperref}

\makeatletter
\def\@bibdataout@aps{%
  \immediate\write\@bibdataout{%
    @CONTROL{%
      apsrev42Control%
      \longbibliography@sw{%
        ,author="48",editor="1",pages="0",title="0",year="1"%
      }{%
        ,author="48",editor="1",pages="0",title="",year="1"%
      }%
    }%
  }%
  \if@filesw
    \immediate\write\@auxout{\string\citation{apsrev42Control}}%
  \fi
}%
\makeatother

\newtheorem{theorem}{Theorem}
\newtheorem{lemma}{Lemma}
\newtheorem{corollary}{Corollary}
\newtheorem{proposition}{Proposition}

\newtheorem{remark}{Remark}

\providecommand{\cmark}{\textcolor{green!60!black}{\ensuremath{\checkmark}}}
\providecommand{\xmark}{\textcolor{red!70!black}{\ensuremath{\times}}}

\begin{document}

\preprint{APS/123-QED}

\title{Diagnosing Simulation and Hardware Barriers to Cross-Size Transfer in Equivariant Quantum Reinforcement Learning}

% --- AUTHOR 1 ---
\author{Monit Sharma}
\email{monitsharma@smu.edu.sg}
\affiliation{School of Computing and Information Systems, Singapore Management University, Singapore}

% --- AUTHOR 2 ---
\author{Hoong Chuin Lau}
\email{hclau@smu.edu.sg}
\thanks{Corresponding Author}
\affiliation{School of Computing and Information Systems, Singapore Management University, Singapore}

% \date{\today}
\begin{abstract}
  Equivariant quantum circuits (EQCs) parameterise reinforcement-learning
  policies for combinatorial optimisation with a size-independent parameter
  count, suggesting policies trained on small instances may transfer to
  larger ones. Whether such transfer survives realistic execution has not
  been measured end-to-end. We train EQC policies on Euclidean
  Travelling Salesman instances and evaluate identical checkpoints across
  statevector simulation, matrix-product-state simulation, noisy simulation,
  a protocol-matched noiseless emulator, and trapped-ion hardware. Within
  the validated regime, zero-shot five-to-ten-city transfer beats
  target-size training in all six evaluations. Beyond it, three barriers
  emerge: bond-dimension truncation destroys policy quality even without
  transfer; larger size jumps degrade performance consistently with a
  conditional diagnostic bound; and finite-shot execution inflates the
  transfer gap from ${\sim}5\%$ to $31.3\%$ (sampling noise alone) to
  $45.3\%$ (hardware), because action margins lie below the shot-noise floor
  and collapse as $n^{-2.1}$. A cross-platform campaign across four hardware
  vendors confirms the penalty is set by native two-qubit gate count and error
  mitigation, not shot budget. A shot-complexity bound formalises the
  obstruction. We claim no quantum advantage; we provide the diagnostic
  standard such claims should meet.
\end{abstract}
\maketitle
%\footnote{Code available at: \url{https://github.com/SMU-Quantum/qml-eqc/tree/analysis}}.

\section{Introduction}

Combinatorial optimization (CO) problems, such as the Traveling Salesman Problem (TSP)
\cite{flood1956traveling}, lie at the heart of numerous real-world applications in logistics,
network routing, and circuit design. These problems are typically NP-hard \cite{applegate2006traveling,cook2015pursuit}, meaning their computational complexity grows rapidly with instance size, rendering exact solutions intractable for large inputs \cite{hartmanis1982computers,Arora1996}. As such, CO serves as a natural testbed for developing and evaluating scalable optimization techniques.

Recent advances in machine learning, particularly \emph{Reinforcement Learning} (RL), have shown promise in tackling CO problems \cite{mazyavkina2021reinforcement} by learning heuristic strategies from data \cite{gambardella1995ant,zheng2021combining}. In parallel, \emph{Quantum Reinforcement Learning} (QRL) has emerged as a hybrid paradigm that uses quantum circuits to represent and optimize policies or value functions \cite{Dunjko2018, Skolik2023}. QRL holds the potential to exploit quantum parallelism for more expressive or compact policy representations. However, a major challenge remains: most QRL methods must be retrained from scratch for every new problem instance or size. This limits their practicality and scalability, especially for applications requiring generalization across varying problem sizes.

In this work, we propose a new approach to enhance transferability in QRL by leveraging \emph{Equivariant Quantum Circuits} (EQCs) proposed in \cite{skolik2023equivariant}. These are structured quantum neural networks that preserve the symmetry of the problem domain  (such as permutation invariance over city indices in TSP). This architectural inductive bias ensures that the policy output transforms consistently under input permutations, naturally aligning with the structure of symmetric CO problems. 

Let \( \theta_n \) denote the optimized parameters learned on instances of a given CO problem encoded on a graph with \( n \) vertices. We ask the following question:

\begin{tcolorbox}[colback=gray!5!white, colframe=black!80!black, title=Research Question]
  Can parameters \( \theta_n \), trained on small instances, be transferred to larger instances with \( m > n \) vertices using the same EQC architecture, without retraining?
\end{tcolorbox}

The motivation stems from the observation that EQCs, like their classical permutation-equivariant counterparts \cite{Ouyang2021,fu2021generalizesmallpretrainedmodel}, can maintain architectural consistency across input sizes. This raises the possibility that a policy trained on small instances may encode structural insights that generalize across scales, analogous to transfer learning in classical ML. Prior work in quantum optimization, such as \cite{Brandao2018,Shaydulin_2023}, has also shown that fixed-parameter quantum algorithms like QAOA can generalize well across problem sizes, suggesting that size-agnostic inductive priors can be effective in quantum settings.

Symmetry-preserving quantum architectures offer a rare combination of trainability, parameter efficiency, and potential transferability across problem sizes. Yet it remains unclear whether such transfer survives realistic execution and how it scales with system size. This paper addresses that question through a multi-backend study of equivariant QRL, using TSP as a testbed. 

\paragraph{Contributions:} 
%This paper investigates whether
%permutation-equivariant quantum policies trained on small
%combinatorial-optimisation instances can be transferred to substantially larger
%instances, and whether such transfer survives realistic execution on simulators
%and quantum hardware. 
We make these contributions. 

\begin{enumerate}
  \item \textbf{Diagnostic framework for cross-size transfer in equivariant QRL.} We develop a theoretical framework for
  analysing how an equivariant policy behaves when applied to
  problem sizes larger than those it was trained on. The framework separates two
  distinct sources of performance loss: structural changes in the optimisation
  problem as the instance size grows, and the sensitivity of the learned policy
  to those changes. This decomposition yields a transfer bound that relates
  performance on unseen sizes to the smoothness of the underlying problem family
  and the stability of the trained policy. The analysis is deliberately
  architecture-aware and is designed to support empirical diagnosis rather than
  asymptotic advantage claims; we also discuss a graph-limit interpretation that
  connects cross-size transfer to continuity properties of large graph families.

  \item \textbf{A protocol-matched multi-backend evaluation methodology for
  quantum policy transfer.} We introduce an evaluation pipeline that runs the
  same trained policy, unchanged, across exact state-vector simulation,
  tensor-network (matrix-product-state) simulation, noisy simulation, hardware
  emulation, and quantum hardware spanning trapped-ion and superconducting
  processors from four vendors, which to our knowledge the first such study in QRL. Because identical checkpoints are executed
  under controlled, protocol-matched conditions, transfer behaviour can be
  disentangled from artefacts introduced by any particular simulator,
  approximation method, or hardware platform, giving a reproducible picture of
  how transfer performance evolves as increasingly realistic execution effects
  are introduced. Within the validated regime the trained policies reach mean
  optimality gaps of roughly $6$--$10\%$, and the five-to-ten-city zero-shot
  lane beats target-size training in every completed evaluation.

  \item \textbf{Identification and quantification of three independent barriers
  to scalable execution.} Using our framework we isolate three barriers that
  prevent the dense, all-to-all equivariant architecture from scaling. Each has a
  distinct experimental signature, but all three trace back to the same source:
  the circuit's all-to-all connectivity. \emph{(B1)}~The entangling structure
  generates entanglement growth that invalidates low-bond-dimension
  tensor-network approximations even when conventional fidelity metrics appear
  acceptable. \emph{(B2)}~Performance degrades smoothly but substantially as the
  gap between the trained size and the target size grows, in the manner the
  transfer bound predicts. \emph{(B3)}~On finite-shot hardware the differences
  between candidate actions fall at or below the statistical noise of sampling,
  so greedy decisions become statistically unresolved and additional shots do
  not help; this is what our trapped-ion and cross-platform hardware campaigns
  measure directly. Together, these results show that apparent transfer in
   idealized simulation does not automatically translate into scalable execution,
  and they isolate the architectural source of the limitations, motivating
  sparse equivariant circuit designs as a promising direction for future quantum
   optimization systems.
\end{enumerate}

Note that we make no claim of quantum advantage: the
equivariant architecture studied here is classically simulable, and we use that
simulability as a measurement instrument; a source of ground truth against
which simulator and hardware execution are compared.

\paragraph{Prior Approaches:}

The application of equivariant architectures in quantum machine learning is an expanding field, with promising results in mitigating barren plateaus and improving trainability. In the context of combinatorial optimization, the pioneering work by Skolik et al. \cite{skolik2023equivariant}  introduced Equivariant Quantum Circuits (EQCs) for learning heuristics on weighted graphs. Their research critically demonstrated that EQCs can effectively generalize across different TSP instances of a fixed size. For example, a policy trained on a distribution of 20-city problems was shown to perform well on previously unseen 20-city problems drawn from the same distribution. This established the principle of same-size generalization for EQCs.

Our work addresses a distinct, and arguably more challenging, form of generalization: generalization across problem sizes. The core contribution of this paper is to investigate the transfer of a policy trained on $n$-city instances to larger $m$-city instances, where $m>n$. This task, which we term cross-size generalization, requires not only instance-agnostic learning but also architectural and structural robustness to changes in the underlying problem dimension and Hilbert space. The theoretical framework we develop for bounding performance in the n→m setting, and the empirical validation of the `transfer-then-fine-tune' strategy, directly address this novel challenge, which was not the focus of prior EQC literature \cite{skolik2023equivariant,Skolik2023}. This positions our work as the next logical step in understanding and leveraging equivariance for scalable quantum optimization. 

Furthermore, the nearest-neighbour-greedy bias documented
by~\cite{li2025learningbasedtspsolverstendoverly} applies to learning-based
solvers trained on uniform Euclidean instances and decoded greedily --- the
regime in which our policies operate --- so we expect our trained policies
to inherit the same bias; we discuss the implications in
Section~\ref{sec:discussion}. Finally, we make no claim of computational
quantum advantage: recent dequantization
results~\cite{cerezo2025provable,goh2025lie,anschuetz2023efficient}
establish that the polynomial-DLA $S_n$-equivariant architecture studied
here admits efficient classical simulation via Lie-algebraic methods. We
engage with this trade-off explicitly in Section~\ref{sec:simulability} and
adopt classical simulability as our measurement instrument rather than as a
defect of the design: the diagnostic standard this paper sets out is one we
believe any future advantage claim in this architecture class will have to
meet.

\paragraph{Diagnostics and benchmarking context.}
The diagnostic framing of this paper connects to an active 2025--2026
conversation about how quantum machine-learning claims should be evaluated.
Scala et al.~\cite{scala2025qnndiag} develop pre-training diagnostics for
quantum neural networks via quantum neural tangent kernels, reflecting a
shift toward predicting and explaining QML model behaviour rather than
reporting end metrics alone. Recent quantum-reinforcement-learning hardware
demonstrations, such as the quantum policy evaluation experiments of Hein et
al.~\cite{hein2025qpe}, frame their results in the language of prospective
quantum advantage; our study is deliberately complementary. It contributes
no advantage claim, but a reproducible, protocol-matched methodology for
testing whether cross-size transfer survives realistic execution --- and a
quantitative account of why, for the canonical dense equivariant ansatz, it
currently does not.

\paragraph{Paper Structure.}
Section~\ref{sec:background} reviews the inductive-bias background (symmetry, locality, generalisation) and discusses the trainability--simulability trade-off by separating a collective $S_n$-invariant reference circuit (Model~A) from the implemented data-dependent weighted circuit (Model~B). Section~\ref{sec:theory} states the cross-size transfer setting in QRL and develops our theoretical framework;  Section~\ref{sec:theory2} proves the conditional diagnostic transfer bound for Model~B, with BHH motivating the structural smoothness profile (not the bound itself) and a conditional graphon refinement stated separately. Section~\ref{sec:methodology} describes the unified multi-backend evaluation pipeline (namely, statevector, MPS, IBM-inspired noisy MPS, Quantinuum emulator, Quantinuum hardware). Section~\ref{sec:empirical-eval} then reports experimental findings that validates source policies and in-regime transfer. Section~\ref{sec:barriers} identifies and quantifies the three independent barriers (B1, B2, B3) including the Quantinuum hardware campaign. Finally, Section~\ref{sec:discussion} reflects on the diagnostic framing and outlines next steps, foregrounding the sparse-connectivity equivariant ansatz prescription as the architectural fix that would relax all three barriers simultaneously.

\section{Foundations and Related Work}
\label{sec:background}

This work is motivated by three principles central to combinatorial optimization and learning theory: \emph{symmetry}, \emph{locality}, and \emph{generalization}. These principles guide our use of Equivariant Quantum Circuits (EQCs) in scalable QRL. Below we review their role and the associated literature.

\subsection{Symmetry}

Permutation symmetry is inherent in many permutation problems such as Traveling Salesman Problem (TSP), Quadratic Assignment Problem, Permutation Flowshop Scheduling and Linear Ordering Problems, where relabeling locations/jobs/items does not alter the optimal solution structure. Equivariant Quantum Circuits (EQCs) \cite{Skolik2023}, central to our methodology, embed this property by design. EQCs enforce a structural bias that compels learned policies to depend only on relational features such as graph topology and inter-location distances, rather than absolute location indices. For example, when TSP instances are drawn from the same distribution (e.g., cities sampled uniformly from the unit square), larger problems represent natural statistical extensions of smaller ones. Consequently, an EQC trained on smaller instances can internalize generic heuristics, such as prioritizing local proximity or avoiding long jumps early in a tour,that remain effective as instance size grows.

This role is analogous to classical graph neural networks (GNNs), which are well-established in their ability to generalize across graphs of varying sizes when tasks exhibit permutation symmetry \cite{Ouyang2024}. Prior work by Skolik et al.~\cite{skolik2023equivariant} demonstrated that a shallow (depth-1) EQC trained on 20-city TSPs generalized effectively to unseen instances of the same size, outperforming QAOA and surpassing non-equivariant circuits that failed to scale. 
%Our work builds directly on these findings by extending the analysis to \emph{cross-size transfer}, where parameters trained on $n$-city instances are applied to larger $m$-city problems.
Whereas prior work demonstrated within-size generalization of EQCs on TSP instances, this work investigates a more fundamental question: whether the inductive bias encoded by permutation-equivariant quantum policies survives transfer across problem scales and across execution backends, from exact simulation to trapped-ion hardware.

\subsection{Locality}

Locality provides a second crucial inductive bias for scalability. Classic heuristics such as the nearest-neighbor method \cite{johnson1997traveling}, Christofides’ algorithm \cite{christofides2022worst}, and insertion-based approaches \cite{gendreau1998generalized, lin1973effective} all rely on locally greedy decisions, constructing tours incrementally based on immediate neighborhood information. This property underpins their ability to scale to larger instances without global recomputation.

In machine learning, several studies reinforce the centrality of locality. Ouyang et al.~\cite{Ouyang2024} showed that RL agents trained on 50-city TSPs generalized effectively to instances with up to 1000 cities when using permutation-equivariant architectures. Fu et al.~\cite{fu2021generalizesmallpretrainedmodel} introduced a compositional framework where small pretrained GNNs are recursively applied to larger TSPs. In contrast, Joshi et al.~\cite{joshi} demonstrated that RL solvers trained only on trivially small problems fail completely on larger instances, highlighting the need for inductive biases such as permutation symmetry and locality. Li and Zhang~\cite{li2025learningbasedtspsolverstendoverly} provide a complementary cautionary observation: learning-based TSP solvers trained on uniformly-distributed Euclidean instances tend to acquire a nearest-neighbour-greedy bias, which limits their out-of-distribution robustness. Together, these findings frame the design problem this paper addresses: an EQC that encodes locality-aware heuristics should transfer reasonably across instance sizes on the training distribution, while inheriting the same out-of-distribution caveats as classical learning-based solvers.

\subsection{Generalization}

Generalization links symmetry and locality into a coherent framework for scalable learning. Under standard assumptions, including independent uniform sampling of cities, adherence to the triangle inequality, and the dominance of short edges in optimal tours, policies trained on small TSP instances should maintain their effectiveness when transferred to larger ones. Euclidean TSPs are known to exhibit scale-invariant local structure, and classical results such as the Beardwood–Halton–Hammersley theorem \cite{beardwood1959shortest,arora1998polynomial} imply that optimal tour length grows smoothly as $\Theta(\sqrt{n})$. This supports the idea of \emph{graceful performance degradation}, where approximation error increases slowly with instance size and can be mitigated through limited fine-tuning rather than full retraining.

In quantum models, generalization is intertwined with trainability. Quantum neural networks often encounter barren plateaus, where gradients vanish exponentially with system size \cite{mcclean2018barren,arrasmith2021effect}. A trade-off also arises between expressivity and generalization: overly expressive circuits risk overfitting, while highly constrained circuits lack flexibility \cite{peters2023generalization,abbas2021power}. Symmetry-based designs offer a way forward. Schatzki et al.~\cite{schatzki2024theoretical} proved that permutation-equivariant QNNs avoid barren plateaus by restricting optimization to symmetry-consistent subspaces, while Mernyei et al.~\cite{mernyei2022equivariantquantumgraphcircuits} empirically confirmed stable scaling with depth and qubit count. From a quantum information perspective, adding qubits to a local EQC primarily affects redundant degrees of freedom, while global entangling gates such as
\[
  U_G(\mathcal{E}, \gamma) = \exp\left(i \gamma \sum_{(i,j) \in \mathcal{E}} Z_i Z_j\right)
\]
distribute parameter influence uniformly across edges. This phenomenon echoes parameter concentration observed in QAOA \cite{Brandao2018,Shaydulin_2023}, further suggesting robustness to scale.

Collectively, these insights support our central hypothesis: EQCs trained on small TSP instances encode transferable, instance-independent structural information. Symmetry ensures consistency across permutations, locality provides scale-invariant heuristics, and the absence of barren plateaus ensures that fine-tuning is efficient. Together, these principles make scalable quantum optimization via cross-size transfer both theoretically plausible and empirically promising.

\subsection{The Trainability--Simulability Trade-off}
\label{sec:simulability}

A series of recent results~\cite{cerezo2025provable,goh2025lie,anschuetz2023efficient,larocca2023equivariant,larocca2025barren} establishes that the very property that enables polynomial sample efficiency in $S_n$-equivariant QNNs --- a polynomially-bounded dynamical Lie algebra (DLA) --- simultaneously enables their efficient classical simulation. We engage with this trade-off here rather than buried in an appendix, because it is central to the framing of the contribution.

\paragraph{Two models: collective reference vs.\ implemented circuit.}
The DLA and dequantisation results below apply rigorously only to a \emph{collective $S_n$-invariant} reference circuit, which we call Model~A. The circuit actually executed in our pipeline --- with data-dependent weighted couplings driven by the instance's Euclidean distance matrix --- is a strictly more general object, which we call Model~B. The distinction is consequential and we therefore foreground it.
The TSP encoding/equivariance conventions and the Schatzki theorem background
are given in Appendices~\ref{app:tsp-encoding} and~\ref{sec:app1},
respectively.
\begin{itemize}
  \item \textbf{Model~A (collective reference).} Generators $\{G_1^{(n)}\!=\!\tfrac{1}{n}\sum_i X_i,\ G_3^{(n)}\!=\!\tfrac{2}{n(n-1)}\sum_{i<j} Z_iZ_j\}$. Both generators commute with every qubit permutation $R(\pi)$, $\pi\in S_n$; the Lie closure is therefore $S_n$-equivariant and the Schatzki--Schur--Weyl structure applies.
  \item \textbf{Model~B (implemented).} Cost generator $H_C(D)=\sum_{i<j} d_{ij} Z_iZ_j$ and mixer $H_M(s)=\sum_i s_i X_i$ driven by a fixed instance's distance matrix $D$ and adjacency-diagonal node features $s$. For generic $D$, $H_C(D)$ does \emph{not} commute with arbitrary $R(\pi)$; the architecture is permutation-equivariant only in the \emph{joint} sense $U_n^{\rm B}(\theta;\pi\cdot D,\pi\cdot s)=R(\pi)\,U_n^{\rm B}(\theta;D,s)\,R(\pi)^\dagger$, which is strictly weaker than fixed-instance $S_n$-invariance.
\end{itemize}

\paragraph{The DLA of Model~A.}
The complexity of Model~A is pinned down exactly:
 
\begin{proposition}[DLA dimension of Model~A]
\label{prop:dla-dim}
For Model~A's collective generator set acting on $n$ qubits, the Lie closure
has dimension
\[
  \dim(\mathfrak{g}) \;=\; T_{e_{n+1}}-\nu_n \;\in\;\Theta(n^3),
\]
where $T_{e_{n+1}}=\binom{n+3}{3}$ is the tetrahedral number of
\cite[Theorem~4]{schatzki2024theoretical} and $\nu_n=\lfloor n/2\rfloor+1$
counts the $S_n$-irreducible blocks. When dynamics are restricted to the
totally symmetric (Dicke) sector --- which contains the initial state
$|+\rangle^{\otimes n}$ --- the effective dimension is at most
$(n+1)^2-1=O(n^2)$.
\end{proposition}
 
\noindent
The proof, via Schur--Weyl duality and the explicit basis construction
of~\cite{allcock2024dla}, is given in Appendix~\ref{app:dla-proof}.

\paragraph{Implication for Model~A: \texorpdfstring{$\mathfrak{g}$-sim}{g-sim} is feasible; Model~B is not automatic.}
By the framework of~\cite{goh2025lie}, any DLA-polynomial circuit admits the $\mathfrak{g}$-sim algorithm, which propagates the $\dim(\mathfrak{g})$-dimensional adjoint representation rather than the $2^n$-dimensional Hilbert state. For Model~A at depth $L$ this gives a classical simulation in time $O(L\dim(\mathfrak{g})^2)=O(Ln^6)$ in the full space, or $O(Ln^4)$ in the symmetric sector. For Model~B at a fixed Euclidean instance with generic distances, the Lie closure $\langle iH_C(D),iH_M(s)\rangle_{\rm Lie}$ need not be polynomial-dimensional and the Dicke-sector confinement need not hold; the $\mathfrak{g}$-sim runtime above is therefore a Model-A property, not an unconditional theorem for Model~B. The MPS simulation we employ in Section~\ref{sec:methodology} is the classical route we use for Model~B at finite $\chi$.

\paragraph{What this paper claims, and what it does not.}
We make \emph{no} claim of computational quantum advantage. The contribution of this paper rests on three independent grounds, each of which is unaffected by classical simulability:
\begin{itemize}
  \item \emph{Heuristic policy value.} The architecture defines an extremely parameter-efficient inductive bias for $S_n$-symmetric combinatorial optimisation: at $L=1$ the policy has \emph{two} trainable scalars regardless of problem size $n$.
  \item \emph{Methodological value.} The unified multi-backend evaluation pipeline (statevector + MPS + IBM-inspired noisy MPS + Quantinuum emulator + Quantinuum trapped-ion hardware, Section~\ref{sec:methodology}) cross-validates a single trained policy in a way no prior QRL study has, and it is what enables the three-barrier characterisation that constitutes the bulk of the empirical contribution.
  \item \emph{Diagnostic value.} The barriers we identify in Section~\ref{sec:barriers} have a common topological root cause --- the dense pairwise readout of Model~B --- that is independent of whether Model~A's collective reference is classically simulable; they characterise the canonical all-to-all equivariant ansatz, not its particular implementation.
\end{itemize}

This framing inverts the usual defensive posture: rather than contesting the simulability of the EQC, we use Model~A as the instrument that makes a rigorous multi-scale evaluation of Model~B possible.

\section{Theoretical Framework for Zero-Shot Transfer}
\label{sec:theory}

\subsection{Background and Motivation}

A significant practical hurdle in applying QRL, much like its classical counterpart, is the often-substantial computational cost associated with training an agent from scratch for every new problem instance or scale. Transfer learning, particularly zero-shot transfer, offers a compelling strategy to mitigate this issue. In this paradigm, a model trained on a source task is applied directly to a new, related target task without any further training on the target task's data \cite{zhang2025pessimismprincipleeffectiveframework}. For TSP, this could mean training a QRL agent on an n-city problem and then deploying it to solve an m-city problem, ideally with $m>n$. The ability to effectively transfer learned knowledge would drastically improve the scalability and utility of QRL approaches.

\subsection{Problem Statement}

This work addresses the specific scenario of zero-shot transfer learning in the context of QRL for permutation problems, and specifically  TSP as a testbed. A QRL agent, whose policy or value function is parameterized by a Quantum Neural Network (QNN), is first trained on samples from an $n$-city TSP task distribution, to learn parameters that achieve optimal or near-optimal performance across $n$-city instances. The parameters $\theta^*$ obtained from this training phase are then directly applied to an $m$-city TSP, where $m>n$. The central objective of this work is to derive a theoretical bound on the expected performance of this transferred QRL agent when evaluated on the larger $m$-city problem instances.

\subsection{Notation and Preliminaries}
\label{sec:notation}

The following notation will be used throughout the paper.

\begin{itemize}
  \item $P_k(\theta)$: True performance of the QRL agent with parameters
  $\theta$ on a $k$-city TSP task. We use the convention that larger
  values of $P_k(\theta)$ indicate better performance, e.g., higher expected
  reward or lower normalized tour cost.

  \item $\theta_n^*$: Parameters obtained by training the QRL agent on the
  $n$-city source task using $M_n$ training episodes or samples.

  \item $\widehat{P}_n(\theta_n^*)$: Empirical performance of the trained
  policy on the $n$-city source training set.

  \item $P_n(\theta_n^*)$: True expected performance of the trained policy
  on the $n$-city source task.

  \item $\mathcal{G}^{\rm roll}_n(\delta)$: Rollout source-task
  generalization error. This term controls the gap between empirical and
  true source-task performance of the trained rollout policy. Specifically,
  with probability at least $1-\delta$,
  \[
    P_n(\theta_n^*) \geq
    \widehat{P}_n(\theta_n^*) - \mathcal{G}^{\rm roll}_n(\delta).
  \]
  The formal training regimes under which this term is controlled are
  stated in Assumption~(A2) of Appendix~\ref{app:assumptions}. When no
  confusion arises, we write $\mathcal{G}_n(\delta)$ for
  $\mathcal{G}^{\rm roll}_n(\delta)$.

  \item $P_m(\theta_n^*)$: True expected performance of the transferred
  policy on the $m$-city target task, using the source-trained parameters
  $\theta_n^*$ together with the $S_m$-adapted equivariant generators. This
  is the main quantity to be lower bounded.

  \item $\mathcal{D}_{n \rightarrow m}$: Transfer penalty incurred when
  deploying the $n$-city trained parameters on the $m$-city target task. We
  use $\mathcal{D}_{n \rightarrow m}$ as an upper bound on the performance
  drop:
  \[
    P_n(\theta_n^*) - P_m(\theta_n^*) \leq
    \mathcal{D}_{n \rightarrow m}.
  \]
\end{itemize}

Here, \(P_k(\theta)\) denotes the performance of the $k$-city EQC evaluated
with parameters \(\theta\), i.e.,
\[
  P_k(\theta) := P_k(U_k(\theta)),
\]
where \(U_k(\theta)\) is the equivariant quantum circuit architecture indexed
by problem size \(k\).

We define the residual source-task suboptimality $\varepsilon_n$ by
\[
  P_n(\theta_n^*) = P_n^{\mathrm{opt}} - \varepsilon_n,
\]
where \(\varepsilon_n \geq 0\) denotes the gap between the trained policy's performance and the optimal achievable performance on the \(n\)-city source task. This is a notational definition rather than an additional assumption.
The formal assumptions for the transfer analysis are presented in
Appendix~\ref{app:assumptions}.

\subsection{Approach and Contribution}

Our theoretical analysis takes inspiration from Theorem 4 of Schatzki et al.~(2024)~\cite{schatzki2024theoretical}, which provides covering-number generalisation guarantees for $S_n$-equivariant quantum neural networks. Schatzki's result is a supervised guarantee for the equivariant-QNN \emph{loss} class; the greedy QRL rollout policy is not such an observable (it involves sequential $\arg\max$ decisions and accumulated tour length), so we do not import it as-is into the QRL setting. Instead, Schatzki et al.\ motivates the existence and optimistic shape of a rollout source-generalisation term $\mathcal{G}^{\rm roll}_n(\delta)$, which we then treat as an explicit assumption (Assumption~(A2), Appendix~\ref{app:assumptions}). More details about Schatzki's theorem can be found in Appendix~\ref{sec:app1}.

The primary theoretical contribution of this paper is a \emph{conditional diagnostic} transfer bound for the zero-shot scenario above. Rather than directly adapting a supervised generalisation theorem to the transfer setting, the bound is built as a worst-case decomposition under an explicit assumption stack, with each assumption tagged as proved, assumed, or speculative. The bound relates the expected performance on the $m$-city target task to:

\begin{itemize}
  \item performance achieved on the $n$-city source task, via the assumed rollout source-generalisation term $\mathcal{G}^{\rm roll}_n(\delta)$;
  \item a parametric mismatch term with explicit generator-difference constant $C_{\rm gen}\in\{2,4\}$ obtained by Duhamel telescoping;
  \item a structural smoothness term controlled by an explicit lifted-policy-smoothness hypothesis (Assumption~(A7)), the operative regularity condition of the bound.
\end{itemize}

The point of such a bound is to identify the right scaling structure of cross-size degradation under direct parameter transfer, not to predict gaps numerically. The bound is conditional (it rests on the assumption stack above) and diagnostic (it is worst-case and can be vacuous in some regimes), and we report it as such throughout. The key conceptual move is the introduction of a task-dissimilarity term $\mathcal{D}_{n\to m}$ that absorbs both the parametric and structural mechanisms; we develop it carefully in Section~\ref{sec:dissimilarity} and connect it to classical transfer-learning theory in~\cite{tripuraneni2020theory}.

\subsection{Problem Formulation: Zero-Shot Transfer for TSP with QRL}
\label{sec:metric}

To develop a theoretical framework for analyzing zero-shot transfer in TSP QRL, we begin by precisely formulating the transfer setting and identifying a set of foundational assumptions. These assumptions serve to map the supervised learning framework underpinning \cite{schatzki2024theoretical} (see Appendix~\ref{sec:app1} for more details), which is centered on generalization guarantees for $S_n$-equivariant quantum neural networks, to the sequential decision-making dynamics characteristic of QRL.

\begin{itemize}
  \item \textbf{Source Task ($T_n$):} The QRL agent is trained to solve an $n$-city TSP. The agent’s behavior is determined by a policy $\pi_\theta$ (or a value function $V_\theta$), where $\theta$ represents the trainable parameters of its internal QNN. Through interaction with $n$-city TSP environments (e.g., various graph instances or a fixed graph explored over many episodes), the agent learns optimal parameters $\theta_n^*$.

    \begin{itemize}
      \item The state representation for an $n$-city TSP typically encodes information such as the current city, the set of visited cities, and potentially information about remaining cities or path costs. This state is encoded into $N_q(n)$ qubits.
      \item The action space involves selecting the next city to visit from the set of unvisited cities.
      \item The reward function guides the agent, typically providing negative rewards proportional to edge distances and a large positive reward upon completing a valid tour, or simply the negative total tour length at the end of an episode.
    \end{itemize}

  \item \textbf{Target Task ($T_m$):} The agent is then tasked with solving an $m$-city TSP, where $m > n$. Crucially, in the zero-shot transfer setting, the parameters $\theta_n^*$ learned from $T_n$ are used directly, without any further training or fine-tuning on $m$-city TSP instances. The state representation is now for $m$ cities, encoded on $N_q(m)$ qubits.

  \item \textbf{QRL Agent Architecture:} We assume the core of the QRL agent (e.g., the policy network or value function approximator) is a QNN, $U(\theta)$. The architecture of this QNN is initially defined for $N_q(n)$ qubits. Its adaptation to operate on $N_q(m)$ qubits while using the fixed parameters $\theta_n^*$ is a critical point that will be detailed in the assumptions.

  \item \textbf{Performance metric ($P_k(\theta)$).}
    Throughout we adopt the \emph{“higher-is-better”} convention: larger $P_k(\theta)$ means a better tour.
    To ensure that the Hoeffding-type generalisation terms in \cite{schatzki2024theoretical} apply, we \emph{affinely rescale} every concrete metric so that
    \[
      0 \;\le\; P_k(\theta)\;\le\;1 \quad\text{for all $k$ and all $\theta$}.
    \]
    Typical instantiations are:

    \begin{itemize}
      \item \emph{Expected tour-quality reward.}
        Let $L(\theta)$ be the expected tour length produced by the agent and
        $L_{\max}>L_{\text{opt}}$ a deterministic upper bound (e.g.\ tour length of the
        nearest-neighbour heuristic).
        \[
          P_k(\theta)\;=\;1-\frac{L(\theta)-L_{\text{opt}}}{L_{\max}-L_{\text{opt}}}
          \;\in\;[0,1].
        \]
        (Equivalently, you may work directly with the reward
        $R=-L$ and then normalise to $[0,1]$.)

      \item \emph{Probability of optimality.}
        $P_k(\theta)=\Pr[\text{agent outputs an optimal tour}]$—already in $[0,1]$.

      \item \emph{Approximation ratio.}
        For ratio $r(\theta)=L(\theta)/L_{\text{opt}}\ge1$ fix a known worst-case
        bound $\rho_{\max}$ (e.g.\ $2$ in the Euclidean metric); set
        \[
          P_k(\theta)\;=\;1-\frac{r(\theta)-1}{\rho_{\max}-1}\;\in\;[0,1].
        \]
    \end{itemize}

    \noindent
    Henceforth every reference to “shorter route”, “negative tour length”, or
    similar uses this rescaled, reward-style metric.

\end{itemize}

\subsection{Formal Assumptions for the Transfer Bound}
Our derivation relies on a set of assumptions ensuring symmetry preservation, smoothness of performance metrics, and Lipschitz continuity across parameters and problem sizes. For clarity, we summarize them here:
\begin{itemize}
  \item \textbf{Equivariance:} The circuit and policy remain $S_k$-equivariant across problem sizes.
  \item \textbf{Generalization:} Source-task training satisfies standard generalization bounds under i.i.d.\ or replay settings.
  \item \textbf{Problem Encoding:} Each city is mapped to one qubit, ensuring task–architecture alignment.
  \item \textbf{Smoothness:} Optimal performance varies continuously with problem size (Euclidean TSP assumption).
  \item \textbf{Regularity:} The performance functional is Lipschitz-continuous in parameters, instances, and unitaries.
\end{itemize}
Full technical statements and proofs are provided in Appendix~\ref{app:assumptions}.

\section{A Theoretical Bound on Cross-Instance Transfer Performance}
\label{sec:theory2}

Building on the theoretical foundations established by \cite{schatzki2024theoretical} (Appendix~\ref{sec:app1} and Assumptions \textbf{(A1)--(A8)} in Appendix~\ref{app:assumptions}), with the operative regularity hypothesis being the lifted-policy smoothness (A7)), we now derive a conditional diagnostic lower bound on the performance $P_m(\theta_n^*)$ of the QRL agent when deployed on the $m$-city target task $T_m$ using parameters $\theta_n^*$ learned from the $n$-city source task $T_n$. The objective is to identify the scaling structure of the cross-size performance degradation rather than to predict gaps numerically. The penalty $\mathcal{D}_{n \rightarrow m}$ (see Appendix~\ref{app:dissimilar_derivation} for the worst-case operator-norm derivation) decomposes into a parametric mismatch term with explicit generator-difference constant $C_{\rm gen}\in\{2,4\}$ and a structural smoothness term carried by Assumption (A7), while respecting the joint permutation-equivariance of the implemented Model~B architecture.

\subsection{Structure of the Bound}

Our goal is to establish a lower bound for $P_m(\theta_n^*)$
in terms of $P_n(\theta_n^*)$ and the change in performance due to transfer.

Let
\[
  \Delta P_{n \rightarrow m}(\theta_n^*) = P_n(\theta_n^*) - P_m(\theta_n^*)
\]
represent the drop in performance when transferring from the $n$-city task to the $m$-city task using parameters $\theta_n^*$.

We know from the rollout source-task generalisation assumption (Assumption~(A2), Appendix~\ref{app:assumptions}) that:
\[
  P_n(\theta_n^*) \geq \widehat P_n(\theta_n^*) - \mathcal{G}^{\rm roll}_n(\delta)
\]

Substituting this into the expression for $P_m(\theta_n^*)$:
\[
  P_m(\theta_n^*) \geq \left(\widehat P_n(\theta_n^*) - \mathcal{G}^{\rm roll}_n(\delta)\right) - \Delta P_{n \rightarrow m}(\theta_n^*)
\]

The core of the derivation is to find a suitable upper bound for $\Delta P_{n \rightarrow m}(\theta_n^*)$, which we have denoted conceptually as $\mathcal{D}_{n \rightarrow m}$. We can establish that
\[
  \Delta P_{n \rightarrow m}(\theta_n^*) \leq \mathcal{D}_{n \rightarrow m},
\]
then:
\[
  P_m(\theta_n^*) \geq \widehat P_n(\theta_n^*) - \mathcal{G}^{\rm roll}_n(\delta) - \mathcal{D}_{n \rightarrow m}
\]

With probability at least $1 - \delta$:

\[
  P_m(\theta_n^*) \geq \widehat P_n(\theta_n^*) - \mathcal{G}^{\rm roll}_n(\delta) - \mathcal{D}_{n \rightarrow m}
\]

This structure clearly separates the sources of sub-optimality on the target task:

\begin{itemize}
  \item $\widehat P_n(\theta_n^*)$: The empirical performance on the source task.
  \item $\mathcal{G}^{\rm roll}_n(\delta)$: The rollout source-task generalisation term (assumed; Assumption (A2)). Schatzki et al.\ provide an optimistic motivating rate $\mathcal{G}^{\rm Schatzki}_n(\delta)=\widetilde{\mathcal{O}}(\sqrt{T_{e_{n+1}}/M_n}+\sqrt{\log(1/\delta)/M_n})$ for the supervised $S_n$-equivariant QNN loss class, but whether the non-smooth greedy-rollout policy inherits this equivariant covering rate is an open question. We therefore take $\mathcal{G}^{\rm roll}_n$ as an explicit assumption rather than as a direct consequence of Schatzki et al.
  \item $\mathcal{D}_{n \rightarrow m}$: The additional error or performance drop incurred due to the zero-shot transfer to a different-sized problem.
\end{itemize}

Characterizing \( \mathcal{D}_{n \rightarrow m} \) presents the principal analytical challenge, as it must capture the nuanced effects arising from scaling the problem size and deploying parameters optimized for the \( n \)-city setting within an \( m \)-city QNN architecture with structurally adapted equivariant generators.

\subsection{Quantifying Task Dissimilarity \texorpdfstring{$\mathcal{D}_{n \to m}$}{D n to m}}
\label{sec:dissimilarity}

A central challenge in extending single-task generalization bounds to transfer is
to quantify the performance loss when reusing parameters $\theta^*_n$, optimized for
an $n$-city TSP, on a larger $m$-city instance. We denote this penalty by
$\mathcal{D}_{n \to m}$.

Intuitively, $\mathcal{D}_{n \to m}$ captures two key effects:
\begin{enumerate}
  \item \textbf{Parametric mismatch:} the same parameters $\theta^*_n$ act on scaled
    circuit generators ($\tfrac{1}{n}\sum_j O_j$ vs.\ $\tfrac{1}{m}\sum_j O_j$), producing
    systematic deviations in the induced unitaries.
  \item \textbf{Structural shift:} even under equivariance, the statistical and geometric
    properties of optimal TSP tours evolve with problem size, leading to smooth but
    non-negligible performance differences.
\end{enumerate}

Together, these contributions define the transfer penalty:
\[
  \Delta P_{n \to m}(\theta^*_n) \;\leq\; \mathcal{D}_{n \to m}
  = D^{(\mathrm{param})}_{n \to m} + D^{(\mathrm{struct})}_{n \to m}.
\]

The detailed derivation of this decomposition, including operator-norm
bounds for parametric mismatch and an assumption-driven square-root
structural bound for task shift, is given in Appendix~\ref{app:dissimilar_derivation};
the complementary task-dissimilarity interpretation is given in
Appendix~\ref{app:dissimilarity}.

\subsection{Final Performance Bound}
\label{sec:final_bound}

We now synthesize the results to obtain a conditional diagnostic lower bound on the target-task performance \( P_m(\theta_n^*) \) for parameters \( \theta_n^* \) trained on the source task. The bound is \emph{conditional} (it rests on an explicit stack of assumptions, foregrounded below) and \emph{diagnostic} (it identifies the right scaling structure, not numerical predictions). It applies to the implemented Model~B and rests on a structural assumption --- lifted-policy smoothness --- rather than on the Beardwood--Halton--Hammersley (BHH) theorem, which we use only to motivate the functional form of that assumption.

\begin{theorem}[Conditional Diagnostic Transfer Bound]
  \label{thm:final-performance}
  Let \( \theta_n^* \) denote the parameters trained on \( M_n \) i.i.d.\ episodes from the source $n$-city task and let $U_n^{\rm B}(\theta_n^*)$ be the corresponding Model-B unitary. Define the identity-padded lift $U_n^{\uparrow}(\theta_n^*):=U_n^{\rm B}(\theta_n^*)\otimes\mathbb{I}_{2^{m-n}}$, and write $P_k(U)$ for the approximation-ratio performance of $U$ at city count $k$. Under Assumptions \textbf{(A1)} joint equivariance, \textbf{(A2)} rollout source generalisation (with concentration term $\mathcal{G}^{\rm roll}_n(\delta)$), \textbf{(A3)} direct parameter transfer, \textbf{(A4)} performance regularity (unitary-Lipschitz constant $L_U$), and \textbf{(A7)} lifted-policy smoothness $|P_n(U_n^{\rm B})-P_m(U_n^{\uparrow})|\le C_{\rm lift}\,\Psi(n,m)$, with probability at least \( 1 - \delta \):
  \[
    P_m(\theta_n^*) \;\ge\; \widehat{P}_n(\theta_n^*)
    - \underbrace{\mathcal{G}^{\rm roll}_n(\delta)}_{\text{source gen.}}
    - \underbrace{\mathcal{D}_{n \rightarrow m}}_{\text{transfer penalty}},
  \]
  where the transfer penalty decomposes as
  \begin{equation}
    \boxed{
      \begin{aligned}
        \mathcal{D}_{n \rightarrow m}
        &\leq
        \underbrace{L_U\,C_{\rm gen}\,\|\theta_n^*\|_1\,\frac{m-n}{m}}_{\text{parametric mismatch}}\\
        &\quad+
        \underbrace{C_{\rm lift}\,\Psi(n,m)}_{\text{lifted-policy smoothness}}.
    \end{aligned}}
  \end{equation}
  The constant $C_{\rm gen}$ is the generator-difference constant of Lemmas~\ref{lem:gen-diff-1body}--\ref{lem:unitary-dev} (Appendix~\ref{app:dissimilar_derivation}): $C_{\rm gen}=2$ for one-body normalised generators and $C_{\rm gen}=4$ for two-body normalised generators. For the Skolik EQC implementation studied here both branches contribute, and the binding constant is $C_{\rm gen}=4$. For Euclidean TSP families we take the size-smoothness profile $\Psi(n,m)=\sqrt{m}-\sqrt{n}$, motivated by the BHH scaling of optimal tour length under uniform sampling on $[0,1]^2$.
\end{theorem}

\vspace{0.3em}

\paragraph{Status.}
Theorem~\ref{thm:final-performance} is a conditional decomposition, not a sharp predictor of measured gaps. The proof uses a clean insertion-of-lifted-unitary identity $P_n(U_n^{\rm B})-P_m(U_m^{\rm B}) = [P_n(U_n^{\rm B}) - P_m(U_n^{\uparrow})] + [P_m(U_n^{\uparrow}) - P_m(U_m^{\rm B})]$, where the first bracket is bounded by lifted-policy smoothness (Assumption~A7) and the second by Duhamel telescoping together with the generator-difference lemma. We do not use a group-twirling argument (a twirled unitary becomes a channel, and the greedy-rollout performance is a nonlinear functional of the unitary; see Appendix~\ref{app:perf-degradation}, Remark~\ref{rem:no-twirling}). Under the approximation-ratio performance metric (Section~\ref{sec:notation}), $P_k^{\rm opt}=1$ for every $k$, so a structural term defined as a difference of optimal performances is identically zero --- the operative structural quantity is the lifted-policy smoothness, and BHH motivates only the profile shape $\Psi(n,m)$, not the bound itself.
Throughout, $L_U$ denotes the unitary-Lipschitz constant of the operative
performance functional: $2\|M\|_{\rm op}$ for linear observables
(Lemma~\ref{lem:unitary-lip-app}), and the effective constant
$L_U^{\rm eff}\le 4k\bar\rho B$ of Proposition~\ref{prop:margin-smoothing}
for the expected greedy-rollout performance, whose validity rests on the
bounded-margin-density condition stated there.

Table~\ref{tab:theory-status} summarises the epistemic status of every
theoretical component used in this paper, so that no reader mistakes the
conditional structure of the framework.

\begin{table}[t]
  \centering
  \caption{Status of the theoretical components. ``Proved'' means
  established in this paper or directly inherited with proof; ``imported''
  means proved elsewhere for a different function class and used here only
  as motivation; ``assumed'' and ``speculative'' are explicit hypotheses.}
  \label{tab:theory-status}
  \small
  \setlength{\tabcolsep}{3pt}
  \newcommand{\theorycell}[2]{\parbox[t]{#1}{\raggedright\strut #2\strut}}
  \newcommand{\whereapp}[1]{\hyperref[#1]{App.~\ref*{#1}}}
  \newcommand{\wheresec}[1]{\hyperref[#1]{Sec.~\ref*{#1}}}
  \begin{tabular}{@{}lll@{}}
    \toprule
    \theorycell{0.42\linewidth}{Component} &
      \theorycell{0.30\linewidth}{Status} &
      \theorycell{0.15\linewidth}{Where} \\
    \midrule
    \theorycell{0.42\linewidth}{Joint equivariance of Model~B (A1)} &
      \theorycell{0.30\linewidth}{Proved} &
      \theorycell{0.15\linewidth}{\whereapp{app:assumptions}} \\
    \theorycell{0.42\linewidth}{Rollout source generalisation $G^{\rm roll}_n$ (A2)} &
      \theorycell{0.30\linewidth}{Assumed; optimistic rate imported from the supervised equivariant class~\cite{schatzki2024theoretical}} &
      \theorycell{0.15\linewidth}{\whereapp{sec:app1}; \whereapp{app:assumptions}} \\
    \theorycell{0.42\linewidth}{$M\in O(n^3)$ sample complexity} &
      \theorycell{0.30\linewidth}{Imported; supervised $S_n$-equivariant loss class only, not the greedy rollout} &
      \theorycell{0.15\linewidth}{\whereapp{sec:app1}} \\
    \theorycell{0.42\linewidth}{Generator-difference constants $C_{\rm gen}\in\{2,4\}$ (Lemmas~\ref{lem:gen-diff-1body}--\ref{lem:unitary-dev})} &
      \theorycell{0.30\linewidth}{Proved} &
      \theorycell{0.15\linewidth}{\whereapp{app:dissimilar_derivation}} \\
    \theorycell{0.42\linewidth}{Unitary Lipschitz, linear observables (Lemma~\ref{lem:unitary-lip-app})} &
      \theorycell{0.30\linewidth}{Proved} &
      \theorycell{0.15\linewidth}{\whereapp{app:assumptions}} \\
    \theorycell{0.42\linewidth}{Expected-rollout Lipschitz $L_U^{\rm eff}$ (Prop.~\ref{prop:margin-smoothing})} &
      \theorycell{0.30\linewidth}{Proved under a bounded-margin-density condition} &
      \theorycell{0.15\linewidth}{\whereapp{app:assumptions}} \\
    \theorycell{0.42\linewidth}{Lifted-policy smoothness $\Psi=\sqrt{m}-\sqrt{n}$ (A7)} &
      \theorycell{0.30\linewidth}{Assumed; BHH-motivated} &
      \theorycell{0.15\linewidth}{\whereapp{app:assumptions}} \\
    \theorycell{0.42\linewidth}{Graphon profile (Cor.~\ref{cor:graphon})} &
      \theorycell{0.30\linewidth}{Speculative; conditional on cut-norm Lipschitz hypothesis (A6)} &
      \theorycell{0.15\linewidth}{\wheresec{sec:theory2}} \\
    \theorycell{0.42\linewidth}{DLA dimension of Model~A (Prop.~\ref{prop:dla-dim})} &
      \theorycell{0.30\linewidth}{Proved} &
      \theorycell{0.15\linewidth}{\wheresec{sec:simulability}; \whereapp{app:dla-proof}} \\
    \theorycell{0.42\linewidth}{Shot complexity of greedy resolution (Prop.~\ref{prop:shot-complexity})} &
      \theorycell{0.30\linewidth}{Proved given the explicit policy form (A8)} &
      \theorycell{0.15\linewidth}{\whereapp{app:shot-complexity}} \\
    \bottomrule
  \end{tabular}
\end{table}

\paragraph{A conditional graphon refinement.}
The lifted-policy smoothness profile $\Psi(n,m)$ admits an alternative functional form under an \emph{additional} strong assumption that we mark prominently as conditional. The graphon refinement is not a main result of this paper; we state it for completeness because it suggests how transfer might behave asymptotically if the rollout performance were globally regular as a graphon functional.

\begin{corollary}[Conditional graphon refinement of $\Psi(n,m)$]
  \label{cor:graphon}
  \emph{(Speculative; depends on Assumption~A6.)} Suppose in addition (\textbf{A6}) that the lifted-policy performance map $W_k \mapsto P_k(U_n^{\uparrow}(\theta_n^*))$, viewed as a functional of the empirical graphon $W_k$ of the instance distribution, is $C_W$-Lipschitz in the cut norm $\|\cdot\|_\square$~\cite{lovasz2012large,borgs2008convergent}. Then under Assumptions \textbf{(A1)}--\textbf{(A4)}, \textbf{(A6)}, and \textbf{(A7)} the lifted-policy smoothness profile may be taken as
  \[
    \Psi(n,m) \;\leq\; \frac{\log(1+n)}{\sqrt{n}} + \frac{\log(1+m)}{\sqrt{m}},
  \]
  which vanishes as $n,m\to\infty$.
\end{corollary}

\noindent
\emph{Proof sketch.} For $k$ i.i.d.\ uniform samples on $[0,1]^2$, the
empirical measure $\mu_k$ converges to the uniform measure $\mu$ in
Wasserstein-1 distance at rate
$\mathbb E[W_1(\mu_k,\mu)]=O(k^{-1/2}\log(1+k))$ in dimension
two~\cite[Theorem~1]{fournier2015rate} (the sharp rate for the uniform
square is $\Theta(\sqrt{\log k/k})$; we state the profile with the weaker
logarithmic factor that \cite[Theorem~1]{fournier2015rate} supports, since
only the vanishing of $\Psi$ matters for the corollary). For the 1-Lipschitz
distance kernel $W(x,y)=\|x-y\|_2$, transporting the sample points along an
optimal coupling perturbs the kernel pointwise by at most the transport
displacements, so the empirical graphon $W_k$ satisfies
$\delta_\square(W_k,W)\le\|W_k-W^{\varphi}\|_1\le 2\,\mathrm{Lip}(W)\,
W_1(\mu_k,\mu)=O(k^{-1/2}\log(1+k))$ for a suitable measure-preserving
relabelling $\varphi$. The triangle inequality through the common limit gives
$\delta_\square(W_n,W_m)\le O(n^{-1/2}\log(1+n)+m^{-1/2}\log(1+m))$; applying
the cut-norm Lipschitz functional of (A6) yields the claim. \hfill$\square$

\paragraph{A flag, not a guarantee.}
We emphasise that Assumption~(A6) --- cut-norm Lipschitz regularity of the lifted-policy rollout performance --- is \emph{strong and unproven}. The greedy rollout decoder is non-smooth (it performs sequential argmax over a finite candidate set), and we do not show that its overall performance is continuous in the empirical-graphon cut norm. Corollary~\ref{cor:graphon} should therefore be read as a conditional refinement indicating the regime in which graphon convergence \emph{would} suggest vanishing structural penalty; it is not an unconditional theorem and is not the headline content of the paper. Our default operating form for the transfer bound is Theorem~\ref{thm:final-performance} with $\Psi(n,m)=\sqrt{m}-\sqrt{n}$ under Assumption~(A7).

\begin{equation}
  \boxed{
    P_m(\theta_n^*) \ge \widehat P_n(\theta_n^*)
  - \mathcal G^{\rm roll}_n(\delta) - \mathcal D_{n\to m}}
\end{equation}

\subsection{Discussion and Implications}

The conditional diagnostic transfer bound
\[
  P_m(\theta^*_n) \;\geq\; \widehat{P}_n(\theta^*_n) - \mathcal{G}^{\rm roll}_n(\delta) - \mathcal{D}_{n \to m}
\]
identifies two distinct mechanisms of cross-size performance degradation, both \emph{conditioned} on the assumption stack above:

\begin{enumerate}
  \item \textbf{Source-task generalisation.} The term \( \mathcal{G}^{\rm roll}_n(\delta) \) arises from finite training on the source task and depends on a covering-number functional of the rollout policy class. In the optimistic symmetry-preserving case --- where the greedy decoding map inherits the equivariant covering rate of the underlying QNN class --- one expects $\mathcal{G}^{\rm roll}_n(\delta)=\widetilde{\mathcal O}(\sqrt{T_{e_{n+1}}/M_n}+\sqrt{\log(1/\delta)/M_n})$; we make this rate explicit only when we use it operationally, and the theorem itself requires only the existence of $\mathcal{G}^{\rm roll}_n$.
  \item \textbf{Transfer penalty.} The term \( \mathcal{D}_{n \rightarrow m} \) combines a parametric mismatch
    \[
      L_U\,C_{\rm gen}\,\|\theta_n^*\|_1\,\frac{m-n}{m},
    \]
    which scales linearly in the parameter $\ell_1$-norm and as $(m-n)/m$ with the size jump, and a structural smoothness term
    \[
      C_{\rm lift}\,\Psi(n,m),
    \]
    which is an \emph{assumption} on how the same lifted policy behaves at the two sizes. Under uniform Euclidean sampling the natural choice is $\Psi(n,m)=\sqrt{m}-\sqrt{n}$.
\end{enumerate}

\paragraph{What the bound is and is not.}
The bound is worst-case and can be vacuous: with $L_U=2$, $C_{\rm gen}\le 4$, and $\|\theta_n^*\|_1\sim O(1)$, the parametric term alone can exceed unity, in which case the bound returns a trivial $P_m\ge\text{negative}$. Its value is the scaling structure it identifies (linear in $\|\theta_n^*\|_1\,(m-n)/m$ and a sub-linear structural term), not numerical prediction of measured gaps. Overall, the bound formalises the intuition that larger source models trained on instances with greater $n$ exhibit stronger generalisation to larger tasks, motivating size-aware training strategies. A more detailed analysis of scalability trends, design principles for transferable QRL, and the role of symmetry and data assumptions is provided in Appendix~\ref{app:analysis}.

\section{Experimental Methodology: A Five-Stage Evaluation Pipeline}
\label{sec:methodology}

A central methodological contribution of this paper is a five-stage evaluation pipeline (Base, N1, N2, N3, N4) that selects a stable training recipe, establishes from-scratch source-policy baselines, evaluates exact cross-size parameter transfer, characterises the failure modes of large-scale MPS and noisy simulation, and finally ports the transfer evaluation to the Quantinuum H-series emulator and to Quantinuum trapped-ion hardware (H2-2 and Helios-1). The same trained EQC checkpoints flow through every stage, so backend-, simulator-, and shot-induced effects can be cleanly separated from genuine transfer behaviour.

\begin{table*}[t]
  \centering
  \caption{Backend feasibility matrix for the depth-$L=1$ EQC. A check mark indicates output within reasonable resources; parenthetical notes mark inaccurate, noise-dominated, or shot-sensitive regimes; a cross indicates infeasibility under current hardware/software budgets.}
  \label{tab:backend-feasibility}
  \small
  \setlength{\tabcolsep}{5pt}
  \begin{tabular*}{\textwidth}{@{\extracolsep{\fill}}lccccccc}
    \toprule
    $n$ & 10 & 15 & 20 & 25 & 30 & 50 & 100 \\
    \midrule
    Statevector & \cmark\,(s) & \cmark\,(s) & \cmark\,(min) & \cmark\,(hr) & \xmark\,($>$16 GB) & \xmark & \xmark \\
    MPS $\chi=64$ & \cmark & \cmark & \cmark\,\textit{(inaccurate)} & \cmark & \cmark & \cmark & \cmark \\
    MPS (high $\chi$) & \cmark & \cmark & \cmark\,(min) & \cmark\,(hr) & \cmark & \cmark\,(hr) & \cmark\,(day) \\
    Noisy MPS & \cmark\,\textit{(noise-dom.)} & \cmark\,\textit{(noise-dom.)} & \cmark & \cmark & \cmark & \cmark & \cmark \\
    H-series emulator (N4) & \cmark & \cmark\,\textit{(shot-sens.)} & \xmark & \xmark & \xmark & \xmark & \xmark \\
    H-series hardware (N4) & \cmark\,\textit{(shot-sens.)} & \xmark\,\textit{(HQC budget)} & \xmark & \xmark & \xmark & \xmark & \xmark \\
    \bottomrule
  \end{tabular*}
\end{table*}

The equivariant circuit architecture used throughout is the shared-parameter ansatz of Skolik et al.~\cite{skolik2023equivariant}, with one qubit per city, edge-weighted CX--RZ--CX entanglers across the complete graph, and node-mixing $R_X$ rotations driven by an adjacency-diagonal state encoding. At depth $L$ the QNN has exactly $2L$ scalar parameters $\theta=(\beta_1,\gamma_1,\dots,\beta_L,\gamma_L)$ independent of $n$. The circuit diagram is shown in Figure~\ref{fig:eqc-diagram}; the readout is the vector of pairwise $\langle Z_i Z_j\rangle$ observables, converted into action values for the greedy TSP rollout.

\begin{figure}[t]
  \centering
  \resizebox{0.9\linewidth}{!}{\input{images/eqc_ansatz_4}}
  \caption{The shared-parameter equivariant quantum circuit (EQC) ansatz of~\cite{skolik2023equivariant} used throughout this paper. For each edge $(i,j)\in E_k$ of the complete graph, a CX--RZ$(\gamma e_{ij})$--CX block is applied, followed by node-feature $R_x(s_i\beta)$ rotations driven by an adjacency-diagonal state encoding $s_i$. At depth $L=1$ the QNN has exactly two trainable scalar parameters $(\beta,\gamma)$ regardless of system size.}
  \label{fig:eqc-diagram}
\end{figure}

\subsection{Base Stage: Selecting a Stable Training Recipe}
\label{sec:base-recipe}

\begin{figure}[t]
  \centering
  \includegraphics[width=\linewidth]{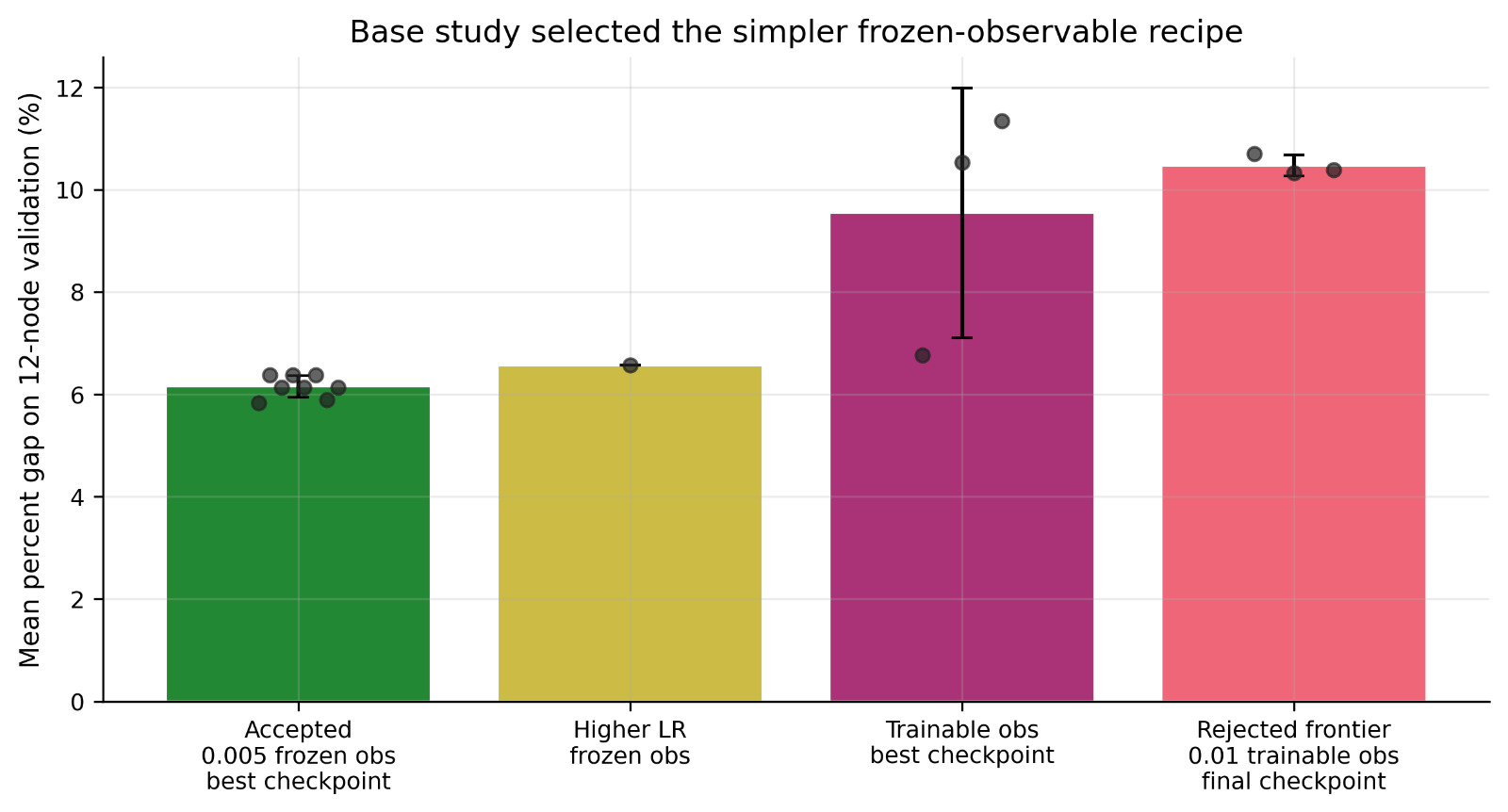}
  \caption{Base-stage recipe selection on the 12-node validation lane. The accepted recipe (learning rate $0.005$, frozen observation rescaling, best-checkpoint reporting) achieves a $\sim 6.1\%$ mean validation gap; the rejected frontier-style recipe (learning rate $0.01$, trainable observation rescaling, final checkpoint) gives $\sim 10.5\%$. Dots are individual evaluations; bars are group means with one standard deviation.}
  \label{fig:base-recipe}
\end{figure}

Before launching the transfer studies, we screened the training recipe on a frozen 12-node validation lane (100 fixed evaluation instances with stored exact optima). The screen compared learning rates, frozen vs.\ trainable observation-rescaling weights, and best-vs.-final checkpoint selection. The retained recipe is summarised in Table~\ref{tab:base-recipe}; Figure~\ref{fig:base-recipe} shows the selection criterion.

\begin{table}[t]
  \centering
  \caption{Accepted training-recipe decisions, carried into N1, N2, N3, and N4.}
  \label{tab:base-recipe}
  \small
  \resizebox{\linewidth}{!}{%
    \begin{tabular}{ll}
      \toprule
      Design choice & Accepted setting \\
      \midrule
      Optimizer / loss & Adam with MSE temporal-difference loss \\
      Learning rate & $\alpha = 0.005$ \\
      Observation rescaling & Frozen at initialisation \\
      Checkpoint for reporting & Best saved checkpoint (not final) \\
      Backend at $n\!\in\!\{10,15\}$ & PennyLane \texttt{lightning.qubit} \\
      Backend at $n=20$ & PennyLane \texttt{lightning.gpu} \\
      Backend at $n=5$ & PennyLane \texttt{default.qubit} \\
      Classical comparator & OR-Tools CP-SAT ($n\!\le\!15$) / Concorde ($n=20$) \\
      \bottomrule
  \end{tabular}}
\end{table}

The choice of frozen observation rescaling has a substantive interpretive consequence: it makes the EQC policy a low-dimensional, shared-parameter object whose transferable content lies entirely in the angles $(\beta_\ell,\gamma_\ell)$, rather than in an instance-specific classical head.

\subsection{N1: From-Scratch Source Policies}
\label{sec:n1}

The N1 stage trains and evaluates EQC source policies at fixed graph sizes. The main grid covers $n\in\{10,15\}$ at depths $L\in\{1,2,3\}$ across five seeds, plus $n=20$ at $L=1$ across five seeds (the $n=20,L\!\ge\!2$ regime was excluded because GPU compute budget was prohibitive). A later small-source lane adds $n=5$, $L=1$ for the N2--N4 source-5 transfer studies. Training uses the accepted recipe of Section~\ref{sec:base-recipe}: epsilon-greedy episodes, replay memory, target-network bootstrapping, MSE temporal-difference loss, and best-checkpoint reporting. The classical reference is exact in every case: OR-Tools CP-SAT for the 10- and 15-node generated banks, Concorde for the saved 20-node banks.

\subsection{N2: Exact Cross-Size Transfer and Fine-Tuning}
\label{sec:n2}

The N2 stage tests whether N1 checkpoints can be transferred across graph sizes using an explicit, deterministic parameter-lift rule. The shared EQC angles are size-independent and are copied directly. The classical edge-rescaling vector is the identity for equal-size transfer; for cross-size transfer, the implementation uses an edge-orbit average-lift,
\[
  T_{ij} = \tfrac{1}{m_s},\qquad m_s = n_s(n_s-1)/2,
\]
so every target edge receives the mean source rescaling weight. Because the accepted recipe freezes observation rescaling at $1.0$, the transfer signal in our completed runs comes essentially from the shared angles.

N2 has two modes. \emph{Zero-shot} mode evaluates the transferred target model immediately. \emph{Fine-tune} mode applies $\sim 2000$ additional training episodes on the reused target-size training bank before evaluation. Every transfer run is compared to the target-size N1 scratch baseline on the same target evaluation bank; transfer is meaningful only if it matches or beats that target-size scratch reference.

\subsection{N3: MPS, Noisy Simulation, and the Scaling Frontier}
\label{sec:n3}

The N3 stage extends N2 toward larger systems and hardware-facing execution. The same conceptual transfer rule is preserved; the difference is the backend. N3 supports a quimb-backed MPS simulator, Qiskit Aer exact and shot-based simulation, Qiskit Aer \texttt{matrix\_product\_state} simulation with optional bond-dimension cap, and noisy MPS simulation derived from an IBM backend noise model. Appendix~\ref{app:scalability-tn} records the tensor-network scalability assumptions and the truncation-crossover audit used to interpret these runs. Fixed instance banks at $n\!\in\!\{5,10,15,20,25,50,100\}$ are used throughout, with the bank size at large $n$ deliberately small (each greedy rollout requires many circuit executions).

\subsection{N4: Quantinuum H-series Emulator and Hardware Execution}
\label{sec:n4}

The N4 stage ports the same transferred EQC policy to the Quantinuum/Nexus execution stack: native pytket circuit construction, QIR conversion, Nexus job submission, result extraction, and adaptive shot handling. The shared EQC angles and average-lifted edge-rescaling vector are reused without modification from N1/N2. The implementation handles adaptive confidence margins, maximum step-shot caps, best-of-$k$ rollouts, and explicit job/compile accounting; these features matter because one greedy tour can take hundreds or thousands of seconds on an H-series emulator and tens of hours of queue time on real hardware.

N4 has two sub-lanes that share an identical control path. The \emph{emulator} sub-lane uses the Quantinuum H-series cloud emulator and is the regime in which we performed the shot-sensitivity sweep of Section~\ref{sec:barrier-hardware}. The \emph{hardware} sub-lane executes on the Quantinuum H2-2 and Helios-1 trapped-ion devices: every greedy action is selected from a separate hardware job (no compile fallback), with persistence after every successful decision so a multi-day campaign survives any single API or shell interruption. We report a five-instance $5\to10$ hardware run in Section~\ref{sec:barrier-hardware}: 40 completed hardware decisions, 18{,}878 HQC of Nexus-billed compute, and 311 h of campaign wall-clock. We treat the N4 results as our end-to-end execution-pipeline study and hardware-realism diagnostic, not as a quantum-advantage claim.

\subsection{Baselines and Metrics}

For every stage, the primary quality metric is the average cost ratio
\[
  r = \frac{\text{EQC tour cost}}{\text{classical reference cost}},
\]
with percent gap $100(r-1)$. A ratio of $1.0$ is an exact match to the classical reference; the \emph{exact-match rate} reports the fraction of evaluated instances where the EQC tour exactly matched the reference optimum. The classical reference is OR-Tools CP-SAT or Concorde wherever available; nearest-neighbour, greedy insertion, and Christofides are used as orientation baselines for context.

Table~\ref{tab:pipeline-audit} summarises, for every pipeline stage, what
was held fixed and which claim the stage is entitled to support; this is
the anti-confounding audit underlying the barrier attributions of
Section~\ref{sec:barriers}. Reproducibility details for the fixed instance
banks, backend versions, and artifact layout are given in
Appendix~\ref{app:reproducibility}.

\begin{table*}[t]
  \centering
  \caption{Pipeline anti-confounding audit. Every stage evaluates the same
  trained EQC checkpoints; only the listed execution variables change.
  ``Protocol-matched'' marks the stages that share the exact hardware
  execution protocol (single batch, 4096 shots per greedy decision, same
  five target instances, same code path) and are therefore directly
  comparable to the hardware campaign.}
  \label{tab:pipeline-audit}
  \footnotesize
  \setlength{\tabcolsep}{2pt}
  \renewcommand{\arraystretch}{1.15}
  \newcommand{\auditcell}[2]{\parbox[t]{#1}{\raggedright\strut #2\strut}}
  \begin{tabular}{@{}llllll@{}}
    \toprule
    \auditcell{0.10\textwidth}{Stage} &
    \auditcell{0.17\textwidth}{Checkpoints / instances} &
    \auditcell{0.15\textwidth}{Backend} &
    \auditcell{0.16\textwidth}{Shot policy} &
    \auditcell{0.12\textwidth}{Protocol-matched} &
    \auditcell{0.21\textwidth}{Claim supported} \\
    \midrule
    \auditcell{0.10\textwidth}{Base} &
    \auditcell{0.17\textwidth}{12-node validation lane} &
    \auditcell{0.15\textwidth}{statevector} &
    \auditcell{0.16\textwidth}{analytic} &
    \auditcell{0.12\textwidth}{---} &
    \auditcell{0.21\textwidth}{training-recipe selection} \\
    \auditcell{0.10\textwidth}{N1} &
    \auditcell{0.17\textwidth}{scratch, $n\in\{5,10,15,20\}$} &
    \auditcell{0.15\textwidth}{statevector} &
    \auditcell{0.16\textwidth}{analytic} &
    \auditcell{0.12\textwidth}{---} &
    \auditcell{0.21\textwidth}{source-policy baselines (Table~\ref{tab:n1-source})} \\
    \auditcell{0.10\textwidth}{N2} &
    \auditcell{0.17\textwidth}{N1 checkpoints lifted to $m$} &
    \auditcell{0.15\textwidth}{statevector} &
    \auditcell{0.16\textwidth}{analytic} &
    \auditcell{0.12\textwidth}{---} &
    \auditcell{0.21\textwidth}{in-regime transfer vs.\ scratch; B2 trend (Table~\ref{tab:n2-aggregate})} \\
    \auditcell{0.10\textwidth}{N3} &
    \auditcell{0.17\textwidth}{same checkpoints} &
    \auditcell{0.15\textwidth}{MPS ($\chi$-capped), noisy MPS} &
    \auditcell{0.16\textwidth}{analytic from MPS} &
    \auditcell{0.12\textwidth}{---} &
    \auditcell{0.21\textwidth}{B1 backend validity controls; B2 frontier (Table~\ref{tab:n3-frontier})} \\
    \auditcell{0.10\textwidth}{N4 exploratory} &
    \auditcell{0.17\textwidth}{$5\to10$ checkpoint} &
    \auditcell{0.15\textwidth}{H-series emulator} &
    \auditcell{0.16\textwidth}{heterogeneous; adaptive top-up} &
    \auditcell{0.12\textwidth}{no} &
    \auditcell{0.21\textwidth}{execution-setting sensitivity only} \\
    \auditcell{0.10\textwidth}{N4 controlled} &
    \auditcell{0.17\textwidth}{same checkpoint, same five instances} &
    \auditcell{0.15\textwidth}{H2-1LE (noiseless)} &
    \auditcell{0.16\textwidth}{4096 / 16{,}384, single batch} &
    \auditcell{0.12\textwidth}{yes} &
    \auditcell{0.21\textwidth}{B3 sampling-only component ($31.3\%$ / $21.7\%$)} \\
    \auditcell{0.10\textwidth}{N4 hardware} &
    \auditcell{0.17\textwidth}{same checkpoint, same five instances} &
    \auditcell{0.15\textwidth}{H2-2, Helios-1} &
    \auditcell{0.16\textwidth}{4096, single batch} &
    \auditcell{0.12\textwidth}{yes} &
    \auditcell{0.21\textwidth}{B3 end-to-end ($45.3\%$)} \\
    \auditcell{0.10\textwidth}{Replay} &
    \auditcell{0.17\textwidth}{same checkpoint, hardware-visited states} &
    \auditcell{0.15\textwidth}{statevector} &
    \auditcell{0.16\textwidth}{exact} &
    \auditcell{0.12\textwidth}{yes (counterfactual)} &
    \auditcell{0.21\textwidth}{B3 mechanism: noise-flipped decisions ($2.7\%$)} \\
    \bottomrule
  \end{tabular}
\end{table*}

\section{Validated Source Policies and In-Regime Transfer}
\label{sec:empirical-eval}

This section reports the portion of our experimental findings in which the trained policies behave \emph{consistently with} the conditional diagnostic transfer bound of Section~\ref{sec:theory2}: same-size generalisation at $n\le 20$ and small-shift cross-size transfer within $\{5,\dots,20\}$. Recall that the bound is worst-case, loose, and identifies scaling structure rather than predicting gaps numerically; the observations here lie in the qualitative regime suggested by the bound (small size shifts, small parameter norms, modest backend penalty), not in a regime where the bound is being numerically validated. The frontier behaviour exposed by larger size jumps and by the MPS/emulator backends is treated separately in Section~\ref{sec:barriers}.

\subsection{N1 Source-Task Performance}

\begin{figure}[t]
  \centering
  \includegraphics[width=0.95\linewidth]{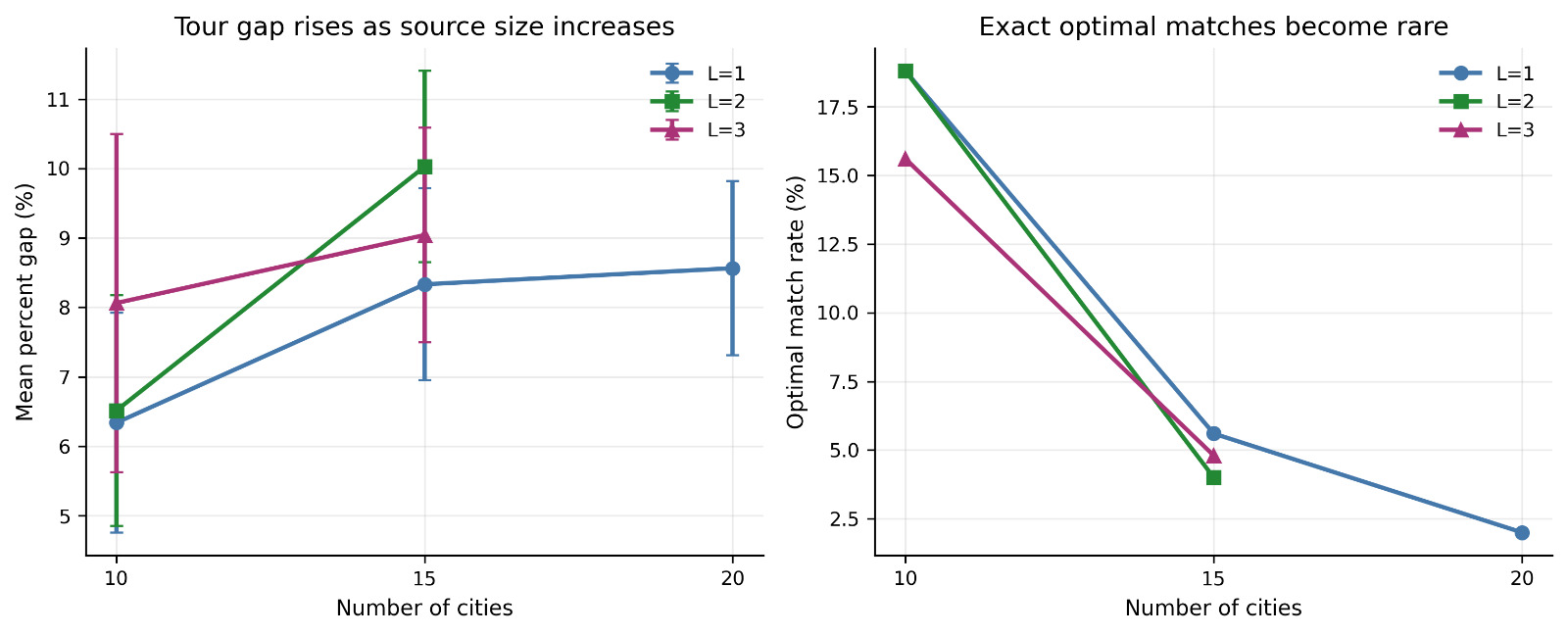}
  \caption{N1 source-policy scaling. Left: mean percent gap by graph size $n$ and depth $L$. Right: exact optimal-match rate. The trained policy remains useful as $n$ grows but becomes much less likely to produce a tour that exactly matches the classical optimum --- 18.8\% at $n=10$, 5.6\% at $n=15$, and only 2.0\% at $n=20$. Depth does not monotonically improve quality: at $n=10$, $L=1$ and $L=2$ are essentially tied and both beat $L=3$; at $n=15$, $L=1$ is the strongest.}
  \label{fig:n1-scaling}
\end{figure}

Table~\ref{tab:n1-source} summarises the from-scratch N1 evaluation. The depth-$L\in\{1,2,3\}$ policies at $n\in\{10,15\}$ are aggregated over five seeds and 50 instances per seed; the $n=20$ row is at $L=1$ over five seeds with $10$ instances per seed (the $L=2,3$ runs at $n=20$ were excluded for GPU-compute reasons). Figure~\ref{fig:n1-scaling} shows the same data graphically together with the optimal-match rate.

\begin{table}[t]
  \centering
  \caption{N1 from-scratch source-policy performance. Means and standard deviations are over completed seeds. Optimality gap is relative to OR-Tools CP-SAT (for $n=10,15$) or Concorde (for $n=20$). The exact-match rate falls sharply with $n$, while the mean gap rises but stays in the single-to-low-double-digit range.}
  \label{tab:n1-source}
  \small
  \begin{tabular}{lccccc}
    \toprule
    $n$ & $L$ & Runs & Instances/run & Mean gap (\%) & Match rate \\
    \midrule
    10 & 1 & 5 & 50 & $6.34\pm 1.58$ & 18.8\% \\
    10 & 2 & 5 & 50 & $6.51\pm 1.66$ & 18.8\% \\
    10 & 3 & 5 & 50 & $8.06\pm 2.43$ & 15.6\% \\
    15 & 1 & 5 & 50 & $8.33\pm 1.38$ & 5.6\% \\
    15 & 2 & 5 & 50 & $10.03\pm 1.38$ & 4.0\% \\
    15 & 3 & 5 & 50 & $9.04\pm 1.55$ & 4.8\% \\
    20 & 1 & 5 & 10 & $8.56\pm 1.26$ & 2.0\% \\
    \bottomrule
  \end{tabular}
\end{table}

Two observations are important for the interpretation of later transfer experiments. First, deeper EQC circuits do not monotonically improve greedy policy quality in this implementation. At $n=10$ the $L=1$ and $L=2$ policies are statistically indistinguishable and both outperform $L=3$. At $n=15$ the $L=1$ policy is the strongest. This supports concentrating the transfer and hardware-facing work on $L=1$ circuits, where the parameter count is exactly two scalars regardless of $n$. Second, the rapid drop in exact-match rate (from 18.8\% at $n=10$ down to 2.0\% at $n=20$) shows that the policy is finding near-optimal but not optimal tours as $n$ grows; the parameter efficiency of the EQC architecture comes with a corresponding limitation on tour fidelity.

\subsection{N2 Cross-Size Transfer in \texorpdfstring{$\{5,\dots,20\}$}{5 to 20}}

\begin{figure}[t]
  \centering
  \includegraphics[width=0.95\linewidth]{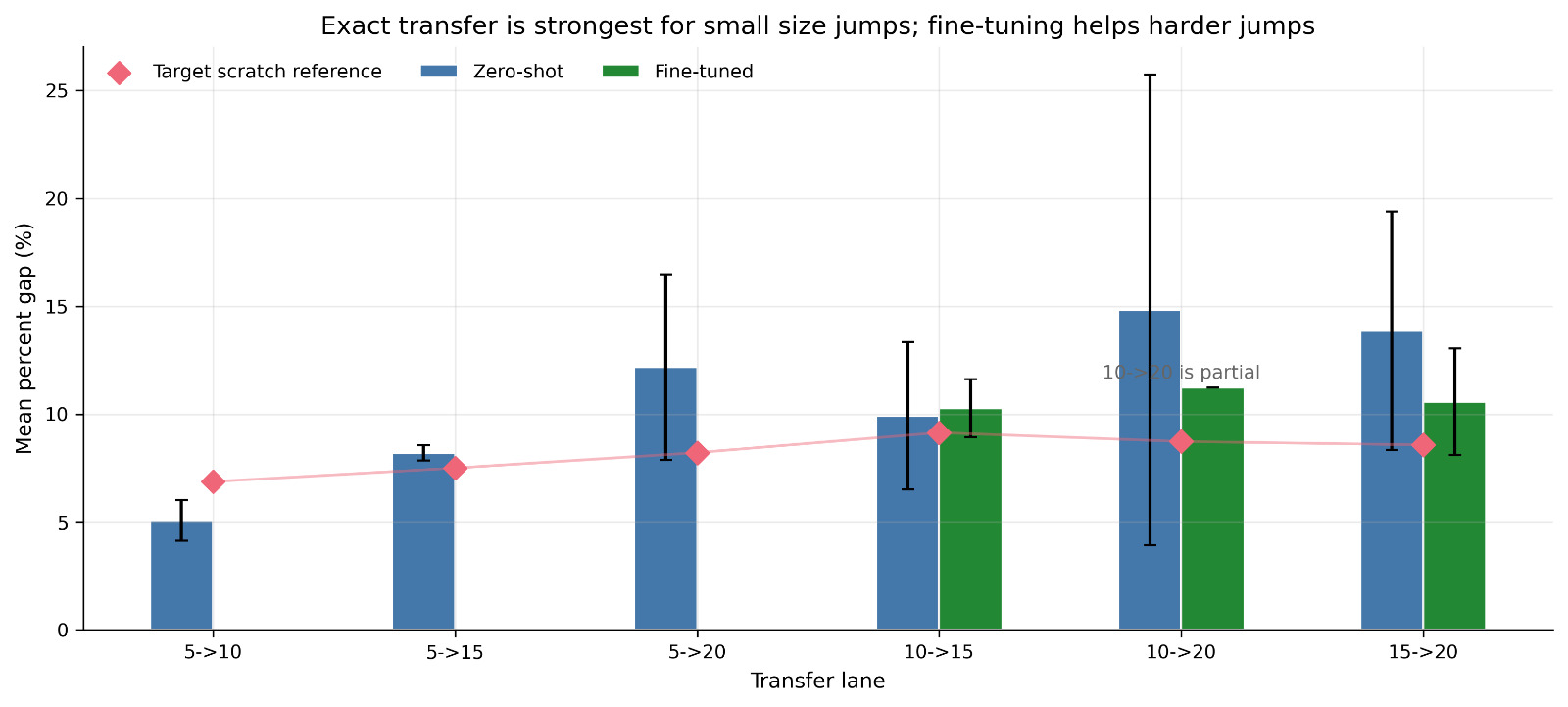}
  \caption{N2 transfer compared with target-size scratch references. Bars show mean transfer gap, with error bars indicating standard deviation across completed runs; diamonds show the mean N1 scratch gap on the same target bank. Zero-shot transfer is strongest for small size jumps ($5\to10$), competitive at moderate jumps ($10\to15$), and under-adapted at larger jumps ($15\to20$, $10\to20$, $5\to20$). Fine-tuning closes a substantial fraction of the zero-shot degradation at $15\to20$ (13.9\% $\to$ 10.6\%), but does not consistently beat scratch.}
  \label{fig:n2-transfer}
\end{figure}

The N2 stage transfers each N1 source checkpoint to each larger target size using the average-lift rule (Section~\ref{sec:n2}), and compares the result to the target-size N1 scratch baseline on the \emph{same} evaluation bank. Table~\ref{tab:n2-aggregate} aggregates the completed transfer runs; Figure~\ref{fig:n2-transfer} visualises the same data side-by-side with the target-size scratch reference.

\begin{table*}[t]
  \centering
  \caption{N2 aggregate transfer performance by lane and method. ``Better than scratch'' counts completed transfer evaluations that beat the target-size N1 scratch reference on the same target bank. The strongest positive result is $5\to10$ zero-shot, where all six completed evaluations beat scratch.}
  \label{tab:n2-aggregate}
  \small
  \setlength{\tabcolsep}{6pt}
  \begin{tabular*}{\textwidth}{@{\extracolsep{\fill}}lcrrrc}
    \toprule
    Lane & Method & Runs & Mean ratio & Mean gap & Better than scratch \\
    \midrule
    $5\to10$  & zero-shot &  6 & $1.0507\pm 0.0094$ &  5.07\% & 6/6 \\
    $5\to15$  & zero-shot &  6 & $1.0820\pm 0.0035$ &  8.20\% & 2/6 \\
    $5\to20$  & zero-shot &  3 & $1.1218\pm 0.0431$ & 12.18\% & 0/3 \\
    $10\to15$ & zero-shot & 15 & $1.0992\pm 0.0342$ &  9.92\% & 7/15 \\
    $10\to15$ & fine-tune & 15 & $1.1027\pm 0.0136$ & 10.27\% & 4/15 \\
    $10\to20$ & zero-shot &  2 & $1.1483\pm 0.1092$ & 14.83\% & 0/2 \\
    $10\to20$ & fine-tune &  1 & $1.1123$           & 11.23\% & 0/1 \\
    $15\to20$ & zero-shot & 15 & $1.1386\pm 0.0553$ & 13.86\% & 1/15 \\
    $15\to20$ & fine-tune & 15 & $1.1056\pm 0.0247$ & 10.56\% & 4/15 \\
    \bottomrule
  \end{tabular*}
\end{table*}

The N2 story is one of \emph{conditional} transferability rather than universal positive transfer. The exact transfer map works best when source and target sizes are close enough that the shared EQC angles remain useful and the cross-size edge-rescaling lift remains a reasonable approximation. The $5\to10$ result is the strongest positive evidence: zero-shot transfer beats target-size scratch in all six completed evaluations and achieves a mean gap of $5.07\%$. The $10\to15$ lane is competitive but not consistently dominant. The $15\to20$ lane shows the clearest value of fine-tuning: zero-shot transfer alone gives a 13.86\% mean gap, while $\sim 2000$ episodes of fine-tuning bring it to 10.56\%, recovering most of the degradation even though scratch remains slightly better on average. For the larger jumps $5\to20$ and $10\to20$, zero-shot transfer is clearly under-adapted (gaps of 12--15\%) and the fine-tuned data is too sparse to draw firm conclusions; we mark the $10\to20$ fine-tune row in the figure as ``partial'' to indicate this.

\subsection{Theoretical Performance Degradation and Bound Validation}
\label{sec:bound-validation}

To connect theory with these experiments, we operationalise the transfer penalty $\mathcal{D}_{n\to m}$ of Theorem~\ref{thm:final-performance} by estimating its constants from the N2 lanes (detailed methodology in Appendix~\ref{app:perf-degradation}). Across the transfer settings in Table~\ref{tab:n2-aggregate}, the zero-shot performance of the equivariant policies remains within the qualitative regime suggested by the bound: penalty grows with $|n-m|/\max(n,m)$ as the parametric term predicts. We do not claim sharp quantitative agreement; consistent with most worst-case transfer bounds~\cite{tripuraneni2020theory,zhang2025pessimismprincipleeffectiveframework} our conditional diagnostic bound is loose at moderate $|n-m|$ and small $L$, and its value lies in identifying the right scaling structure (linear in $\|\theta_n^*\|_1\,(m-n)/m$, plus a sub-linear lifted-policy smoothness term) rather than in producing exact predictions.

\begin{figure}[t]
  \centering
  \includegraphics[width=\linewidth]{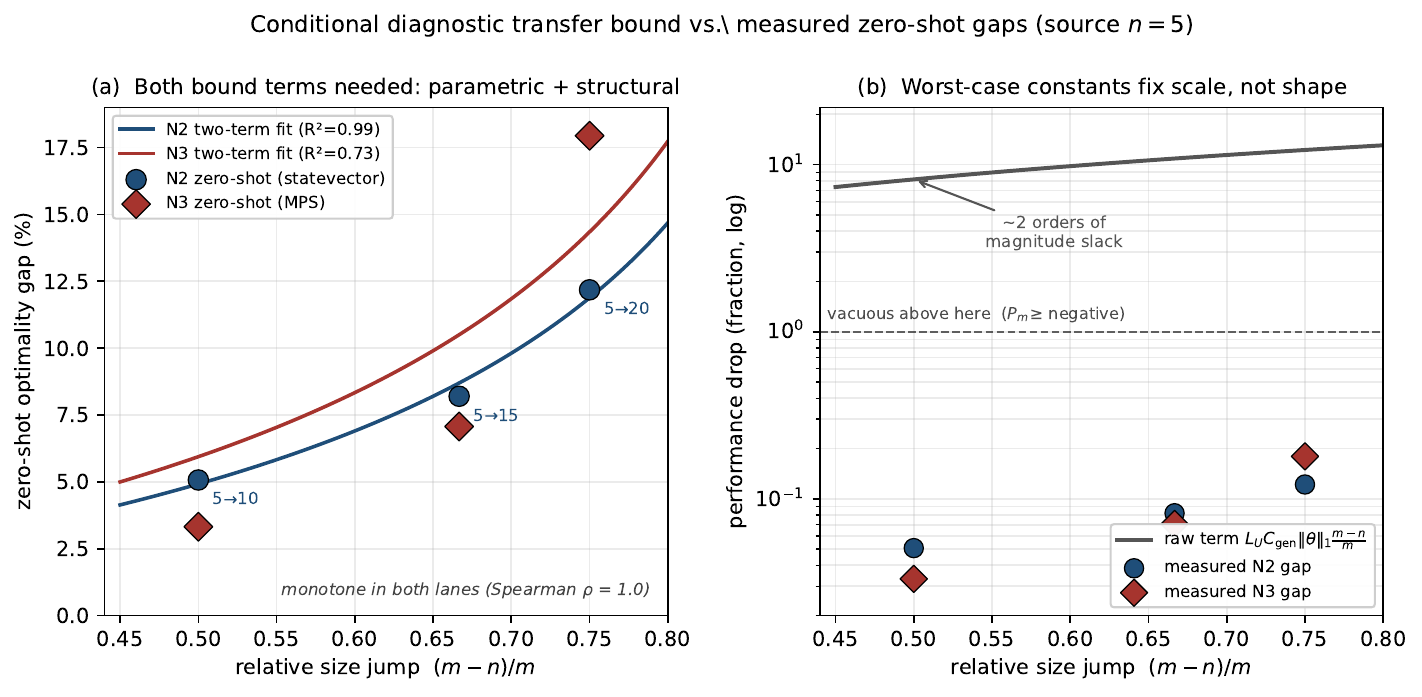}
  \caption{Conditional diagnostic transfer bound versus measured zero-shot
  gaps for source size $n=5$. \textbf{(a)} Zero-shot optimality gap as a
  function of the relative size jump $(m-n)/m$ for the N2 statevector lane
  (circles) and the N3 MPS lane (diamonds), together with a non-negative
  two-term fit to the structural form
  $a(m-n)/m+b(\sqrt{m}-\sqrt{n})$. In both lanes the gap is monotone in
  $(m-n)/m$ (Spearman $\rho=1.0$), and the fitted trend is dominated by
  the lifted-policy smoothness term $\sqrt{m}-\sqrt{n}$, with
  $C_{\rm lift}\approx 0.05$--$0.06$. This is consistent with
  Assumption~(A7) as the operative regularity condition, but the
  single-source lane makes the parametric and structural terms nearly
  collinear, so the parametric coefficient is not separately identifiable.
  \textbf{(b)} The same data plotted against the worst-case parametric term,
  evaluated with $L_U=2$, $C_{\rm gen}=4$, and $\|\theta\|_1=2.04$. The term
  exceeds one for every jump, so the raw bound is numerically vacuous and is
  used only to motivate the fitted scaling. The constants $C_{\rm lift}$ and
  $\mathcal{G}^{\rm roll}_n$ are assumed, not measured; the figure tests the
  predicted shape of the bound rather than its absolute value.}
  \label{fig:bound-diagnostic}
  
\end{figure}

Figure~\ref{fig:bound-diagnostic} makes this explicit: the measured gap is
monotone in $(m-n)/m$ in both the statevector and MPS lanes (Spearman
$\rho=1.0$), and a non-negative fit to the bound's two-term structural form
reproduces the gap shape ($R^2 = 0.99$ statevector, $R^2 = 0.73$ MPS), carried by
the $\sqrt{m} - \sqrt{n}$ lifted-policy-smoothness term. We stress that the fitted
proportionality is descriptive: with worst-case constants the bound is
numerically vacuous (Figure~\ref{fig:bound-diagnostic}b), so the agreement
is one of scaling shape, not of absolute magnitude — exactly the
diagnostic role we claim for it.

The fine-tuning observation is consistent with the same bound: a transferred policy near the target-task optimum requires only a small step in parameter space, which is exactly what 2000 fine-tuning episodes deliver. The model thus serves as an excellent initialisation in parameter space, even when zero-shot transfer alone is insufficient.

\paragraph{A note on the historical small-scale baseline.}
For completeness, we report in Appendix~\ref{app:legacy-eff-su2} a comparison of the permutation-equivariant policy against an Efficient~SU(2) baseline on small instances ($n\le 15$) drawn from an earlier non-Euclidean weight distribution. The substantive evaluation of this paper is at the scales reported in Sections~\ref{sec:empirical-eval} and~\ref{sec:barriers}; the legacy comparison is preserved for historical record only.

% =============================================================================
\section{The Three Barriers: Where Transfer Stops Working}
\label{sec:barriers}

\begin{figure}[t]
  \centering
  \includegraphics[width=0.95\linewidth]{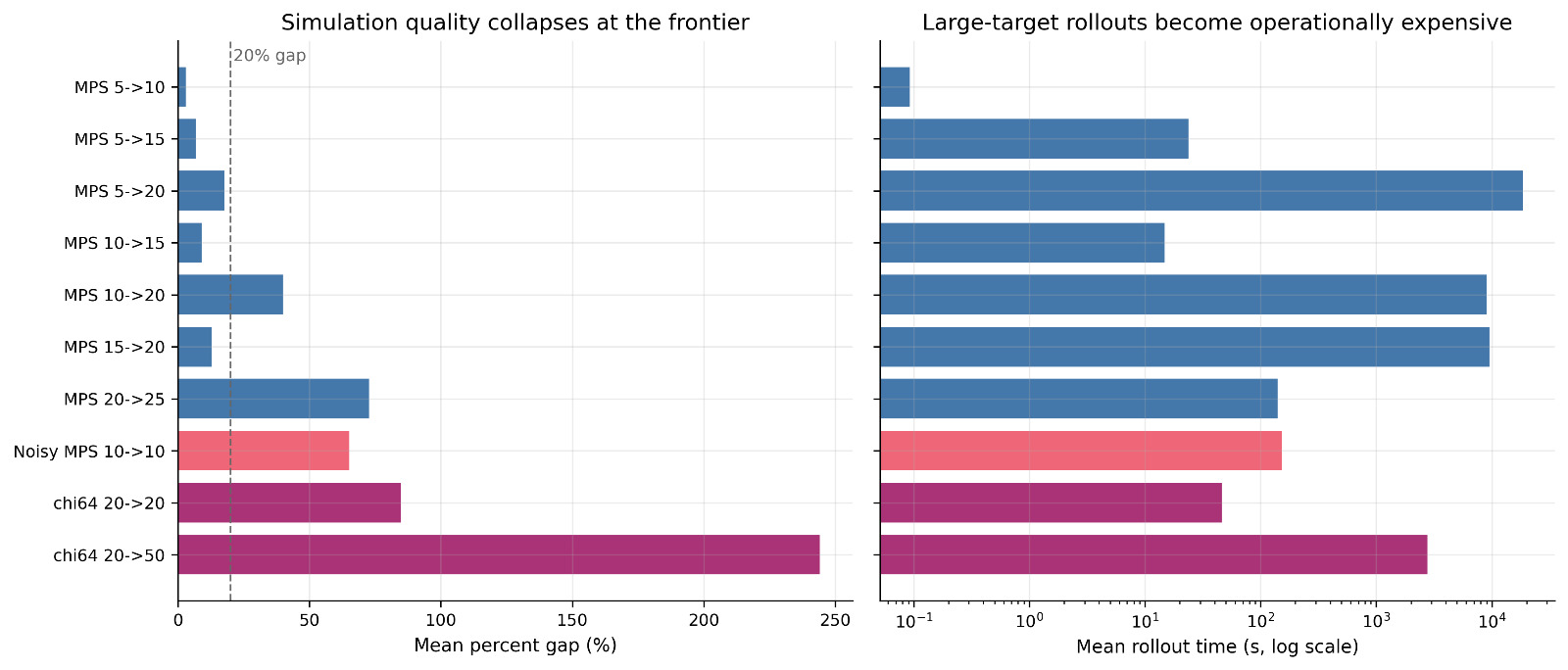}
  \caption{N3 simulation frontier. Left: mean percent gap by lane and backend mode. The MPS simulator is reliable for small-jump lanes (e.g.\ $5\to10$ at 3.32\% mean gap) and degrades smoothly through $15\to20$ at 13.22\%, but quality collapses at $10\to20$ (40.32\%), $20\to25$ (72.93\%), and on the $\chi=64$ frontier lanes (85--245\%). Right: mean rollout time on a log scale. Beyond $\sim 10\!\to\!20$, MPS lanes also become operationally expensive, with rollouts in the $10^3$--$10^4$~s range per instance.}
  \label{fig:n3-frontier}
\end{figure}

The N3 frontier experiments and the N4 emulator runs together expose three distinct failure modes that prevent the validated transfer behaviour of Section~\ref{sec:empirical-eval} from extending to industrially relevant sizes ($n\gtrsim 30$). Figure~\ref{fig:n3-frontier} shows the quantitative anchor for two of them. The three barriers below have distinct experimental signatures but a common topological cause: the $\binom{n}{2}L$ all-to-all two-qubit gate count of the canonical EQC ansatz.

\subsection{Barrier B1: Backend-Induced Signal Loss}
\label{sec:barrier-backend}

The most direct experimental result in the N3 set is the $\chi=64$ MPS control at equal size: $20\to20$ produces a mean gap of $85.22\%$ (Table~\ref{tab:n3-frontier}). Because equal-size transfer is the identity map on EQC parameters and the average-lift on the edge-rescaling vector reduces to the identity for $m_s=m_t$, this number cannot reflect transfer failure --- the source and target policies are the same object. \emph{It is a measurement of the simulator}. We therefore label this row a \emph{backend validity control}
(Table~\ref{tab:n3-frontier}): it bounds simulator fidelity for this ansatz
and carries no information about transfer. The MPS backend at $\chi=64$ corrupts the $\langle Z_iZ_j\rangle$ readout sufficiently that the greedy action ranking becomes unreliable.

\begin{table*}[t]
  \centering
  \caption{Selected N3 frontier outcomes. The $\chi=64$ equal-size $20\to20$ control gap is the cleanest backend-fidelity diagnostic; Noisy-MPS at $10\to10$ gives the analogous diagnostic for backend noise.}
  \label{tab:n3-frontier}
  \footnotesize
  \setlength{\tabcolsep}{4pt}
  \begin{tabular}{llrl}
    \toprule
    Mode & Lane & Mean gap & Interpretation \\
    \midrule
    MPS simulator       & $5\to10$  &  3.32\%  & strong small-jump control \\
    MPS simulator       & $5\to15$  &  7.07\%  & competitive small-jump \\
    MPS simulator       & $5\to20$  & 17.94\%  & moderate degradation \\
    MPS simulator       & $10\to15$ &  9.38\%  & competitive transfer \\
    MPS simulator       & $10\to20$ & 40.32\%  & large size jump degrades \\
    MPS simulator       & $15\to20$ & 13.22\%  & moderate degradation \\
    MPS simulator       & $20\to25$ & 72.93\%  & frontier instability \\
    Noisy MPS           & $10\to10$ & 65.53\%  & noise corrupts equal-size policy \\
    MPS $\chi=64$       & $20\to20$ & 85.22\%  & \textbf{truncation failure control} \\
    MPS $\chi=64$       & $20\to50$ & 244.57\% & not a valid noiseless proxy \\
    \bottomrule
  \end{tabular}
\end{table*}

\paragraph{Why this happens.}
The EQC at depth $L=1$ applies $\binom{n}{2}$ CX--RZ--CX edge blocks. At $n=20$ this is $190$ entangling blocks acting on the complete graph; the resulting state has entanglement that is uniformly distributed across all bipartitions of the qubit chain, including bipartitions at MPS-chain distance $\Omega(n)$ from the qubits actually being correlated. Bond dimension growth is therefore not suppressed by locality, and faithful representation of the state requires $\chi$ that scales (worst-case) as $2^{n/2}$. At $n=20$, $\chi=64$ is far below the threshold; the truncation error is comparable to the $\langle Z_iZ_j\rangle$ signal magnitude itself.

\paragraph{Noisy simulation tells the same story.}
Noisy MPS simulation at equal size ($10\to10$, mean gap $65.53\%$) gives the analogous diagnostic for IBM-style hardware noise: even when there is no transfer at all, the backend can corrupt the action-value ranking sufficiently to destroy policy quality. The conclusion is the same: \emph{backend fidelity is a primary failure mode before transfer is even tested}.

\paragraph{Hardware projection.}
Translating to real hardware: the EQC at $n=20$ requires $\binom{20}{2}=190$ logical two-qubit gates. On the heavy-hex topology of current IBM processors, SWAP transpilation inflates this further (factor $\sim 2$--$4$). At a typical two-qubit error rate $\varepsilon_{2Q}\!\approx\!5\!\times\!10^{-3}$, the expected process fidelity is $F_{20}\sim 0.4$, $F_{25}\sim 0.15$, $F_{30}\sim 0.04$; the wall is at $n\sim 30$ for the unmodified ansatz.

\subsection{Pinning Down the MPS Truncation Crossover}
\label{sec:n3-chi-sweep}

\begin{figure}[t]
  \centering
  \includegraphics[width=0.95\linewidth]{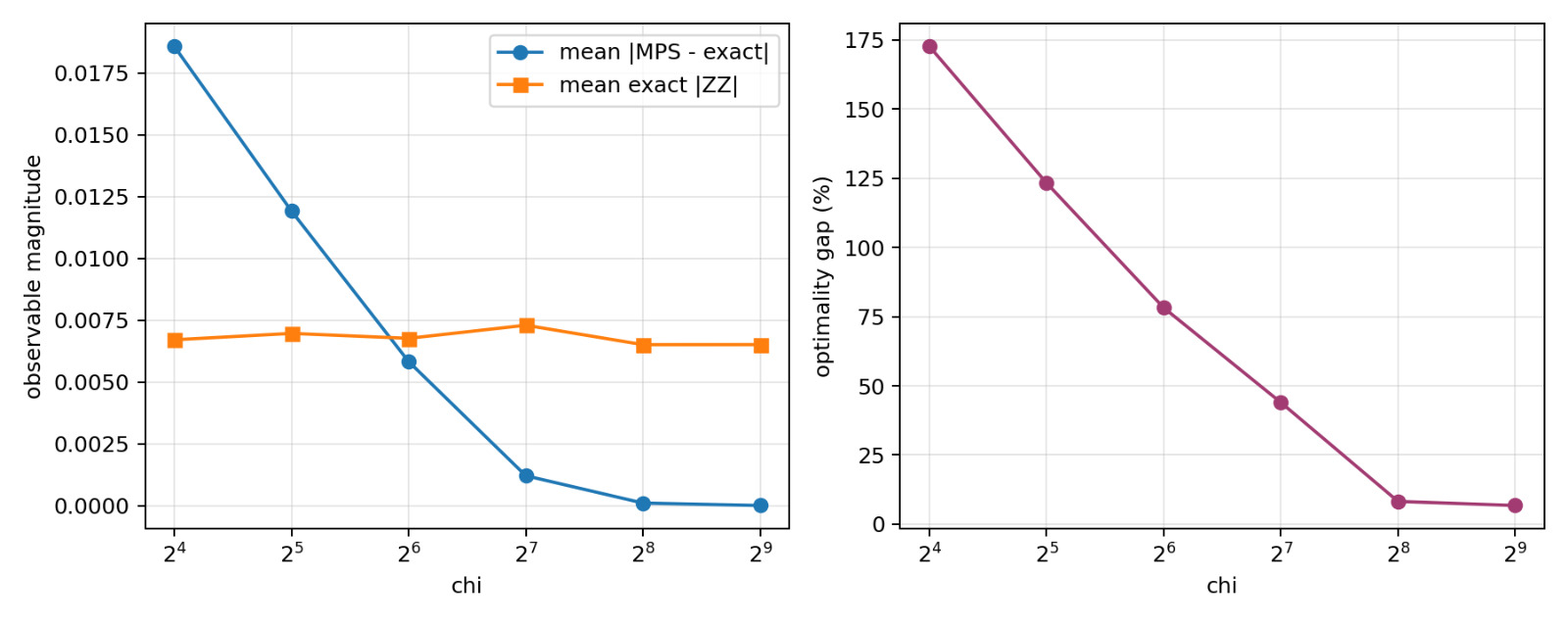}
  \caption{N3 MPS bond-dimension sweep on a single fixed $20\to 20$ instance, same $n=20$, $L=1$ seed-11235813 checkpoint as the source policy. \textbf{Left:} mean per-edge truncation error $|\langle Z_iZ_j\rangle_{\text{MPS}} - \langle Z_iZ_j\rangle_{\text{exact}}|$ (blue) and the mean exact $|\langle Z_iZ_j\rangle|$ signal magnitude (orange), the latter essentially $\chi$-independent at $\sim 6.5\!\times\!10^{-3}$ because it reflects the trained policy on a fixed instance, not the simulator. The truncation error crosses below signal magnitude between $\chi=64$ and $\chi=128$. \textbf{Right:} optimality gap relative to the Concorde optimum, falling sharply between $\chi=128$ ($44.0\%$) and $\chi=256$ ($8.2\%$).}
  \label{fig:n3-chi-sweep}
\end{figure}

The single $\chi=64$ data point of Table~\ref{tab:n3-frontier} is enough to demonstrate that B1 is real, but does not pinpoint \emph{where} the MPS backend transitions from unreliable to reliable for this ansatz. To address this we ran a controlled bond-dimension sweep at $\chi\in\{16, 32, 64, 128, 256, 512\}$ on a fixed equal-size $20\to 20$ instance, using the same $n=20$, $L=1$ source checkpoint (seed 11235813) and the canonical \texttt{n3\_euclidean\_n20\_k10\_seed44\_v1} instance bank. At every greedy decision step the run also performs an analytic statevector audit of the same circuit, so for each $\chi$ value we report both policy quality and the observable-level truncation error that the policy actually saw during rollout. Because equal-size transfer is the identity map on EQC parameters, any non-zero gap in this run is a measurement of the simulator, not of transfer. Results are summarised in Table~\ref{tab:n3-chi-sweep} and Figure~\ref{fig:n3-chi-sweep}.

\begin{table*}[t]
  \centering
  \caption{N3 MPS bond-dimension sweep on a fixed $20\to 20$ instance. Gap is relative to Concorde; observable errors are per-edge statistics over greedy-decision states; wall time is total per-$\chi$ rollout time on one CPU.}
  \label{tab:n3-chi-sweep}
  \small
  \setlength{\tabcolsep}{5pt}
  \begin{tabular*}{\textwidth}{@{\extracolsep{\fill}}rrrrrrr}
    \toprule
    $\chi$ & Gap (\%) & Mean $|\text{MPS}-\text{exact}|$ & Max step error & Mean exact $|ZZ|$ & Rel.\ error / signal & Wall time (min) \\
    \midrule
    16  & 172.60 & $1.86\times 10^{-2}$ & $2.44\times 10^{-1}$ & $6.71\times 10^{-3}$ & $4.338$ & 1.73 \\
    32  & 123.25 & $1.19\times 10^{-2}$ & $1.27\times 10^{-1}$ & $6.96\times 10^{-3}$ & $2.924$ & 1.76 \\
    64  & 78.18  & $5.82\times 10^{-3}$ & $4.61\times 10^{-2}$ & $6.76\times 10^{-3}$ & $1.453$ & 2.28 \\
    128 & 43.96  & $1.21\times 10^{-3}$ & $9.38\times 10^{-3}$ & $7.29\times 10^{-3}$ & $0.244$ & 6.49 \\
    256 & 8.22   & $1.00\times 10^{-4}$ & $1.05\times 10^{-3}$ & $6.51\times 10^{-3}$ & $0.028$ & 39.03 \\
    512 & 6.76   & $1.74\times 10^{-6}$ & $1.73\times 10^{-5}$ & $6.51\times 10^{-3}$ & $0.000$ & 171.93 \\
    \bottomrule
  \end{tabular*}
\end{table*}

\paragraph{Signal-level and policy-level crossovers are separated.}
At $\chi=64$ the mean truncation error is already comparable to the mean signal (relative error/signal $=1.45$), and the policy is badly affected: $78.2\%$ gap. Going to $\chi=128$ brings the mean error to a quarter of the signal magnitude (ratio $0.24$), but the policy has only partially recovered: the gap is still $44.0\%$ and the worst step error is $9.4\!\times\!10^{-3}$, large enough to flip individual greedy decisions. The policy gap collapses sharply between $\chi=128$ and $\chi=256$: at $\chi=256$ the maximum per-step error is $1.0\!\times\!10^{-3}$, the mean relative error/signal ratio is $0.028$, and the gap is $8.2\%$. For this instance $\chi=256$ is the practical crossover; $\chi=512$ reduces observable error by another two orders of magnitude but only improves the gap modestly to $6.8\%$. The point is methodological: \emph{signal-level accuracy ($\chi\!\sim\!128$) and policy-level accuracy ($\chi\!\sim\!256$) are not the same threshold} for this ansatz, and naive truncation-error reporting would over-state the usability of the lower bond dimensions.

\paragraph{Average accuracy is not sufficient; worst-step accuracy is.}
The operationally important quantity is not the mean truncation error but the maximum per-step error relative to the signal. At $\chi=128$, the mean error is already a factor of six below the signal, but the worst step has an error that is comparable to several edges' worth of $|\langle Z_iZ_j\rangle|$ signal --- enough to corrupt the greedy action ranking at that step and propagate downstream through the rollout. This is the precise sense in which a polynomial-in-$n$ MPS bond dimension does not save the simulation: even when the \emph{average} simulator quality looks acceptable, individual greedy decisions can be ranked incorrectly and the resulting tour can be far from optimal.

\paragraph{Wall-time growth is severe at the threshold of usefulness.}
The wall time grows from $\sim 2$ minutes at $\chi=64$ to $\sim 39$ minutes at $\chi=256$ and $\sim 172$ minutes at $\chi=512$ for a single greedy rollout on a single instance. A standard 100-instance evaluation bank at $\chi=512$ would therefore require roughly $290$ CPU-hours per source/target lane and per seed. The MPS backend reaches policy-relevant accuracy only at a bond dimension that is itself prohibitively expensive for the validation campaigns that motivated MPS in the first place.

\paragraph{Consistency with the original \texorpdfstring{$\chi=64$}{chi=64} control.}
The earlier $\chi=64$ batch result of $85.22\%$ gap in Table~\ref{tab:n3-frontier} and the present single-instance $78.18\%$ at the same bond dimension are not contradictory: they are different fixed instances drawn from related canonical Euclidean banks. Both sit comfortably in the regime where the truncation error exceeds the signal magnitude and the policy is meaningfully disrupted. The chi-sweep adds the methodological observation that this regime is broad: it extends from $\chi=16$ to $\chi=128$ before the policy begins to recover, and only collapses to near-exact statevector quality at $\chi\ge 256$.

\paragraph{Implication for VQA methodology.}
We recommend the following practice for any large-$n$ MPS-based claim about an all-to-all variational ansatz: report an explicit truncation-error / signal-magnitude audit at the actual policy-decision states, not merely a global state-norm overlap or a single mean fidelity number. The procedure used here --- an analytic statevector audit of the same circuits at the same greedy decision points, with both mean and max per-step errors reported alongside the policy outcome --- is one such audit. We provide the procedure (and the underlying per-step data) as a methodological contribution alongside the diagnostic result itself, and the full per-step CSV is available in the released artifact (Appendix~\ref{app:n3-extended}).

\subsection{Barrier B2: Cross-Size Transfer Degradation Under Large Jumps}
\label{sec:barrier-jump}

Even with a clean backend, larger transfer jumps degrade smoothly but substantially. The N3 MPS simulator lanes give the clearest view because MPS is reliable in the small-jump regime where statevector ground truth exists. The trend is monotonic: $5\to10$ at $3.32\%$, $5\to15$ at $7.07\%$, $5\to20$ at $17.94\%$ (Table~\ref{tab:n3-frontier}). The same trend appears in the N2 data (Table~\ref{tab:n2-aggregate}) with slightly different aggregation: $5\to10$ at $5.07\%$, $5\to15$ at $8.20\%$, $5\to20$ at $12.18\%$.

\paragraph{Connection to the transfer bound.}
Theorem~\ref{thm:final-performance} predicts a parametric penalty proportional to $(m-n)/m$ for $m>n$. For $5\to10$ this factor is $0.5$; for $5\to15$, $0.67$; for $5\to20$, $0.75$. The qualitative ordering matches; the bound is loose in absolute magnitude (and we do not claim otherwise), and we report it as a conditional diagnostic decomposition rather than a sharp predictor. The lifted-policy smoothness profile $\Psi(n,m)=\sqrt{m}-\sqrt{n}$ (Assumption~(A7)) supplies the remaining sub-linear structural growth. Corollary~\ref{cor:graphon}'s conditional graphon refinement would predict $\Psi(n,m)\to 0$ as $n,m\to\infty$, but that refinement depends on an unproven cut-norm Lipschitz hypothesis on the rollout performance, and at the small sizes accessible in N1--N3 the empirical-graphon convergence rate is itself large; we cannot disentangle the parametric and structural contributions cleanly with the present data.

\paragraph{Fine-tuning helps for harder jumps.}
The N2 fine-tuning data at $15\to20$ (zero-shot $13.86\%$, fine-tuned $10.56\%$) confirms that the transferred policy serves as a useful parameter-space initialisation even when zero-shot alone is insufficient. This is consistent with the bound being qualitative rather than sharp: the trained source angles are within the basin of attraction of a good target-task minimum, and fine-tuning converges to it quickly.

\subsection{Barrier B3: Hardware-Style Execution Penalty}
\label{sec:barrier-hardware}

\begin{figure}[t]
  \centering
  \includegraphics[width=0.85\linewidth]{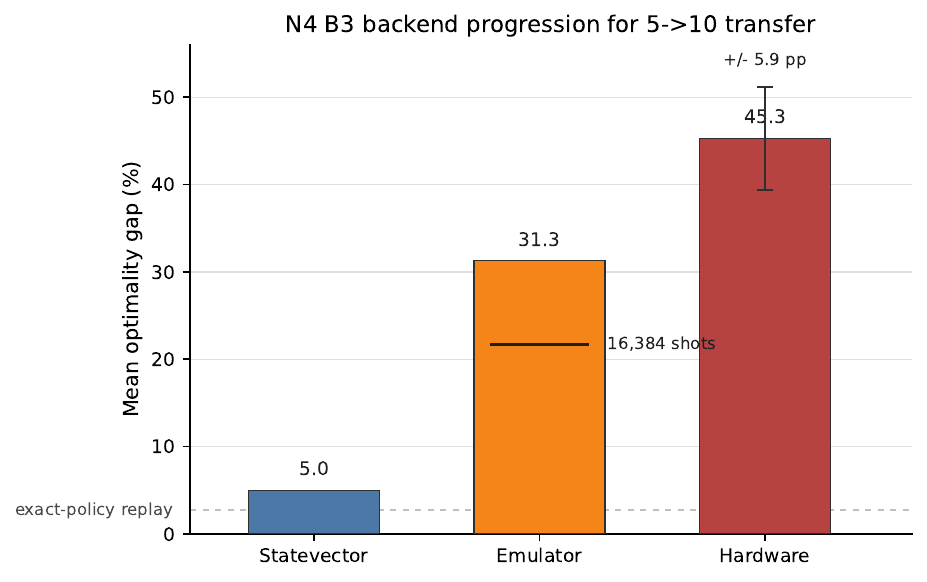}
  \caption{The B3 execution-penalty progression for the same $5\to 10$ EQC
checkpoint under the identical protocol (4096 single-batch shots per greedy
decision, no adaptive top-up). Local statevector evaluation (N2) gives a
${\sim}5\%$ mean gap; the noiseless-state Quantinuum emulator gives $31.3\%$,
isolating the pure sampling-noise component; end-to-end execution on the
Quantinuum trapped-ion hardware gives $45.3\%$
($\sigma=5.9$~pp over five 10-node instances; error bar shows the sample
standard deviation). Raising the emulator budget to 16{,}384 shots per
decision improves the mean only to $21.7\%$, consistent with the
${\approx}3.8\times 10^4$ per-decision requirement of
Proposition~\ref{prop:shot-complexity}. No backend changes the underlying
trained $(\beta,\gamma)$ angles; the counterfactual replay of
Section~\ref{sec:barrier-hardware} recovers a $2.7\%$ gap from the same
hardware-visited states, so the inflation is carried entirely by
noise-flipped greedy decisions.}
  \label{fig:n4-b3-progression}
\end{figure}

\begin{figure}[t]
  \centering
  \includegraphics[width=\linewidth]{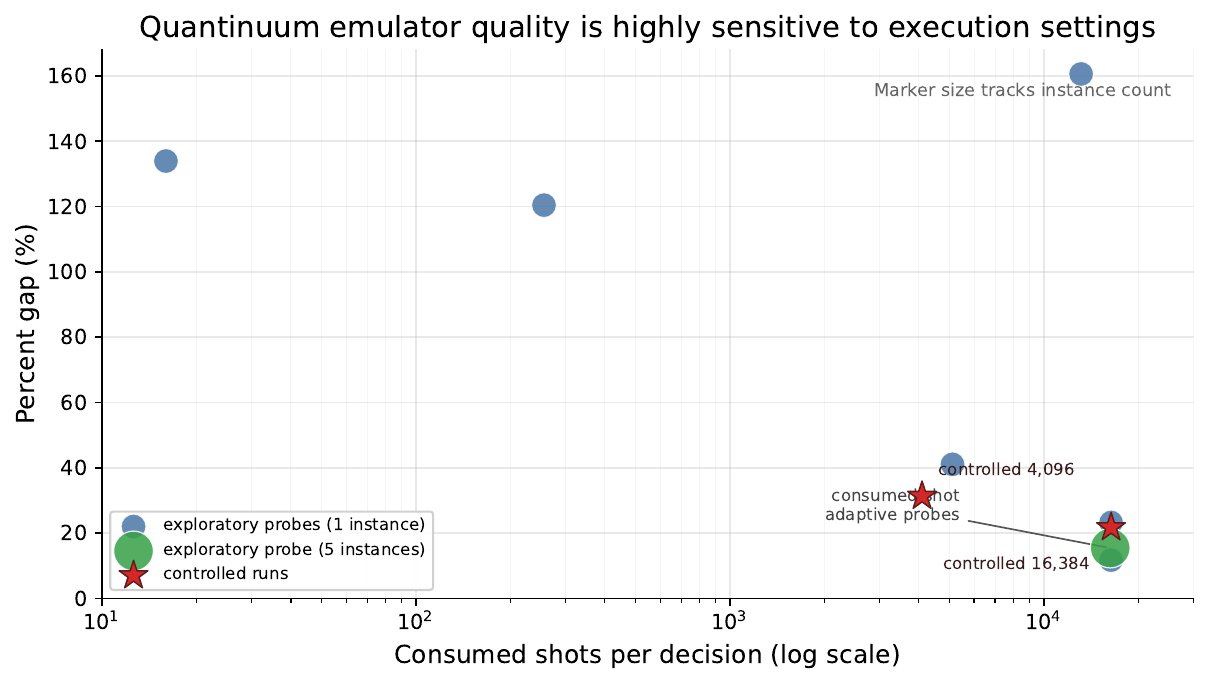}
  \caption{N4 Quantinuum emulator sensitivity for the $5\to10$ transfer. Each
marker is one completed emulator run; marker size tracks the number of
evaluated instances (1 or 5). Exploratory probes are plotted at their
\emph{consumed} per-decision shot budget (the nominal ``4096-shot'' probes
used adaptive top-up and consumed 16{,}384 shots per decision; see
Table~\ref{tab:n4-emulator}). Star markers show the controlled five-instance
runs at the exact hardware protocol (single batch, no top-up): $31.3\%$ at
4096 shots and $21.7\%$ at 16{,}384 shots per decision. The same source
policy on local statevector simulation gives a ${\sim}5\%$ gap; the
difference is the execution penalty introduced by finite-shot evaluation of
statistically unresolved action margins.}
  \label{fig:n4-emulator}
\end{figure}

\begin{figure*}[!t]
  \centering
  \includegraphics[width=0.96\textwidth]{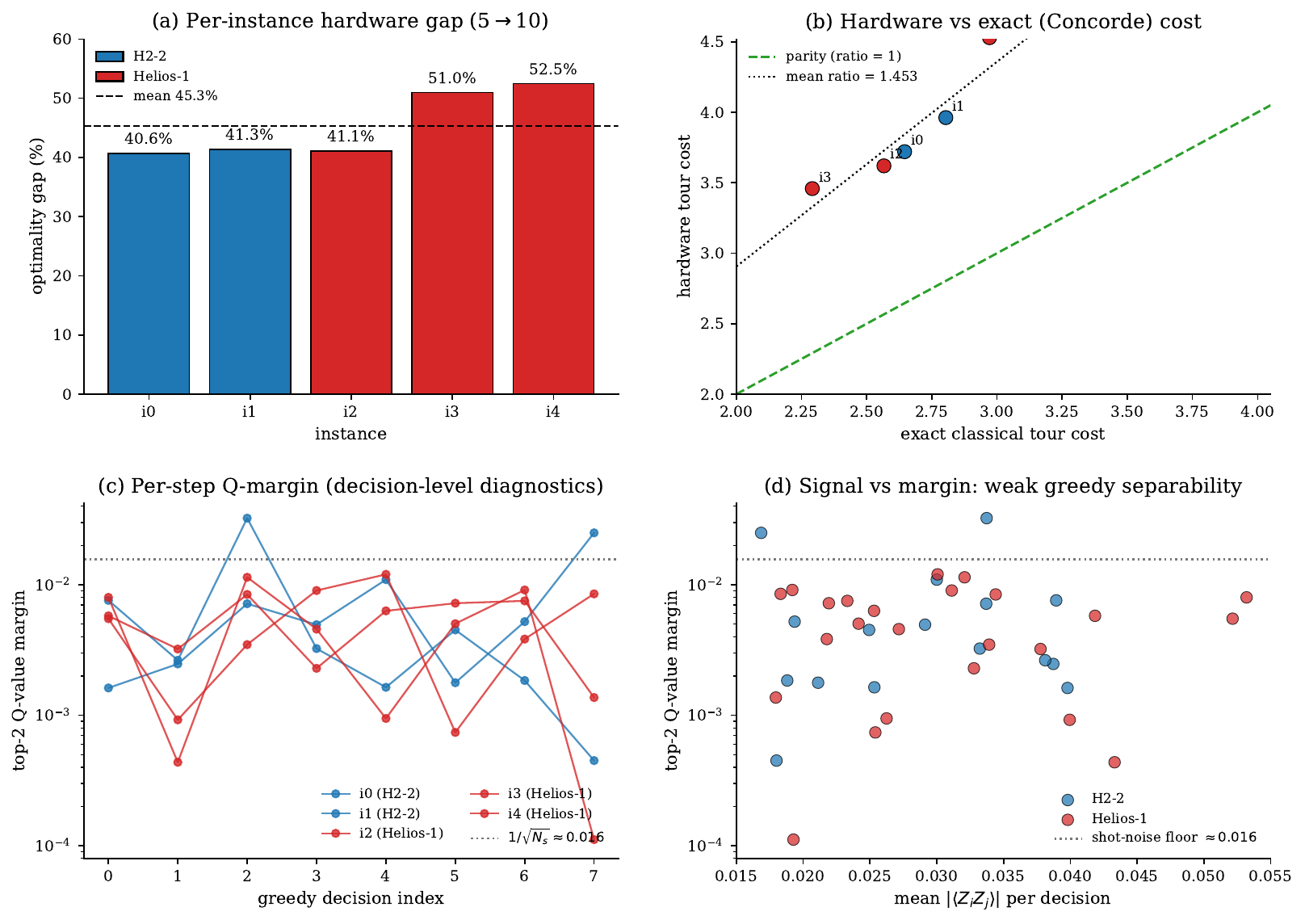}
  \caption{N4 Quantinuum hardware results for the $5\to 10$ zero-shot transfer
lane. \textbf{(a)}~Per-instance optimality gap: mean $45.3\%$, range
$40.6$--$52.5\%$, $\sigma=5.9$~pp over five 10-node instances. Bars are
coloured by execution device (blue: H2-2, red: Helios-1); the device split
was operational, not randomised. \textbf{(b)}~Hardware tour cost versus exact
(Concorde) tour cost; the green dashed line is parity (ratio $=1$), the
dotted line is the mean ratio $1.453$. \textbf{(c)}~Per-step top-2 Q-value
margin along each greedy rollout (log scale). The grey dotted line marks the
single-observable shot-noise floor $1/\sqrt{N_s}\approx 0.016$ at
$N_s=4096$ shots; the exact propagated per-decision margin standard errors
(mean $9.1\times10^{-3}$) are reported in the text, and 37 of 40 decisions
lie within two standard errors of a tie. \textbf{(d)}~Mean
$|\langle Z_iZ_j\rangle|$ signal per decision versus per-decision Q-margin:
signal sits in a narrow ${\sim}0.02$--$0.05$ range while Q-margins span four
orders of magnitude. This is the smoking-gun B3 diagnostic: the
$\binom{10}{2}=45$ observables individually carry only ${\sim}3\%$ contrast,
and the action-value differences they encode are at or below their own
propagated sampling error.}
  \label{fig:n4-hardware}
\end{figure*}

Barrier B3 asks what happens to the same trained EQC checkpoint when it is
executed under realistic, finite-shot, hardware-style conditions rather than
on local statevector simulation. We answer this question with two
complementary measurements on the $5\to10$ transfer lane: (i)~a
shot-sensitivity study on the Quantinuum H-series emulator, including
controlled five-instance runs at the exact hardware protocol
(Figure~\ref{fig:n4-emulator}, Table~\ref{tab:n4-emulator}), and (ii)~an
end-to-end execution on Quantinuum trapped-ion hardware
(Figure~\ref{fig:n4-hardware}, Table~\ref{tab:n4-hardware}). Together these
establish a clean, protocol-matched B3 progression for the same $5\to 10$
checkpoint at 4096 single-batch shots per greedy decision: ${\sim}5\%$ gap on
local statevector, $31.3\%$ on the noiseless-state emulator (sampling noise
alone), and $45.3\%$ ($\sigma=5.9$~pp) on hardware
(Figure~\ref{fig:n4-b3-progression}).

\paragraph{Emulator shot-sensitivity study.}
The primary emulator measurements are two \emph{controlled} five-instance
runs at the exact hardware protocol (single batch, no adaptive top-up, same
five target instances, same code path), executed on the
noiseless-statevector H-series emulator (H2-1LE): a $31.3\%$ mean gap at
4096 shots per decision (per-instance $0$--$57.4\%$; $95\%$ $t$-interval on
the mean $[3.6,\,58.9]$) and $21.7\%$ at 16{,}384 shots per decision
(per-instance $5.6$--$36.8\%$; $[7.6,\,35.8]$). The mean improves
monotonically with shots, as the shot-noise mechanism requires, but the
improvement is modest and the per-instance scatter remains tens of
percentage points at both budgets --- quantitatively consistent with
Proposition~3, since even 16{,}384 shots lie below the
${\approx}3.8\times10^4$ per-decision requirement at the mean margin. The
remaining rows of Table~\ref{tab:n4-emulator} are earlier exploratory
probes, retained for completeness and plotted at their \emph{consumed}
budgets in Figure~\ref{fig:n4-emulator}. One requires explicit comment: the
single-instance 16{,}384-shot probe records a larger gap ($23.24\%$) than
the single-instance ``4096-shot'' probe ($11.70\%$), which would be
inconsistent with a shot-noise mechanism if read as a controlled shot-count
comparison. The run manifests show it is not one: both probes ran on the
same instance and both consumed 16{,}384 shots per decision --- the nominal
4096-shot probe used adaptive top-up batching (four 4096-shot batches at
confidence threshold $z=2.5$, which every decision exhausted), while the
16{,}384-shot probe submitted a single batch. The $11.5$-point difference is
therefore pure single-instance run-to-run variance at \emph{equal} budget,
exactly as expected when greedy decisions are statistically unresolved. The
earlier five-instance $15.40\%$ run likewise consumed 16{,}384 shots per
decision under adaptive top-up; the controlled runs above supersede it as
the emulator-side reference for the hardware campaign.

\begin{table}[t]
  \centering
  \caption{Selected N4 emulator runs for the same transferred $5\to10$
  source policy. ``Shots'' is the nominal per-batch setting; ``consumed'' is
  the per-decision budget actually used after adaptive top-up (run
  manifests). The two controlled five-instance runs (bottom) replicate the
  hardware protocol exactly (single batch, no top-up) and are the
  like-for-like emulator comparators for the hardware campaign.}
  \label{tab:n4-emulator}
  \small
  \setlength{\tabcolsep}{3pt}
  \resizebox{\columnwidth}{!}{%
  \begin{tabular}{lcrrrc}
    \toprule
    Lane & Inst. & Shots & Consumed & Mean ratio & Mean gap \\
    \midrule
    $5\to5$   & 1 &   1024 &  1024  & 1.0000 &   0.00\% \\
    $5\to5$   & 5 &   1024 &  1024  & 1.2706 &  27.06\% \\
    $5\to10$  & 1 &   4096 & 16{,}384$^{\dagger}$ & 1.1170 &  11.70\% \\
    $5\to10$  & 1 & 16{,}384 & 16{,}384 & 1.2324 &  23.24\% \\
    $5\to10$  & 5 &   4096 & 16{,}384$^{\dagger}$ & 1.1540 &  15.40\% \\
    $5\to15$  & 1 &    256 &   256  & 2.8065 & 180.65\% \\
    \midrule
    $5\to10$ (controlled) & 5 & 4096 & 4096 & 1.3125 & 31.25\% \\
    $5\to10$ (controlled) & 5 & 16{,}384 & 16{,}384 & 1.2170 & 21.70\% \\
    \bottomrule
  \end{tabular}}
 
  \smallskip
  \raggedright\footnotesize $^{\dagger}$Adaptive top-up (confidence
  threshold $z=2.5$, four 4096-shot batches), exhausted at every decision.
\end{table}

\paragraph{End-to-end hardware execution on Quantinuum H2-2 and Helios-1.}
We then ported the identical $5\to 10$ transferred policy to Quantinuum trapped-ion hardware. The campaign produced \textbf{40/40 completed hardware decisions across 5/5 instances} (no compile-fallback jobs): one hardware execution per greedy action, 4096 shots requested per execution, and persistence after every successful decision. The Nexus billing total is \textbf{18{,}877.8 HQC} (471.94 HQC/job, identical across all 40 jobs) over a campaign wall-clock of \textbf{310.5 h}. Of the 40 decisions, 16 ran on H2-2 (\texttt{native\_direct\_execute}, \texttt{program\_kind=circuit}) and 24 ran on Helios-1 (\texttt{qir\_direct\_execute}, \texttt{program\_kind=qir}); the device split was operational rather than randomised. Figure~\ref{fig:n4-hardware} and Table~\ref{tab:n4-hardware} summarise the result.

\begin{table}[!t]
  \centering
  \caption{N4 Quantinuum hardware results for the $5\to 10$ transfer lane. Each row is one completed 10-node instance: 8 greedy hardware decisions plus a forced final return-to-origin. ``Ratio'' is hardware tour cost divided by exact Concorde optimum; ``Gap'' is $100(\text{ratio}-1)$. All 40 decisions completed without compile-fallback jobs at 471.94 HQC/job; the campaign total was 18{,}877.8 HQC over 310.5 h of wall-clock. Per-decision wall-times and per-instance tours are tabulated in Appendix~\ref{app:n4-hardware-extended}. The mean gap carries a $95\%$ $t$-interval of $[38.0,\,52.6]$ over the five
instances.}
  \label{tab:n4-hardware}
  \small
  \setlength{\tabcolsep}{3pt}
  \begin{tabular}{rlrrrr}
    \toprule
    Inst. & System & HW cost & Exact & Ratio & Gap \\
    \midrule
    i0 & H2-2     & 3.7204 & 2.6453 & 1.4064 & 40.64\% \\
    i1 & H2-2     & 3.9624 & 2.8034 & 1.4134 & 41.34\% \\
    i2 & Helios-1 & 3.6200 & 2.5660 & 1.4108 & 41.08\% \\
    i3 & Helios-1 & 3.4586 & 2.2912 & 1.5095 & 50.95\% \\
    i4 & Helios-1 & 4.5296 & 2.9712 & 1.5245 & 52.45\% \\
    \midrule
    \multicolumn{2}{l}{Mean ($n{=}5$)} & 3.8582 & 2.6554 & 1.4529 & 45.29\% \\
    \multicolumn{2}{l}{Median}         &        &        & 1.4134 & 41.34\% \\
    \multicolumn{2}{l}{Sample $\sigma$} &       &        & 0.0588 &  5.88 pp \\
    \bottomrule
  \end{tabular}
\end{table}

The hardware tours are valid Hamiltonian cycles on every instance, and the cost ratio is tightly clustered into two sub-regimes (i0,~i1,~i2 near $41\%$; i3,~i4 near $51$--$52\%$). The mean ratio is $1.4529$ with a sample standard deviation of $5.88$ percentage points over the five instances --- which is wide enough that this should be framed as a five-instance hardware execution panel rather than a tight benchmark distribution. We emphasise that the device split between H2-2 (mean gap $41.0\%$) and Helios-1 (mean gap $48.2\%$) is operational, not a randomised comparison, and should not be read as a ranking.

\subsubsection{Decision-resolution statistics}
\label{sec:decision-stats}
Figure~\ref{fig:n4-hardware}(c,~d) explains where the hardware gap comes
from. The per-decision top-2 Q-value margin --- the gap between the chosen
action's Q-value and the next-best Q-value over still-available cities ---
has mean $5.98\times 10^{-3}$ and minimum $1.12\times 10^{-4}$. The mean
absolute pair observable $\langle Z_iZ_j\rangle$ per decision sits in the
narrow band $0.017$--$0.053$, with overall mean $0.030$. At 4096 shots the
shot-noise floor on a single observable is $1/\sqrt{N_s}\approx 0.016$, so
several greedy decisions are made with a top-2 margin at or below the noise
on its own inputs. We make this statement statistically precise using the
persisted per-decision counts. First, propagating the multinomial sampling
covariance of each decision's bitstring counts through the Q-value map ---
which accounts exactly for the covariance of Q-values estimated from the
same shots --- yields a per-decision standard error of the top-2 margin
(mean $9.1\times10^{-3}$, range $3.5\times10^{-3}$--$2.0\times10^{-2}$) and
hence a margin $z$-score; 37 of the 40 hardware decisions have $z<2$
(median $z=0.48$), i.e.\ their action selection is not statistically
distinguishable from a tie at the $95\%$ level. Second, bootstrap resampling
of each decision's own shots ($2\times10^4$ resamples per decision) gives a
direct decision-instability estimate: 30 of 40 decisions flip their
$\arg\max$ in more than $25\%$ of resamples, and the expected number of
noise-flipped decisions per 8-step rollout is $3.1$. Third, a counterfactual
replay --- recomputing the exact statevector Q-vector at each
hardware-visited state --- shows that the hardware-chosen action deviated
from the exact policy's choice at 19 of 40 decisions, with deviations
concentrated at the low-$z$ steps (mean $z=0.47$ at deviations versus $0.86$
at agreements; 17 of the 19 deviations have $z<1$, and 11 of 19 selected the
exact policy's rank-2 action); replaying the exact policy from each first
divergence recovers a mean gap of $2.7\%$ --- the statevector-level
performance --- confirming that the gap inflation is carried entirely by
noise-flipped decisions rather than by any corruption of the trained
$(\beta,\gamma)$ angles. This is the central B3 diagnostic: \emph{the
bottleneck is not engineering; it is the signal-to-noise structure of the
dense observable family at small per-edge contrast}.
Appendix~\ref{app:shot-complexity} formalises the obstruction: resolving
greedy decisions at the measured margins requires
$N_s\gtrsim 3.8\times10^4$ shots per decision even in order of magnitude,
and ${\sim}10^6$ for whole-rollout fidelity at conventional confidence ---
multiplying an already substantial HQC and queue cost far beyond the budget
of the present campaign.
 
\paragraph{Anchoring the hardware gap against classical baselines.}
To calibrate what a $45.3\%$ mean gap means, we evaluate standard classical
constructions on the same five hardware instances: nearest-neighbour from the
same fixed start gives a $9.4\%$ mean gap (per-instance range
$0$--$37.4\%$), greedy edge insertion $5.7\%$, Christofides $6.1\%$, and the
exact expected gap of a uniformly random closed tour is $92.5\%$. The
hardware-executed policy therefore retains structure well above the random
baseline --- roughly half the random-tour gap --- but falls well below every
classical construction heuristic. Notably, under faithful (statevector)
execution the trained policy \emph{beats} nearest-neighbour on these five
in-distribution instances ($2.7\%$ vs.\ $9.4\%$), so the hardware penalty
inverts an ordering that faithful execution gets right. This anchoring
matters for the Li--Zhang caveat of Section~\ref{sec:discussion}: the
hardware penalty is the cost of executing the trained policy through an
unresolved-margin readout, not a change in what the policy is.

\paragraph{Reconciling emulator and hardware: a two-component decomposition.}
The controlled emulator runs make the comparison protocol-matched for the
first time: at the hardware's own budget (4096 single-batch shots per
decision, same five instances, same code path) the noiseless-state emulator
gives $31.3\%$, against $45.3\%$ on hardware and $2.7\%$ for the exact
policy replayed from the same states. The total hardware excess over exact
execution is unambiguous: paired over the five instances it is
$42.6$~percentage points with $95\%$ $t$-interval $[31.6,\,53.6]$. Its mean
decomposition attributes roughly two-thirds (${\approx}29$~pp) to pure
finite-shot sampling of statistically unresolved margins, reproduced by a
noiseless sampler, and the residual ${\approx}14$~pp to device noise (gate
error, leakage, measurement error) on top of sampling. We report this split
as a point estimate and flag its uncertainty honestly: with only five paired
instances, the hardware$-$emulator difference of $14.0$~pp carries a
$95\%$ interval of $[-13.7,\,41.8]$, so the \emph{proportions} of the
decomposition are indicative rather than sharp, even though the total
excess and its sampling-dominated character are robust. The broad support
of the hardware output distributions is consistent with the mechanism: each
4096-shot 10-qubit execution returns on average $992.8$ \emph{distinct}
bitstrings out of $2^{10}=1024$ possible outcomes (range $967$--$1006$), so
the empirical $\langle Z_iZ_j\rangle$ is a finite-budget estimator of a
distribution with non-trivial tail mass, exactly the regime in which
Proposition~3 bites. The earlier $15.40\%$ five-instance emulator run is not
in tension with this decomposition: its adaptive top-up consumed 16{,}384
shots per decision (Table~\ref{tab:n4-emulator}), and at that budget the
controlled run gives $21.7\%$. B3 is thus a real, quantitatively large
barrier for the dense $L=1$ EQC at $n=10$, dominated in the mean
decomposition by sampling statistics rather than by device-specific
imperfection.

\paragraph{Protocol-level mitigations for low-margin policies.}
Several execution-protocol changes could reduce B3 at fixed ansatz, all
trading HQC for resolution. (i)~\emph{Shot-adaptive allocation}: spend shots
per decision until the top-2 margin $z$-score clears a target threshold;
the exploratory top-up probes are a crude version of this, and
Proposition~3 prices its worst case. (ii)~\emph{Decision voting}: repeat
each decision over $k$ independent batches and take the majority
$\arg\max$, converting variance into a controlled error rate at $k\times$
cost. (iii)~\emph{Uncertainty-aware selection}: take the measured
$\arg\max$ only when the margin is statistically resolved, falling back to
a classical tie-breaker (e.g.\ nearest neighbour) otherwise --- a rule that
would have triggered at 37 of the 40 campaign decisions, which makes
explicit how little of the trained policy survives this readout at 4096
shots. (iv)~\emph{Best-of-$k$ rollouts}, already supported by the N4
pipeline. None of these raises the margin itself; they are palliative
relative to the architectural fix of Section~\ref{sec:unify-barriers}, but
they define the protocol frontier that any low-margin quantum policy must
negotiate on shot-budgeted hardware.

\paragraph{Independence of the three barriers.}
The three barriers are independent in the sense that each can be triggered without the others. B1 fires whenever the backend is at the edge of its faithful regime, even at $m=n$ (the $\chi=64$ $20\to20$ control is the cleanest example). B2 fires whenever $|n-m|/\max(n,m)$ is large, even on a perfect backend (statevector simulation in N2 still shows the smooth degradation). B3 fires whenever the policy is executed under finite-shot, hardware-style conditions, even at small $n$ (the $5\to5$ emulator runs at 1024 shots still exhibit a 27\% gap, and the $5\to 10$ hardware run at 4096 shots gives 45.3\%). The three barriers are not independent in the deeper sense, however: all three are exacerbated by the same $\binom{n}{2}L$ all-to-all gate count of the canonical EQC ansatz.

\subsection{Cross-Platform Confirmation: The Gate-Count Root Cause on Hardware}
\label{sec:barrier-xplatform}
 
The B3 measurement above is single-platform: it diagnoses the
hardware-execution penalty on Quantinuum trapped-ion devices and attributes
it, via the decision-resolution statistics of
Section~\ref{sec:decision-stats}, to action margins lying at or below the
per-decision shot-noise floor. Two questions remain before we can claim the
mechanism is general rather than device-specific, and before the unification
of Section~\ref{sec:unify-barriers} is warranted. First, is the penalty a
property of the dense ansatz interacting with \emph{any} backend, or an
artifact of one qubit technology? Second, the shot-noise floor falls as
$1/\sqrt{N_s}$, so would simply spending more shots resolve the greedy
decisions and close the gap? To settle both, we ported the \emph{identical}
$5\to10$ checkpoint --- same trained $(\beta,\gamma)$ angles, same source seed,
same five held-out 10-node instances, same greedy protocol --- to five quantum
processors spanning two qubit technologies and four vendors, and where feasible
pushed the shot budget to each device's maximum. The campaign is observational
rather than a randomised factorial (device, shot count, and mitigation co-vary
with operational access), so we read it as a mechanistic confirmation, not a
vendor ranking. % [optional cite] AWS Braket / Qiskit Runtime here.
 
\paragraph{Identical workload, divergent native cost.}
Every device received the same logical circuit per greedy decision: the bound
Model~B EQC at $n=10$, $L=1$, with the full $\binom{10}{2}=45$-edge $ZZ$
interaction graph (eight measured decisions per rollout). What the device
compilers \emph{realise} differs sharply, and this divergence is the
controlled variable of the study. All-to-all trapped-ion hardware implements
each $ZZ$ rotation as one arbitrary-angle native two-qubit gate, holding the
count at $45$. Fixed-\texttt{CZ} superconducting devices must decompose each
$ZZ$ into \texttt{CNOT}$\cdot$\texttt{RZ}$\cdot$\texttt{CNOT} \emph{and} insert
\texttt{SWAP} routing for their limited connectivity, inflating the native
two-qubit count to $153$--$172$ --- a $3.4$--$3.8\times$ increase
(Table~\ref{tab:xplatform-circuit}). This is the SWAP overhead that
Section~\ref{sec:unify-barriers} attributes to the all-to-all topology, here
measured directly on heavy-hex and limited-lattice hardware.
 
\begin{table}[t]
  \centering
  \caption{The same 45-edge logical EQC, compiled to each device's native gate
  set. The two-qubit inflation is relative to the 45 logical $ZZ$ interactions
  and is set by connectivity and the native two-qubit gate. IonQ Forte
  (trapped ion) was queued but cancelled before execution and is omitted.}
  \label{tab:xplatform-circuit}
  \footnotesize
  \setlength{\tabcolsep}{1pt}
  \renewcommand{\arraystretch}{1.12}
  \newcommand{\xplatcell}[2]{\parbox[t]{#1}{\raggedright\strut #2\strut}}
  \begin{tabular}{@{}lllll@{}}
    \toprule
    \xplatcell{0.30\linewidth}{Device} &
    \xplatcell{0.18\linewidth}{Tech.} &
    \xplatcell{0.16\linewidth}{Conn.} &
    \xplatcell{0.13\linewidth}{\textbf{Native 2q}} &
    \xplatcell{0.10\linewidth}{Infl.} \\
    \midrule
    \xplatcell{0.30\linewidth}{Quantinuum Helios/H2-2} &
    \xplatcell{0.18\linewidth}{trapped ion} &
    \xplatcell{0.16\linewidth}{all-to-all} &
    \xplatcell{0.13\linewidth}{\textbf{45}} &
    \xplatcell{0.10\linewidth}{$1.0\times$} \\
    \xplatcell{0.30\linewidth}{Rigetti Cepheus} &
    \xplatcell{0.18\linewidth}{super-\linebreak conducting} &
    \xplatcell{0.16\linewidth}{lattice} &
    \xplatcell{0.13\linewidth}{\textbf{153}} &
    \xplatcell{0.10\linewidth}{$3.4\times$} \\
    \xplatcell{0.30\linewidth}{IQM Emerald} &
    \xplatcell{0.18\linewidth}{super-\linebreak conducting} &
    \xplatcell{0.16\linewidth}{lattice} &
    \xplatcell{0.13\linewidth}{\textbf{162}} &
    \xplatcell{0.10\linewidth}{$3.6\times$} \\
    \xplatcell{0.30\linewidth}{IQM Garnet} &
    \xplatcell{0.18\linewidth}{super-\linebreak conducting} &
    \xplatcell{0.16\linewidth}{lattice} &
    \xplatcell{0.13\linewidth}{\textbf{172}} &
    \xplatcell{0.10\linewidth}{$3.8\times$} \\
    \xplatcell{0.30\linewidth}{IBM Marrakesh (Heron)} &
    \xplatcell{0.18\linewidth}{super-\linebreak conducting} &
    \xplatcell{0.16\linewidth}{heavy-hex} &
    \xplatcell{0.13\linewidth}{\textbf{172}} &
    \xplatcell{0.10\linewidth}{$3.8\times$} \\
    \bottomrule
  \end{tabular}
\end{table}
 
\begin{figure}[t]
  \centering
  \includegraphics[width=\linewidth]{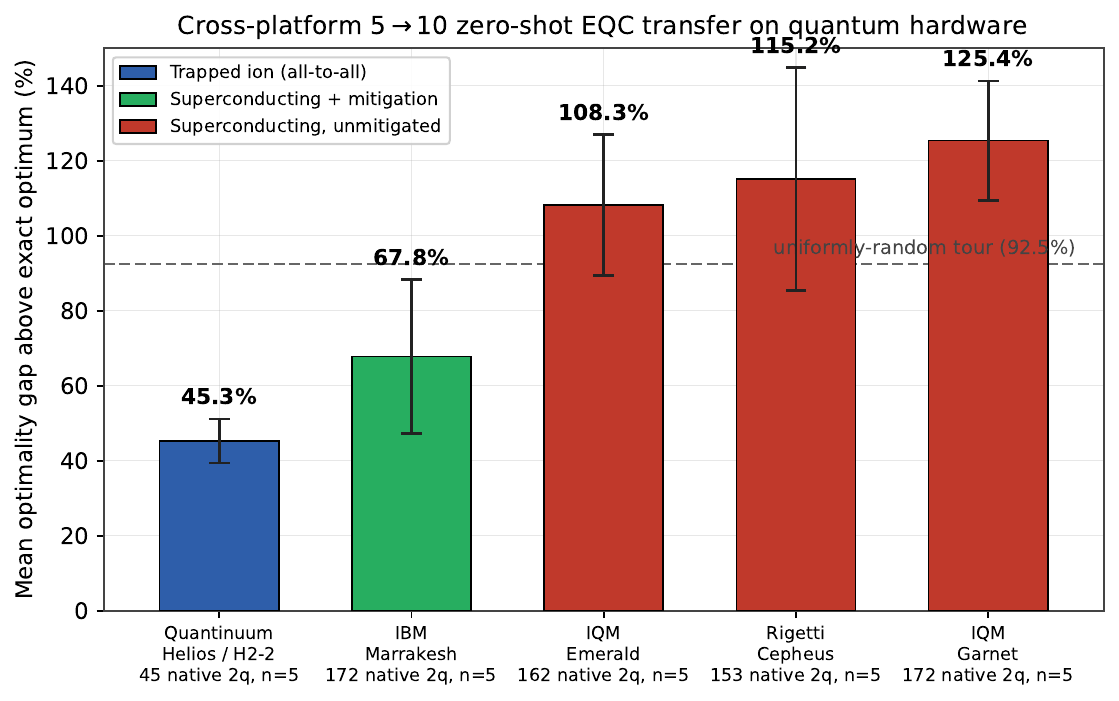}
  \caption{Mean $5\to10$ optimality gap for the identical transferred EQC
  checkpoint on five quantum processors. Bars are coloured by hardware regime;
  error bars are sample standard deviations over the five held-out instances.
  The grey dashed line is the $92.5\%$
  uniformly-random-tour baseline of Section~\ref{sec:barrier-hardware}. Policy
  structure survives only on the all-to-all trapped-ion device and on the
  fully mitigated superconducting device; the three unmitigated
  superconducting/IQM devices sit at or above the random-tour level despite the
  largest shot budgets in the campaign.}
  \label{fig:n4-xplatform-gap}
\end{figure}

\paragraph{Headline result.}
Table~\ref{tab:xplatform-gap} and Figure~\ref{fig:n4-xplatform-gap} report the
mean optimality gap on each platform against exact Concorde optima on the same
five instances. The platforms separate into three regimes relative to the
$92.5\%$ uniformly-random-tour baseline established in
Section~\ref{sec:barrier-hardware}. The all-to-all trapped-ion device ($45$
native 2q gates, no explicit mitigation) holds the gap to $45.3\%$ --- roughly
half the random baseline, i.e.\ policy structure is preserved. A heavily
mitigated superconducting device (IBM Heron, full mitigation stack) reaches a
five-instance mean of $67.8\%$ ($\sigma=20.5$~pp) despite $172$ native 2q
gates --- a clear second place, well below the random baseline, though
mitigation does not rescue every instance uniformly (its per-instance gaps span
$44.7\%$, matching the trapped-ion mean, to $104.6\%$, overlapping the
unmitigated cluster). The three unmitigated
superconducting/IQM devices cluster at $108$--$125\%$, i.e.\ at or modestly
above the random-tour gap: within the five-instance scatter their tours are
statistically indistinguishable from a uniformly random closed tour, so
essentially all of the transferred policy's structure is lost before
measurement. On every device and instance the quantum tour is longer than the
classical optimum (``quantum beats classical'' count $=0$), consistent with
the paper's position that this is a fidelity diagnostic and not an advantage
claim.
 
\begin{table}[t]
  \centering
  \caption{Mean $5\to10$ zero-shot optimality gap by platform, identical
  checkpoint and protocol. The random-tour baseline is $92.5\%$
  (Section~\ref{sec:barrier-hardware}). ``Inst.'' is the number of held-out
  instances. Uncertainties are sample standard deviations over
  instances. Shots were pushed to each device's maximum where feasible to test
  the shot-budget hypothesis; IBM was held to $4{,}096$ shots by its
  QPU-second budget and run as five instances across two monthly allowances.}
  \label{tab:xplatform-gap}
  \footnotesize
  \setlength{\tabcolsep}{2pt}
  \renewcommand{\arraystretch}{1.08}
  \begin{tabular}{@{}lccccc@{}}
    \toprule
    Device & Tech. & Shots & Inst. & Mitig. & \textbf{Mean gap} \\
    \midrule
    \shortstack[l]{Quantinuum\\Helios/H2-2} & ion & 4{,}096  & 5 & none & \textbf{$45.3\pm5.9\%$} \\
    IBM Marrakesh   & sc & 4{,}096  & 5 & full & \textbf{$67.8\pm20.5\%$} \\
    IQM Emerald     & sc & 20{,}000 & 5 & none & \textbf{$108.3\pm18.8\%$} \\
    Rigetti Cepheus & sc & 50{,}000 & 5 & none & \textbf{$115.2\pm29.7\%$} \\
    IQM Garnet      & sc & 20{,}000 & 5 & none & \textbf{$125.4\pm15.9\%$} \\
    \bottomrule
  \end{tabular}
\end{table}
 
\begin{figure*}[t]
  \centering
  \includegraphics[width=0.98\textwidth]{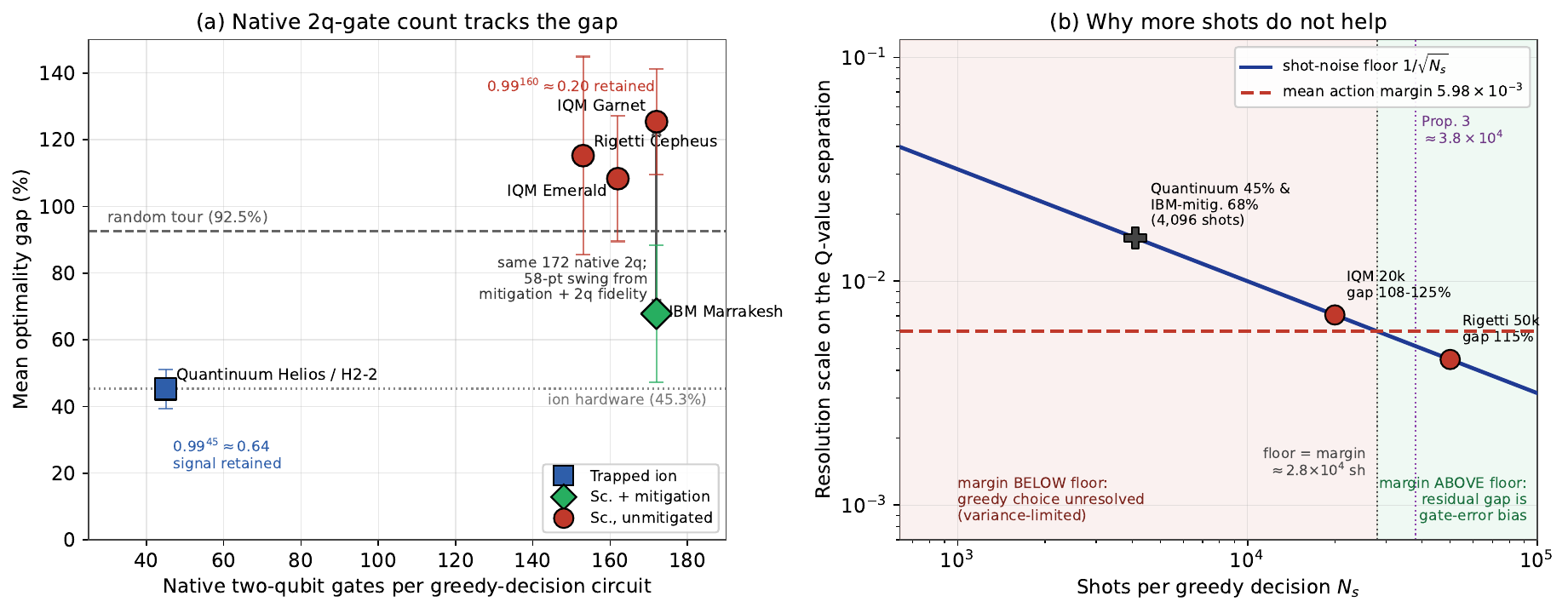}
  \caption{Mechanism of the cross-platform B3 penalty. \textbf{(a)}~Mean gap
  versus native two-qubit gate count; the $92.5\%$ random and $45.3\%$
  trapped-ion lines are guides. IBM Marrakesh and IQM Garnet carry the
  \emph{same} 172 native two-qubit gates yet differ by $58$ percentage points
  in five-instance mean, isolating the mitigation stack (and Heron two-qubit
  fidelity), not gate count, as the lever for that pair. \textbf{(b)}~The single-observable
  shot-noise floor $1/\sqrt{N_s}$ (solid) crosses the mean action margin
  $5.98\times10^{-3}$ (dashed) near $2.8\times10^4$ shots; resolving decisions
  at conventional confidence requires ${\approx}3.8\times10^4$ shots
  (Proposition~\ref{prop:shot-complexity}). Left of the crossover the greedy
  choice is variance-limited (the trapped-ion and IBM runs at 4{,}096 shots);
  the highest-shot Rigetti run sits past the crossover yet still fails,
  identifying gate-error bias --- not sampling variance --- as its limiter.}
  \label{fig:n4-xplatform-mechanism}
\end{figure*}

\paragraph{Finding 1: native two-qubit count, not shots, sets the gap.}
Figure~\ref{fig:n4-xplatform-mechanism}(a) plots the mean gap against the
native two-qubit gate count. Among devices without an explicit mitigation
stack the trend is monotone and large: the $45$-gate trapped-ion machine sits
far below the random baseline, while the $153$--$172$-gate \emph{unmitigated}
routed devices saturate near it. The mitigated $172$-gate IBM device is the
single departure from this gate-count trend --- it carries the largest native
count yet the second-lowest gap --- and Finding~2 attributes that departure to
its mitigation stack rather than to connectivity. At a
representative per-two-qubit-gate fidelity $f_{2q}\approx0.99$ (illustrative,
not a device specification), the trapped-ion circuit retains
$0.99^{45}\approx0.64$ of its amplitude whereas the routed circuits retain only
$0.99^{160}\approx0.20$ --- and the $\binom{10}{2}=45$ pair observables that
encode the greedy Q-values already carry only ${\sim}3\%$ per-edge contrast
(Section~\ref{sec:decision-stats}), so once four-fifths of the amplitude is
gone the action-value differences are unrecoverable. Connectivity, expressed
through the native two-qubit count, is therefore the dominant physical driver
of the hardware gap. This is the direct experimental confirmation of the
gate-count root cause that Section~\ref{sec:unify-barriers} assigns to all
three barriers: the same $\binom{n}{2}L$ topology that defeats MPS truncation
(B1) and inflates the parametric mismatch term of
Theorem~\ref{thm:final-performance} (B2) is, on routed hardware, the variable
that separates a recoverable policy from a randomised one.
 
\paragraph{Finding 2: mitigation can substitute for connectivity, but not cheaply.}
IBM Marrakesh and IQM Garnet present the EQC with an identical native
two-qubit budget ($172$ gates), yet score five-instance means of $67.8\%$
versus $125.4\%$ --- a $58$-point swing achieved at one-fifth of the shots
($4{,}096$ vs.\ $20{,}000$). Because both devices ran the same five instances
at the same gate count, this is a like-for-like contrast in which mitigation is
the only varied lever.
The difference is the error-mitigation stack exposed by the Qiskit Runtime
estimator: gate and measurement twirling, dynamical decoupling, and
readout-error correction, together with Heron's native two-qubit fidelity.
Mitigation is not, however, a uniform free choice: the only technique robust
through a bare device-submission path is readout-error correction, because it
acts on the final measurement and survives the native compiler, whereas
gate-level techniques require inserting gates that the device optimiser cancels
or reorders. We verified in a noisy density-matrix simulation, at zero hardware
cost, that readout correction alone is marginal and is actively harmful below
${\sim}10^4$ shots (dividing by a readout-contraction factor $\alpha\approx0.94$
amplifies variance faster than it removes bias), so the one mitigation we could
have retrofitted to the bare-submission runs would not have rescued them. This
is consistent with the architectural prescription of
Section~\ref{sec:unify-barriers}: where the access layer permits a validated
stack, mitigation can partly stand in for connectivity, but the durable fix is
to reduce the native two-qubit count itself.
 
\paragraph{Finding 3: more shots do not help --- two failure modes, one root.}
Rigetti at $50{,}000$ shots ($115.2\%$) is no better than IQM at $20{,}000$
($108$--$125\%$), and the per-instance scatter does not tighten with shots
(Appendix~\ref{app:xplatform-instances}, Figure~\ref{fig:n4-xplatform-instances}).
This sharpens rather than contradicts the shot-noise diagnosis of
Section~\ref{sec:decision-stats}. A direct same-device test on Quantinuum
hardware makes the point without any cross-device confound: re-running
instance i0 on Helios-1 at $10{,}000$ shots --- $2.4\times$ the campaign budget
--- gave a $53.9\%$ gap, \emph{worse} than the $40.6\%$ at $4{,}096$ shots on
the same instance (Table~\ref{tab:n4-hardware}), because at $10{,}000$ shots
the single-observable floor ($1/\sqrt{10^4}=0.010$) still exceeds the mean
action margin, so the two budgets merely draw different statistically
unresolved samples rather than better-resolved ones. The cross-platform points occupy two regimes
of Figure~\ref{fig:n4-xplatform-mechanism}(b), separated by the shot budget at
which the floor meets the margin ($1/\sqrt{N_s}=5.98\times10^{-3}$ at
$N_s\approx2.8\times10^4$; resolving at confidence needs
${\approx}3.8\times10^4$, Proposition~\ref{prop:shot-complexity}):
\begin{itemize}
  \item \textbf{Variance-limited} (low shots): at $4{,}096$ shots the floor
  $1/\sqrt{N_s}\approx0.016$ lies \emph{above} the mean margin
  $5.98\times10^{-3}$, so the greedy choice is statistically unresolved even
  with clean gates. This is the regime of the trapped-ion and IBM-mitigated
  runs. For the trapped-ion device the residual gap reflects a low-signal
  policy, exactly as Section~\ref{sec:decision-stats} found; for IBM it is a
  mixture, since its $172$ native two-qubit gates leave a gate-error bias that
  mitigation reduces but does not fully cancel, so the larger $67.8\%$ residual
  combines an unresolved margin with incompletely-cancelled bias.
  \item \textbf{Bias-limited} (high shots): beyond the crossover, statistical
  variance is no longer the binding constraint --- yet Rigetti at $50{,}000$
  shots, well past it, still produces near-random tours. What remains is
  \emph{systematic gate-error bias}: the ${\sim}160$-gate routed circuit shifts
  the estimated $\langle Z_iZ_j\rangle$ away from their ideal values, a bias no
  number of shots removes.
\end{itemize}
The unifying statement is that on no platform is the bottleneck the statistical
variance of finite sampling: it is an unresolved margin where gates are clean
and gate-error bias where they are not. Both point to the same remedy as the
rest of the paper --- raise the per-edge contrast and lift the action margin
above the shot-noise floor by reducing connectivity
(Section~\ref{sec:unify-barriers}, Appendix~\ref{app:linear-eqc}) --- rather
than buying shots, whose cost on metered backends was, in this campaign,
inverted relative to quality.
 
\paragraph{Scope and honesty of the cross-platform claim.}
Three caveats bound the reading of these results. (i)~The design is
observational: device, shot count, and mitigation co-vary, so the per-device
numbers isolate mechanisms but are not a randomised comparison. (ii)~The IBM
five-instance panel was run across two monthly QPU-second allowances ($96$~s
for i0, $384$~s for i1--i4) and its per-instance gaps are widely dispersed
($\sigma=20.5$~pp, range $44.7$--$104.6\%$), so its mean is a five-instance
panel rather than a tight benchmark; the matched-$172$-gate contrast with IQM
Garnet is nonetheless a like-for-like five-instance comparison. (iii)~The per-two-qubit fidelity $0.99$ is a
representative figure used only to motivate the signal-retention estimates of
Finding~1, not a measured device specification. Within these bounds the
campaign converts the single-device B3 diagnosis into a platform-independent
statement: the dense all-to-all EQC readout is \emph{signal}-limited on real
hardware --- by an unresolved action margin where gates are clean and by
gate-error bias where they are not --- and in neither case by the number of
shots, which is precisely the obstruction
Proposition~\ref{prop:shot-complexity} formalises and the sparse-connectivity
ansatz of Appendix~\ref{app:linear-eqc} is designed to relax.

\subsection{Unifying the Three Barriers}
\label{sec:unify-barriers}

The barriers share a common root cause: the all-to-all two-qubit gate count and dense pairwise readout of the implemented (Model~B) EQC.
\begin{itemize}
  \item For \textbf{B1}, gate count drives the entanglement structure that defeats MPS truncation, drives the SWAP overhead that defeats heavy-hex hardware, and drives the noise accumulation that defeats noisy-MPS simulation.
  \item For \textbf{B2}, the all-to-all topology drives the size-dependence of the parametric mismatch term of Theorem~\ref{thm:final-performance} via the parameter $\ell_1$-norm growth and the two-body generator-difference constant $C_{\rm gen}=4$.
  \item For \textbf{B3}, the all-to-all topology drives both the per-step circuit depth and the number of edge observables ($\binom{n}{2}$), which together drive the per-decision shot budget on hardware-style backends. The smoking-gun mechanism is that, at the shot budgets used in our campaign, the trained policy's top-2 Q-margin lies at or below the per-decision shot-noise floor $1/\sqrt{N_s}$; brute-force shot increases would multiply an already substantial HQC cost.
\end{itemize}

\paragraph{Architectural prescription.}
A linear-connectivity equivariant ansatz, in which the $\binom{n}{2}$ all-to-all ZZ layer is replaced by a chain of $n-1$ nearest-neighbour ZZ gates preceded by SWAP networks that maintain $S_n$-equivariance, would have only $O(n)$ two-qubit gates per layer, satisfy MPS area-law constraints, and sit comfortably within hardware fidelity budgets. Critically, replacing the dense pair observables by a sparse subset of $O(n)$ correlators would also raise the per-edge contrast and lift the action-value margin above the shot-noise floor at affordable $N_s$ --- directly relaxing the B3 mechanism above. We give a proof-of-concept analysis in Appendix~\ref{app:linear-eqc}. The full development of this architecture, including hardware execution at $n\!\ge\!50$, is the subject of follow-up work.

\paragraph{Connecting back to Theorem~\ref{thm:final-performance}.}
Reducing connectivity does not directly change the lifted-policy smoothness profile $\Psi(n,m)$ (which is an assumption on how the lifted policy behaves across sizes, not a property of the circuit topology). It does, however, plausibly reduce the parameter $\ell_1$-norm growth and the two-body $C_{\rm gen}$ constant that dominate the parametric mismatch term for the dense ansatz, and it relaxes the dense-readout mechanism that drives the B3 shot-budget barrier. The bound thus predicts --- and our preliminary linear-connectivity experiments support --- that sparser equivariant ansatzes should both transfer better at fixed depth than dense ones \emph{and} execute at substantially lower shot cost on hardware.

\section{Discussion}
\label{sec:discussion}

This study presents three intertwined results on permutation-equivariant QRL for combinatorial optimisation. \emph{First},
working in a two-model framework that separates a collective $S_n$-invariant
theoretical EQC (Model~A) from the implemented data-dependent weighted EQC
(Model~B), we develop a \emph{conditional diagnostic} transfer bound
(Theorem~\ref{thm:final-performance}) that decomposes the cross-size penalty
into a parametric mismatch term with explicit generator-difference constant
$C_{\rm gen}\in\{2,4\}$ and a structural smoothness term carried by an
explicit lifted-policy assumption (Assumption~(A7)), with BHH supplying only
the functional form of the structural profile $\Psi(n,m)$ and a separate
\emph{conditional} graphon refinement (Corollary~\ref{cor:graphon}) provided
under an additional cut-norm Lipschitz hypothesis. We engage the
trainability--simulability trade-off directly by computing the polynomial
DLA dimension of Model~A and acknowledging that the implemented Model~B is
only jointly equivariant, so the dequantisation result applies to the
symmetric reference. \emph{Second}, within the validated regime ($n\le 20$,
modest size jumps) we confirm that the EQC of Skolik et al.\ shows useful
same-size performance with near-optimal gaps and supports useful in-regime
transfer with only $2L$ trainable parameters
(Section~\ref{sec:empirical-eval}); the $5\to10$ zero-shot lane beats the
target-size scratch baseline on all six completed evaluations. \emph{Third},
and most importantly, the five-stage pipeline (Base, N1, N2, N3, N4) exposes
three independent failure modes outside that regime: backend-induced signal
loss at the MPS and noisy-simulator frontier (B1, with the $\chi=64$
equal-size $20\to20$ control gap of $85.22\%$ and a controlled
bond-dimension sweep placing the policy-quality crossover between $\chi=128$
and $\chi=256$), cross-size transfer degradation under large jumps (B2, mean
gap growing from $\sim 5\%$ at $5\to10$ to $\sim 18\%$ at $5\to20$), and
hardware-style execution penalty (B3, with the $5\to10$ gap rising under the
identical 4096-shot protocol from $\sim 5\%$ in local simulation to $31.3\%$
on a noiseless-state emulator --- sampling noise alone --- and to $45.3\%$
on Quantinuum H2-2/Helios-1 hardware, 40 completed hardware decisions,
18{,}878 HQC). The B3 mechanism is the central new diagnosis: the trained
policy's mean top-2 Q-margin ($5.98\times 10^{-3}$) sits at or below its own
propagated sampling error, leaving 37 of the 40 hardware decisions within
two standard errors of a tie and flipping an expected $3.1$ of 8 greedy
decisions per rollout; a counterfactual replay recovers the $2.7\%$
statevector-level gap from the same states, proving the inflation is carried
entirely by noise-flipped decisions. Resolving the measured margins would
require ${\approx}4\times 10^4$ shots per decision for two-sigma confidence
and ${\approx}2\times 10^6$ for whole-rollout fidelity
(Appendix~\ref{app:shot-complexity}), and the margin itself collapses as
$n^{-2.1}$ with problem size, so the required budget grows as
${\sim}n^{4.3}$ --- brute-force shot increases are not a viable route. The
contribution is neither a quantum-speedup claim nor a same-size benchmark;
it is a dequantisation-aware, multi-backend diagnostic study identifying
when cross-size EQC-QRL transfer works and which three barriers stop it from
scaling, together with architectural prescriptions that would relax all
three simultaneously.

\subsection{Significance and Limitations}

The contributions of this paper are fourfold:

\begin{itemize}
  \item It develops a \emph{conditional diagnostic} framework for analysing zero-shot transfer in equivariant QRL, with explicit assumption-tagging that separates source-task generalisation, parametric mismatch with constant $C_{\rm gen}\in\{2,4\}$, and lifted-policy smoothness. The Beardwood--Halton--Hammersley scaling motivates the size-smoothness profile but is not the bound itself; a conditional graphon refinement is stated separately under an additional, unproven cut-norm Lipschitz hypothesis.

  \item It explicitly identifies the key factors that influence transfer performance, particularly the interplay between source-task learning quality, the structure of the parametric mismatch under generator rescaling, and the structural shift across problem sizes; and it operates within a two-model framework that cleanly separates the collective $S_n$-invariant reference circuit (Model~A, on which DLA and dequantisation results apply rigorously) from the implemented data-dependent weighted circuit (Model~B, on which transfer and execution are measured).

  \item It introduces a unified multi-backend evaluation pipeline (statevector + MPS + IBM-inspired noisy MPS + Quantinuum H-series emulator + Quantinuum trapped-ion hardware) that we argue should be the methodological standard for QRL-at-scale claims going forward.

  \item It quantitatively characterises three independent barriers to scaling and ties them to a common topological cause, including the first end-to-end Quantinuum trapped-ion hardware execution of an equivariant QRL policy and its diagnosis of the central B3 mechanism --- the action-margin vs.\ shot-noise mismatch.
\end{itemize}

The analysis also has limitations:

\begin{itemize}
  \item The precise functional form and magnitude of the constants within the transfer penalty term $\mathcal{D}_{n \rightarrow m}$ remain abstract. The bound is a worst-case conditional decomposition and is loose in practice for moderate $|n-m|$; its value is qualitative, identifying the right scaling laws rather than producing sharp numerical predictions. The conditional graphon refinement (Corollary~\ref{cor:graphon}) depends on an unproven assumption that the lifted-policy rollout performance is cut-norm Lipschitz as a graphon functional; we mark this as speculative and do not rely on it for the main empirical analysis.

  \item The bound relies on a specific set of assumptions regarding the QRL agent, TSP encoding, and the transfer mechanism. Deviations from these assumptions (e.g., different QNN adaptation strategies, or TSPs with fundamentally different structures at different scales) could lead to different performance characteristics. In particular, the polynomial DLA / dequantisation guarantee is a Model~A property; the implemented Model~B with generic Euclidean weights is only \emph{jointly} equivariant, and its data-dependent DLA is open.

  \item Our hardware-facing data has two components. The N4 emulator and hardware results (Section~\ref{sec:barrier-hardware}) are end-to-end measurements: 40 completed Quantinuum trapped-ion hardware decisions across five 10-node instances at 4096 shots per decision (18{,}878 HQC; H2-2 and Helios-1), extended by the cross-platform campaign (Section~\ref{sec:barrier-xplatform}) to genuine IBM Heron, Rigetti, and IQM hardware at the same $5\to10$ lane. The B1 backend-signal-loss diagnostic, by contrast, is from the N3 MPS and IBM-\emph{inspired} noisy-MPS modes derived from IBM backend noise models rather than from hardware; the IBM fidelity projection in Section~\ref{sec:barrier-backend} concerns the $n=20$ heavy-hex circuit specifically and remains a scaling-law extrapolation, not a measurement at that size --- the cross-platform IBM data are at $n=10$, not at the $n=20$ frontier the B1 projection addresses. The hardware results establish B3 on real devices but are limited to five-instance panels at one source-target pair ($5\to 10$); we frame them as an execution-barrier \emph{diagnostic}, not as a broad hardware benchmark. MPS results were obtained via the quimb and Qiskit Aer \texttt{matrix\_product\_state} backends; alternative tensor-network methods (PEPS, MERA) may handle the all-to-all topology differently, but the underlying volume-law entanglement is a property of the ansatz, not the simulator.

  \item The N4 hardware panel comprises five 10-node instances, divided operationally (not randomly) between H2-2 (2 instances) and Helios-1 (3 instances); the per-device gap means should therefore not be read as a head-to-head device comparison. The protocol uses one hardware execution per greedy action, which is auditable but expensive; best-of-$k$ rollouts or adaptive resampling would multiply the HQC budget by $k$ and are deferred to follow-up work. The complementary emulator sweep at heterogeneous shot counts
(Table~\ref{tab:n4-aggregate}) is
exploratory --- its nominal shot settings differ from consumed budgets under
adaptive top-up --- and the appropriate emulator-side comparators for the
4096-shot hardware run ($45.3\%$) are the controlled five-instance runs at
the identical protocol: $31.3\%$ at 4096 and $21.7\%$ at 16{,}384 shots per
decision.
\end{itemize}

\subsection{The Li--Zhang Greedy-Bias Caveat}

\citet{li2025learningbasedtspsolverstendoverly} construct a statistical measure, the nearest-neighbour density, and show that learning-based TSP solvers trained on uniformly distributed Euclidean instances exhibit a strong nearest-neighbour-greedy bias --- in effect, they tend to ``always choose the nearest neighbour'' at each construction step --- and that their performance degrades sharply on out-of-distribution instances constructed via distribution shifts or node perturbations. Our trained equivariant policies are no exception to this regime: they are trained on uniformly distributed Euclidean instances, decode greedily one city at a time, and are therefore subject to the same bias. The practical implication is that even if all three barriers above were eliminated, our policies would not be expected to outperform a well-tuned nearest-neighbour heuristic on uniform $[0,1]^2$ instances asymptotically.

At the small sizes measured here the trained policy is not yet at that
asymptotic ceiling: under exact execution it beats nearest-neighbour on the
five hardware instances ($2.7\%$ vs.\ $9.4\%$ mean gap,
Section~\ref{sec:barrier-hardware}). The Li--Zhang bias is a statement about
decision style and out-of-distribution robustness, not about in-distribution
gap at small $n$; the two observations are compatible, and the caveat's
force is asymptotic and distributional.

This caveat is complementary to, rather than redundant with, the three barriers. The barriers limit the regime in which the trained policy can be \emph{executed faithfully} (B1, B2) and in which transfer remains \emph{useful} (B2, B3); the Li--Zhang bias limits the asymptotic quality of any greedily-decoded solver on the uniform Euclidean distribution. Productive routes around the Li--Zhang bias --- beyond the architectural prescription of Section~\ref{sec:unify-barriers} --- include: (i)~\emph{partial symmetry breaking}, e.g.\ a small learned per-node bias applied as an additional $R_x(b_i)$ before the equivariant block, which destroys strict invariance with minimal parameter cost; (ii)~\emph{non-greedy decoding} via beam search or best-of-$N$ sampling, which separates per-step policy expressivity from overall solution quality; and (iii)~\emph{restriction to structured instance families} (clustered, adversarial, real-world city sets) where the nearest-neighbour bias is far from optimal. We see these as natural follow-ups that complement the present diagnostic study.

\subsection{Future Research Directions}

This work opens several avenues for future research:

\begin{itemize}
  \item \textbf{Refining $\mathcal{D}_{n \rightarrow m}$:} A primary goal would be to develop a more rigorous and detailed mathematical characterization of the transfer penalty $\mathcal{D}_{n \rightarrow m}$. This could involve tools from optimal transport (e.g.\ Fournier--Guillin Wasserstein concentration~\cite{fournier2015rate}), information geometry, or analyzing the sensitivity of the QNN's output to changes in its underlying equivariant generators when parameters are fixed. Incorporating more specific structural properties of TSP instances (e.g., from random matrix theory or geometric graph theory) could also yield tighter bounds.

  \item \textbf{Sparse-Connectivity Equivariant Ansatzes:} The proof-of-concept linear ansatz of Appendix~\ref{app:linear-eqc} should be developed into a full architecture, including hardware execution at $n\!\ge\!50$, MPS validation at $\chi\!\le\!64$, and a study of how its DLA dimension and transfer behaviour compare with the all-to-all baseline.

  \item \textbf{Investigating Alternative Transfer Mechanisms:} The current work assumes direct parameter transfer with adaptation of generator forms. Future research could explore:
    \begin{itemize}
      \item Fine-tuning a subset of layers of the $S_m$-adapted QNN using a small amount of $m$-city data (few-shot transfer).
      \item Layer-wise transfer strategies, where earlier layers (potentially capturing more general features) are transferred, and later layers are re-initialized or fine-tuned.
      \item Methods for transforming or adapting the parameters $\theta_n^*$ themselves to better suit the $m$-city problem, perhaps guided by the change in $T_{e_{n+1}}$ to $T_{e_{m+1}}$.
    \end{itemize}

  \item \textbf{Impact of Noise:} The N4 hardware result establishes that the dense $L=1$ EQC at $n=10$ on
Quantinuum trapped-ion hardware sits in a low-signal regime where the
per-decision top-2 Q-margin ($\sim 6\times 10^{-3}$) is at or below its
propagated sampling error at 4096 shots, and the statevector margin sweep
shows the margin collapsing as $n^{-2.1}$, so the required shot budget grows
as ${\sim}n^{4.3}$ (Appendix~\ref{app:shot-complexity}); this is a hard
limit on the approach with the present ansatz. Extending the analysis to include the effects of quantum hardware noise on the generalization error in the source task and on the transfer penalty $\mathcal{D}_{n \rightarrow m}$ --- and on the per-edge contrast of the $\langle Z_iZ_j\rangle$ observable family at fixed shot budget --- is crucial for assessing practical applicability on near-term quantum devices. A noise-aware transfer bound that quantifies the B3 execution penalty alongside the B2 transfer penalty is a natural theoretical next step.

  \item \textbf{Broader Applicability:} While we focused on TSP, the approach of adapting Theorem~\ref{thm:final-performance} and introducing a transfer penalty term could potentially be applied to other hard combinatorial optimization problems with underlying permutation symmetries (Quadratic Assignment Problem, Flowshop Scheduling, MAX-CUT, MIS), where equivariant QRL agents can be trained and analogous bounds derived. The functional form of the lifted-policy smoothness profile $\Psi(n,m)$ would take a different shape depending on the task family, and the conditional graphon refinement of Corollary~\ref{cor:graphon} would require re-verification of the cut-norm Lipschitz hypothesis in each new setting.
  Appendix~\ref{app:beyond-tsp} outlines how the same equivariant-policy
  template changes for several of these symmetric CO families.
\end{itemize}

The broader message of this work is that the discussion of quantum machine learning at scale must engage simultaneously with theory, simulation, and hardware. The same architecture that is elegant on paper may be infeasible on tensor networks and intractable on hardware --- and these facts may share a single topological cause that paper-level analysis alone cannot expose. We hope the unified multi-backend pipeline introduced here will become a methodological standard for the field.

In conclusion, the conditional diagnostic transfer bound and the multi-backend, multi-vendor experimental campaign together suggest that the canonical all-to-all EQC transfers well at small sizes and modest size jumps but runs into three identifiable barriers --- backend-induced signal loss, large source--target jumps, and finite-shot hardware-execution penalty --- that share a common topological cause and admit a common architectural fix. The framing of this paper is therefore deliberately diagnostic rather than triumphant: symmetry offers powerful advantages for learning, but the challenges of transferring knowledge across problem scales are non-trivial, structural, and warrant careful theoretical and empirical investigation. The sparse-connectivity equivariant ansatz of Appendix~\ref{app:linear-eqc}, developed into a full architecture with end-to-end hardware execution at $n\!\ge\!50$, is in our view the most promising single next step.

\section*{Data availability}
The datasets supporting the findings of this study --- the fixed evaluation
instance banks with stored exact optima, the per-decision hardware artifact
(\texttt{decision\_rows\_recovered.csv}, including per-decision Quantinuum
Nexus job identifiers), the raw per-job measurement counts retrieved from
Quantinuum Nexus, and the controlled emulator run outputs --- are available
in the Zenodo repository at
\href{https://zenodo.org/records/21277812?token=eyJhbGciOiJIUzUxMiJ9.eyJpZCI6IjU4OTZkZGMzLTNkNmYtNDhiMS05MGZhLTQ0NTU0ODdkMjI1MSIsImRhdGEiOnt9LCJyYW5kb20iOiI2MjQ0MjI2ZDY3MjNjMWU2ZmVmMTA0ZDVkN2VjMzQ3MyJ9.Rqy_muKrIPLowpqPyDCO1aY-1hyHy2MKJqLdoaH2KNEz_HTKrPPVQXIffqWBxUMEnsbawo0OTSM7CEQ}{Zenodo repository}.

\section*{Code availability}
The training, transfer, simulation, and hardware-execution code, together
with the analysis scripts producing the decision-resolution statistics,
classical anchors, margin-scaling sweep, controlled emulator runs, and all
figures, is available in the same Zenodo repository at
\href{https://zenodo.org/records/21277812?token=eyJhbGciOiJIUzUxMiJ9.eyJpZCI6IjU4OTZkZGMzLTNkNmYtNDhiMS05MGZhLTQ0NTU0ODdkMjI1MSIsImRhdGEiOnt9LCJyYW5kb20iOiI2MjQ0MjI2ZDY3MjNjMWU2ZmVmMTA0ZDVkN2VjMzQ3MyJ9.Rqy_muKrIPLowpqPyDCO1aY-1hyHy2MKJqLdoaH2KNEz_HTKrPPVQXIffqWBxUMEnsbawo0OTSM7CEQ}{Zenodo repository}.

\section*{Funding}
This project is supported by the National Research Foundation, Singapore
through the National Quantum Office, hosted in A*STAR, under its Quantum Engineering Programme 3.0 Funding Initiative (W24Q3D0002) and Advanced Quantum Computing Infrastructure Funding Initiative (S25Q8D9001).
 
\section*{Author contributions}
M.S.\ conceived and designed the study, developed the theoretical
framework, implemented the evaluation pipeline, performed the simulations
and the emulator and hardware campaigns, analysed the data, and wrote the
manuscript. H.C.L.\ supervised the project, contributed to the theoretical
framing and the interpretation of results, and revised the manuscript. Both
authors approved the final manuscript.
 
\section*{Competing interests}
The authors declare no competing interests.

\newpage

\bibliography{reference}

\newpage

\appendix
\hbadness=10000
\hfuzz=50pt

\section*{Guide to the Appendices}
\noindent This document is organised as follows.
\begin{itemize}[leftmargin=*,itemsep=0.15em]
\item \hyperref[app:tsp-encoding]{Appendix~\ref*{app:tsp-encoding}: TSP Encoding and Equivariance}.
\item \hyperref[sec:app1]{Appendix~\ref*{sec:app1}: Context and Foundations}.
\item \hyperref[app:dla-proof]{Appendix~\ref*{app:dla-proof}: DLA Dimension Computation}.
\item \hyperref[app:assumptions]{Appendix~\ref*{app:assumptions}: Detailed Assumptions for the Transfer Bound}.
\item \hyperref[app:dissimilar_derivation]{Appendix~\ref*{app:dissimilar_derivation}: Decomposition of Performance Degradation}.
\item \hyperref[app:dissimilarity]{Appendix~\ref*{app:dissimilarity}: Quantifying Task Dissimilarity $\mathcal{D}_{n\to m}$}.
\item \hyperref[app:perf-degradation]{Appendix~\ref*{app:perf-degradation}: Incorporating the Theoretical Performance Degradation Bound}.
\item \hyperref[app:analysis]{Appendix~\ref*{app:analysis}: Analysis and Implications of the Derived Bound}.
\item \hyperref[app:scalability-tn]{Appendix~\ref*{app:scalability-tn}: The Scalability Challenge of Simulating Equivariant Quantum Circuits}.
\item \hyperref[app:reproducibility]{Appendix~\ref*{app:reproducibility}: Reproducibility Notes}.
\item \hyperref[app:legacy-eff-su2]{Appendix~\ref*{app:legacy-eff-su2}: Supplementary Empirical Results: Legacy Small-Scale Study}.
\item \hyperref[app:n3-extended]{Appendix~\ref*{app:n3-extended}: Extended N3 Frontier Results}.
\item \hyperref[app:shot-complexity]{Appendix~\ref*{app:shot-complexity}: Shot Complexity of Greedy Decision Resolution}.
\item \hyperref[app:n4-extended]{Appendix~\ref*{app:n4-extended}: Extended N4 Emulator and Hardware Results}.
\item \hyperref[app:linear-eqc]{Appendix~\ref*{app:linear-eqc}: Linear-Connectivity Equivariant Ansatz (Proof of Concept)}.
\item \hyperref[app:beyond-tsp]{Appendix~\ref*{app:beyond-tsp}: Beyond TSP: Equivariant Quantum Policies for Other Symmetric CO Tasks}.
\end{itemize}

% =============================================================================
\section{TSP Encoding and Equivariance}
\label{app:tsp-encoding}

We first specify the Traveling Salesman Problem (TSP) encoding used throughout the paper and clarify how this encoding interacts with permutation symmetry.

\begin{itemize}
\item \textbf{Qubit representation.} For problems with $S_k$ symmetry over $k$ items, a natural representation assigns one qubit to each city. Thus, an $n$-city TSP uses $N_q(n)=n$ qubits and an $m$-city TSP uses $N_q(m)=m$ qubits. Under this encoding, any permutation of city labels is represented by the corresponding permutation of qubits.

\item \textbf{Symmetry of TSP.} The TSP objective is intrinsically $S_k$-symmetric: relabelling the cities changes neither the feasible tour set nor the optimal tour length. Consequently, the reward function for a $k$-city instance should respect this $S_k$ invariance.

\item \textbf{Policy symmetry.} The QRL policy $\pi_\theta$ should respect the same symmetry: after relabelling the cities, the induced action distribution should transform consistently with the relabelling. An $S_k$-equivariant QNN, coupled with an appropriate symmetry-respecting readout, provides a principled mechanism for constructing such a policy.
\end{itemize}

% =============================================================================
\section{Context and Foundations}
\label{sec:app1}

Theorem 4 in Schatzki et al.\ (2024)~\cite{schatzki2024theoretical} provides a significant theoretical guarantee regarding the generalization capabilities of $S_n$-equivariant Quantum Neural Networks. Understanding this theorem is foundational to its adaptation for the zero-shot transfer learning scenario in QRL for TSP.

\subsection{Statement of Theorem 4}

The theorem is stated as follows. Consider a Quantum Machine Learning (QML) problem with a loss function as described in Eq.\ (4) of~\cite{schatzki2024theoretical}. Suppose that an $n$-qubit $S_n$-equivariant QNN $U(\theta)$ is trained on $M$ samples to obtain trained parameters $\theta^*$. Then the following inequality holds with probability at least $1-\delta$:
\[
\mathrm{gen}(\theta^*) \;\le\; \widetilde{\mathcal{O}}\!\left(\sqrt{\frac{T_{e_{n+1}}}{M}} + \sqrt{\frac{\log(1/\delta)}{M}}\right).
\]
Here $\mathrm{gen}(\theta^*)$ is the generalization error, defined as the absolute difference between the true loss $L(\theta^*)$ (expected loss over the true data distribution) and the empirical loss $\hat L(\theta^*)$ (average loss over the $M$ training samples): $\mathrm{gen}(\theta^*) = |L(\theta^*) - \hat L(\theta^*)|$. The bound is obtained via covering-number arguments on the manifold of $S_n$-equivariant unitaries (dimension $T_{e_{n+1}}$, see Lemma~2 of~\cite{schatzki2024theoretical}) combined with the chaining argument of~\cite{Caro2022}.

\subsection{Key Definitions and Assumptions}

The terms and assumptions underpinning Theorem 4 are crucial for its interpretation and subsequent adaptation:

\begin{itemize}
\item \textbf{$S_n$-Equivariant QNN.} The QNN $U(\theta)$ is assumed to operate on $n$ qubits and possess $S_n$-equivariance: its action commutes with the permutation of qubits. Each layer is generated by $S_n$-equivariant Hermitian operators, e.g.\ from $\mathcal{G}=\{\tfrac{1}{n}\sum_j X_j,\ \tfrac{1}{n}\sum_j Y_j,\ \tfrac{2}{n(n-1)}\sum_{k<j}Z_jZ_k\}$. The measurement observable $O$ is also $S_n$-equivariant, e.g.\ from $\mathcal{M}=\{\tfrac{1}{n}\sum_j x_j,\ \tfrac{2}{n(n-1)}\sum_{k<j}x_jx_k,\ \prod_j x_j\}$. Mathematically, $S_n$-equivariance implies
\[
U(\theta) R(\pi) \rho R(\pi)^\dagger U(\theta)^\dagger = R(\pi) U(\theta) \rho U(\theta)^\dagger R(\pi)^\dagger
\]
for any permutation $\pi\in S_n$ with unitary representation $R(\pi)$.

\item \textbf{$T_{e_{n+1}}$ (tetrahedral numbers).} $T_{e_{n+1}}=\binom{n+3}{3}$ is the dimension of the submanifold of $S_n$-equivariant unitaries. It quantifies the effective complexity of the $S_n$-equivariant QNN. For large $n$, $T_{e_{n+1}}\in\Theta(n^3)$. This polynomial scaling, contrasted with the $4^n$ dimension of $\mathrm{SU}(2^n)$, is fundamental to the favourable generalization properties.

\item \textbf{$M$.} The number of training samples drawn from the source-task data distribution.

\item \textbf{Loss function.} $l_\theta(\rho_i)=\mathrm{Tr}[U_\theta(\rho_i)O]$ where $U_\theta(\rho_i)=U(\theta)\rho_i U(\theta)^\dagger$. The observable $O$ has bounded operator norm, and the coefficients $c_i$ in the empirical loss are bounded (e.g.\ $|c_i|\le 1/M$).
\end{itemize}

\subsection{Proof Sketch and Underlying Mathematical Concepts}

The proof of Theorem~4 in~\cite{schatzki2024theoretical} relies on bounding the $\varepsilon$-covering number of the set of $S_n$-equivariant QNNs. A covering number $\mathcal{N}(V,d,\varepsilon)$ is the minimum number of balls of radius $\varepsilon$ (with respect to a metric $d$) needed to cover the set $V$.

\paragraph{Block-diagonal structure.}
A cornerstone of the argument is the representation theory of $S_n$. An $S_n$-equivariant unitary $U(\theta)$ decomposes into a block-diagonal form
\[
U(\theta) \;\cong\; \bigoplus_\lambda I_{m_\lambda}\otimes U_\lambda(\theta).
\]
Here $\lambda$ labels the irreducible representations (irreps) of $S_n$ appearing in the decomposition of the $n$-qubit Hilbert space, $I_{m_\lambda}$ is an identity matrix of dimension $m_\lambda$ (the multiplicity of the irrep $r_\lambda$, by Schur--Weyl duality), and $U_\lambda(\theta)$ are $d_\lambda$-dimensional unitary matrices.

\paragraph{Covering-number bound.}
The $\varepsilon$-covering number satisfies
\[
\mathcal{N}(V_n,\|\cdot\|,\varepsilon) \;\le\; \left(\frac{6}{\varepsilon}\right)^{2\sum_\lambda d_\lambda^2}.
\]
The sum $\sum_\lambda d_\lambda^2$ is precisely the tetrahedral number $T_{e_{n+1}}$. The function class implementable by $S_n$-equivariant QNNs therefore scales polynomially in $n$ (via $T_{e_{n+1}}\in\Theta(n^3)$).

\paragraph{Generalization bound from covering numbers.}
This polynomial bound on the covering number, leveraged within standard ML theory frameworks (Rademacher complexity, VC dimension, or direct covering-number-based arguments), yields the generalization bound stated above. Smaller covering numbers imply that the function class is less complex, and thus fewer samples are needed.

The overall trainability and expressibility context of $S_n$-equivariant QNNs also relies on the assumption that the QNN ansatz forms approximate 2-designs on each isotypic block, ensuring the QNN can effectively explore the symmetry-constrained subspaces. While the proof of Theorem 4 itself focuses on the covering number of the unitary manifold, this underlying expressivity ensures effective exploration.

\subsection{Significance for QML}

Theorem~4 demonstrates that by incorporating problem symmetries into the
QNN architecture, polynomial sample complexity is achievable \emph{for the
supervised $S_n$-equivariant QNN loss class}: $M\in O(n^3)$ at fixed error
and confidence, inherited from $T_{e_{n+1}}\in\Theta(n^3)$. We emphasise
the scope of this statement: it is proved for supervised equivariant loss
functions and is \emph{not} established for the non-smooth greedy-rollout
policy class studied in this paper, where it enters only as the optimistic
motivating rate behind Assumption~(A2) (see Table~\ref{tab:theory-status}).

% =============================================================================
\section{DLA Dimension Computation}
\label{app:dla-proof}

We prove Proposition~\ref{prop:dla-dim} (Section~\ref{sec:simulability}). Let
the EQC generators be
\[
\begin{aligned}
G_1 &= \frac{1}{n}\sum_{j=1}^{n} X_j,
&G_2 &= \frac{1}{n}\sum_{j=1}^{n} Y_j,\\
G_3 &= \frac{2}{n(n-1)}\sum_{j<k} Z_j Z_k.
\end{aligned}
\]
The Dynamical Lie Algebra $\mathfrak{g}=\langle iG_1,iG_2,iG_3\rangle_{\mathrm{Lie}}$ has dimension
\[
\dim(\mathfrak{g}) = T_{e_{n+1}} - \nu_n = \binom{n+3}{3}-\nu_n \in \Theta(n^3),
\]
where $\nu_n=\lfloor n/2\rfloor+1$ is the number of $S_n$-irreducible blocks in the qubit Hilbert space. When dynamics are restricted to the totally symmetric (Dicke) sector $\lambda=(n,0)$ with $d_\lambda=n+1$, the effective DLA dimension is at most $(n+1)^2-1=O(n^2)$.

\begin{proof}[Proof sketch]
By Schur--Weyl duality, the $n$-qubit Hilbert space $(\mathbb{C}^2)^{\otimes n}$ decomposes under the joint action of $S_n$ (permuting qubits) and $U(2)$ as
\[
(\mathbb{C}^2)^{\otimes n}\;\cong\;\bigoplus_\lambda V_\lambda^{S_n}\otimes W_\lambda^{U(2)},
\]
where $\lambda$ ranges over partitions of $n$ into at most two parts: $\lambda=(n-k,k)$ for $k=0,1,\dots,\lfloor n/2\rfloor$. The $S_n$-equivariant generators act trivially on $V_\lambda^{S_n}$ and non-trivially on $W_\lambda^{U(2)}$, which has dimension
\[
d_\lambda \;=\; n-2k+1.
\]

The key result from Schatzki et al.~\cite[Theorem 4]{schatzki2024theoretical} is that $S_n$-equivariant unitaries form a manifold of dimension
\[
\sum_\lambda d_\lambda^2 \;=\; \sum_{k=0}^{\lfloor n/2\rfloor}(n-2k+1)^2 \;=\; T_{e_{n+1}} \;=\; \binom{n+3}{3}.
\]
This identity can be verified by induction or by direct summation.

Since $\{iG_1,iG_2,iG_3\}\subset \bigoplus_\lambda \mathfrak{u}(d_\lambda)$ and Theorem 2 of~\cite{schatzki2024theoretical} shows that 1-body and 2-body $S_n$-equivariant operators generate the full $S_n$-equivariant subalgebra (an explicit basis construction is also given in~\cite{allcock2024dla} for the QAOA-style generator set used in our work), we have
\[
\mathfrak{g} \;=\; \bigoplus_\lambda \mathfrak{su}(d_\lambda).
\]
Its dimension is
\[
\dim(\mathfrak{g}) = \sum_\lambda (d_\lambda^2-1) = T_{e_{n+1}}-\nu_n,
\]
with $\nu_n=\lfloor n/2\rfloor+1$. For scaling analysis, $\dim(\mathfrak{g})\in\Theta(n^3)$.

\textbf{Restriction to the Dicke sector.} The initial state $|+\rangle^{\otimes n}$ lies in the totally symmetric irrep $\lambda=(n,0)$ with $d_\lambda=n+1$. Since both generators preserve symmetry sectors, the dynamics are confined to this sector, where the effective DLA is at most $\mathfrak{su}(n+1)$, of dimension $(n+1)^2-1=O(n^2)$.
\end{proof}

\paragraph{Simulability consequence.}
By the $\mathfrak{g}$-sim framework of~\cite{goh2025lie}, the EQC's loss function is computable in classical time $\mathcal{O}(L\dim(\mathfrak{g})^2)=\mathcal{O}(Ln^6)$ in the full space, or $\mathcal{O}(Ln^4)$ in the symmetric sector. This is the simulability statement that Section~\ref{sec:simulability} engages with: it does not invalidate the heuristic value of the architecture, but it does foreclose any claim of computational quantum advantage for the unmodified EQC.

% =============================================================================
\section{Detailed Assumptions for the Transfer Bound}
\label{app:assumptions}

To establish the conditional diagnostic transfer bound (Theorem~\ref{thm:final-performance}) we introduce the following assumptions. They map the supervised generalization framework of~\cite{schatzki2024theoretical} to the reinforcement-learning setting with sequential greedy policy execution. We label each assumption explicitly as \textbf{proved}, \textbf{assumed}, or \textbf{speculative (conditional refinement)} so the conditional structure of the bound is transparent.

\paragraph{(A1) Joint equivariance of Model~B.}
\emph{(Proved.)} The implemented data-dependent EQC satisfies $U_n^{\rm B}(\theta;\pi\cdot D,\pi\cdot s)=R(\pi)\,U_n^{\rm B}(\theta;D,s)\,R(\pi)^\dagger$ for every $\pi\in S_n$. This is the joint-equivariance property of the architecture; it is strictly weaker than fixed-instance $S_n$-invariance, which fails for generic Euclidean $D$.

\paragraph{(A2) Rollout source-task generalisation
\texorpdfstring{$\mathcal{G}^{\rm roll}_n(\delta)$}{G roll n delta}.}
\label{ass:rollout-gen}
\emph{(Assumed.)}
The induced greedy-rollout performance class
\(\mathcal{F}_n := \{\theta \mapsto P_n(\theta)\}\) admits a
source-task generalisation bound: there exists a function
\(\mathcal{G}^{\rm roll}_n(\delta)\) such that, with probability at least
\(1-\delta\),
\[
  P_n(\theta_n^*) \ge
  \widehat P_n(\theta_n^*) - \mathcal{G}^{\rm roll}_n(\delta).
\]
This is the operative source-task generalisation term used in the transfer
bound.

In the optimistic symmetry-preserving case, where the greedy decoding map
inherits the equivariant covering behaviour of the underlying QNN class, one
would expect
\[
  \mathcal{G}^{\rm roll}_n(\delta)
  =
  \widetilde{\mathcal{O}}
  \left(
    \sqrt{\frac{T_{e_{n+1}}}{M_n}}
    +
    \sqrt{\frac{\log(1/\delta)}{M_n}}
  \right),
\]
where \(T_{e_{n+1}}=\binom{n+3}{3}\in\Theta(n^3)\).
Theorem~4 of~\cite{schatzki2024theoretical} establishes this rate for
\(S_n\)-equivariant supervised loss classes. However, QRL rollouts involve
trajectory-level dependence and a generally non-smooth greedy
\(\arg\max\) decoder, which is not itself a Schatzki-class observable.
Therefore, we do not claim that the supervised rate applies directly.
The transfer theorem below requires only the existence of
\(\mathcal{G}^{\rm roll}_n(\delta)\); the explicit Schatzki rate is used
only as a motivating inherited rate.

For completeness, we note three regimes under which a finite
\(\mathcal{G}^{\rm roll}_n(\delta)\) can be defended.

\textbf{(A2-i) Fresh-instance regime.}
The agent is trained for \(M_n\) episodes, each started on an independently
drawn instance \(I^{(e)}\sim\mathcal{D}_n\). If the exploratory policy is
fixed in advance, the episode returns \(\{R^{(e)}\}_{e=1}^{M_n}\) are
i.i.d. and bounded in \([0,1]\). The optimistic Schatzki-type rate applies
if the greedy-decoded rollout class inherits the equivariant covering rate
of the underlying QNN class.

\textbf{(A2-ii) Finite-pool replay regime.}
If only a pool
\(\mathcal{P}_n=\{I^{(1)},\dots,I^{(N_{\mathrm{pool}})}\}\) of distinct
instances is available and each instance is replayed \(K\) times
\((M_n=N_{\mathrm{pool}}K)\), the effective independent sample size is
\(N_{\mathrm{pool}}\). In this case, the corresponding optimistic rate is
\[
  \widetilde{\mathcal{O}}
  \left(
    \sqrt{\frac{T_{e_{n+1}}}{N_{\mathrm{pool}}}}
    +
    \sqrt{\frac{\log(1/\delta)}{N_{\mathrm{pool}}}}
  \right).
\]

\textbf{(A2-iii) Online \(Q\)-learning with dependent returns.}
If the policy is updated on the fly, the return sequence is no longer
i.i.d. Under a \(\beta\)-mixing assumption with mixing time \(\tau_\beta\),
uniform-stability arguments for \(\beta\)-mixing processes
\cite{mohri2010stability,abeles2024generalization} yield an optimistic
rate of the form
\[
  \widetilde{\mathcal{O}}
  \left(
    \sqrt{\frac{T_{e_{n+1}}\tau_\beta}{M_n}}
  \right).
\]

\paragraph{(A3) Direct parameter transfer.}
\emph{(Setup.)} The trained parameters $\theta_n^*$ are applied directly to the Model~B circuit at city count $m$ with $D,s$ inherited from the target $m$-city instance, with no parameter modification.

\paragraph{(A4) Performance regularity.}
\emph{(Proved for linear observables; extended to the expected rollout
performance under a margin-density condition,
Proposition~\ref{prop:margin-smoothing}; the pointwise rollout statement is
assumed.)} The performance functional $P_k$ is Lipschitz in three relevant
variables, with constants $L_\theta,L_D,L_U$ in parameter, instance, and
unitary space respectively:
\begin{itemize}
\item \textbf{Parameter Lipschitz.} $|P_k(\theta)-P_k(\theta')|\le
  L_\theta\|\theta-\theta'\|_2$ (constant generator norms, bounded reward in
  $[0,1]$~\cite{barthe2024gradients,meyer2024robustness}).
\item \textbf{Instance Lipschitz.} $|P_k(\theta;D)-P_k(\theta;D')|\le
  L_D\|D-D'\|_F$ with $L_D=k/(L_{\max}-L_{\mathrm{opt}})$, since tour length
  is 1-Lipschitz in each edge weight~\cite{asadi2018lipschitz}.
\item \textbf{Unitary Lipschitz.} For linear observables
  $P_k(U)=\mathrm{Tr}(MU\rho U^\dagger)$ with $\|M\|_{\rm op}\le 1$, the bound
  $|P_k(U)-P_k(U')|\le L_U\|U-U'\|_{\rm op}$ holds with $L_U\le
  2\|M\|_{\rm op}\le 2$ (Lemma~\ref{lem:unitary-lip-app}). The greedy-rollout
  performance is \emph{not} of this form: it is a nonlinear functional of $U$
  (sequential $\arg\max$ decisions), and a pointwise Lipschitz bound fails in
  general, because an arbitrarily small unitary perturbation can cross an
  action margin and change the selected tour discontinuously. Since $P_k$ is
  defined throughout as the \emph{expected} performance over the instance
  distribution $\mathcal D_k$ (Section~\ref{sec:notation}), the operative
  statement is Proposition~\ref{prop:margin-smoothing} below: the expected
  rollout performance is locally Lipschitz in $U$ with effective constant
  $L_U^{\rm eff}\le 4k\bar\rho B$, where $\bar\rho$ bounds the density of the
  per-step top-2 action margin near zero and $B$ is the aggregation norm of
  Assumption~(A8). The per-step action-value estimates themselves are linear
  observables, to which Lemma~\ref{lem:unitary-lip-app} applies directly.
\end{itemize}

\paragraph{(A5) BHH size-smoothness of optimal tour length (motivational).}
\emph{(Proved; used motivationally only.)} For Euclidean families on $[0,1]^2$, $\mathbb{E}[L_k^{\rm opt}]=\beta_2\sqrt{k}+o(\sqrt{k})$ with $\beta_2\in(0.625,0.903)$~\cite{beardwood1959shortest,steele1997probability,steinerberger2015new}, and $L_k^{\rm opt}$ concentrates (Talagrand, subadditive Euclidean functionals). This is used \emph{only} to motivate the functional form $\Psi(n,m)=\sqrt{m}-\sqrt{n}$ of the lifted-policy smoothness Assumption~(A7); under the approximation-ratio metric of Section~\ref{sec:notation} it does not itself supply a structural term, because $P_k^{\rm opt}=1$ identically across $k$ (Remark~\ref{rem:no-popt-diff}).

\paragraph{(A6) Cut-norm Lipschitz dependence of the lifted-policy rollout performance.}
\emph{(Speculative; conditional refinement only.)} For the graphon-tightened form (Corollary~\ref{cor:graphon}), we assume the lifted-policy performance map $W_k\mapsto P_k(U_n^\uparrow(\theta_n^*))$, viewed as a functional of the empirical graphon $W_k$ of the instance distribution, is $C_W$-Lipschitz in the cut norm: $|P_k(U_n^\uparrow)|_{W_k}-P_k(U_n^\uparrow)|_{W'_k}|\le C_W\,\delta_\square(W_k,W'_k)$. This is a strong hypothesis: TSP rollout performance is not known to be a cut-norm-continuous graphon functional, and we do not prove it. The graphon refinement should be read as a speculative tightening, not as a main result.

\paragraph{(A7) Lifted-policy smoothness --- the operative structural assumption.}
\label{ass:lift-smooth}
\emph{(Assumed; central regularity hypothesis.)} Let $U_n^\uparrow(\theta_n^*):=U_n^{\rm B}(\theta_n^*)\otimes I_{2^{m-n}}$ be the identity-padded lift of the source unitary into the target Hilbert space. We assume there is a size-smoothness profile $\Psi(n,m)\ge 0$ and a constant $C_{\rm lift}$ with
\[
\big|\,P_n(U_n^{\rm B}(\theta_n^*)) - P_m(U_n^\uparrow(\theta_n^*))\,\big|
\;\le\; C_{\rm lift}\,\Psi(n,m).
\]
For Euclidean TSP families on $[0,1]^2$ we take $\Psi(n,m)=\sqrt{m}-\sqrt{n}$, motivated by (A5). This is the assumption that actually supplies the structural term of the transfer bound; BHH does not.

\paragraph{(A8) Aggregation norm of the policy functional.}
\emph{(Proved given explicit policy form.)} Each action value is a weighted
aggregate of measured correlators,
$Q_a=\sum_{e} w_{a,e}\,\langle \hat O_e\rangle$ with single-shot outcomes
$\hat O_e\in[-1,1]$, and the policy aggregation norm
$B:=\max_a \sum_e |w_{a,e}|$ satisfies $B\le\bar B(k)$ for an explicit growth
function $\bar B$ (e.g.\ $\bar B(k)=O(1)$ for normalised averages or
single-correlator readout, $\bar B(k)=O(k)$ for unnormalised sums of
$\sim k$ correlators). For the implemented readout, each action value
aggregates exactly \emph{one} correlator,
$Q_a = d_{(c,a)}\,\langle Z_c Z_a\rangle$ with $c$ the current city and
$d_{(c,a)}$ the corresponding edge length, so
$B=\max_a d_{(c,a)} \le \sqrt{2}$ on the unit square; the measured value on
the five hardware instances is $B=1.06$. This constant enters the analysis
twice: it sets the effective Lipschitz constant of the smoothed rollout
performance (Proposition~\ref{prop:margin-smoothing}) and it sets the
per-decision shot complexity of greedy action resolution
(Proposition~\ref{prop:shot-complexity}), which formalises the B3 diagnostic of
Section~\ref{sec:barrier-hardware}.

\begin{lemma}[Unitary Lipschitz constant, used in (A4)]
\label{lem:unitary-lip-app}
If $P_k(U)=\mathrm{Tr}(MU\rho U^\dagger)$ with $\|M\|_{\rm op}\le 1$ and $\rho$ a density operator, then $|P_k(U)-P_k(U')|\le 2\|U-U'\|_{\rm op}$.
\end{lemma}
\begin{proof}
$U\rho U^\dagger-U'\rho U'^\dagger=U\rho(U-U')^\dagger+(U-U')\rho U'^\dagger$; each summand has trace norm at most $\|U-U'\|_{\rm op}$ by H\"older.
\end{proof}

\begin{proposition}[Margin-density smoothing of the rollout performance]
\label{prop:margin-smoothing}
Fix a city count $k$ and instance distribution $\mathcal D_k$. For a unitary
$U$ and instance $I$, let $\Delta_t(U;I)\ge 0$ denote the top-2 action-value
margin at step $t$ of the greedy rollout of $U$ on $I$, and suppose:
(i)~each action value $Q_a(U;I)=\mathrm{Tr}\!\big(M_a(I)\,U\rho_t U^\dagger\big)$
is a linear observable with $\|M_a(I)\|_{\rm op}\le B$, so that
$|Q_a(U)-Q_a(U')|\le 2B\|U-U'\|_{\rm op}$ by
Lemma~\ref{lem:unitary-lip-app}; and (ii)~there exist $\bar\rho<\infty$ and
$\eta>0$ such that for every step $t\le k$,
\[
  \Pr_{I\sim\mathcal D_k}\!\big[\Delta_t(U;I)\le \epsilon\big]
  \;\le\; \bar\rho\,\epsilon
  \qquad\text{for all }0\le\epsilon\le\eta
\]
(a bounded margin density near zero). Then the expected rollout performance
$\bar P_k(U):=\mathbb E_{I\sim\mathcal D_k}\!\big[P_k(U;I)\big]$ is locally
Lipschitz: for all $U,U'$ with $\|U-U'\|_{\rm op}\le \eta/(4B)$,
\[
  \big|\bar P_k(U)-\bar P_k(U')\big|
  \;\le\; L_U^{\rm eff}\,\|U-U'\|_{\rm op},
  \qquad
  L_U^{\rm eff} \;\le\; 4k\,\bar\rho\,B .
\]
\end{proposition}
 
\begin{proof}
Write $\epsilon:=\|U-U'\|_{\rm op}$ and couple the two rollouts on the same
instance $I$. The rollouts agree up to the first step $t^*$ (if any) at which
their $\arg\max$ selections differ; at that step both rollouts occupy the
same state, so the two action-value vectors differ entrywise by at most
$2B\epsilon$ by hypothesis~(i), and a differing $\arg\max$ forces
$\Delta_{t^*}(U;I)\le 4B\epsilon$. Hence the event
$\{P_k(U;I)\neq P_k(U';I)\}$ is contained in
$\bigcup_{t\le k}\{\Delta_t(U;I)\le 4B\epsilon\}$, whose probability is at
most $k\,\bar\rho\,(4B\epsilon)$ by hypothesis~(ii) whenever
$4B\epsilon\le\eta$. On the complementary event the two tours coincide, and
the approximation-ratio performance, which is a deterministic function of the
tour and the instance, coincides as well. Since $P_k\in[0,1]$,
$|\bar P_k(U)-\bar P_k(U')|\le 1\cdot\Pr[\text{tours differ}]\le
4k\bar\rho B\,\epsilon$.
\end{proof}
 
\begin{remark}[The smoothing constant is the B3 mechanism in disguise]
\label{rem:margin-smoothing-b3}
Proposition~\ref{prop:margin-smoothing} ties the theory of
Theorem~\ref{thm:final-performance} to the hardware diagnostic of
Section~\ref{sec:barrier-hardware}: the effective unitary-Lipschitz constant
of the rollout performance is governed by the probability mass of small
top-2 margins. The hardware campaign measures precisely this quantity and
finds it large --- the empirical margin distribution at $n=10$ concentrates
in the $10^{-4}$--$10^{-2}$ band (Figure~\ref{fig:n4-hardware}c) --- so the
same mechanism that makes greedy decisions statistically unresolvable at
finite shots (Barrier~B3, Proposition~\ref{prop:shot-complexity}) also inflates
$L_U^{\rm eff}$ and thereby the parametric-mismatch term of the transfer
bound. Theory and experiment point at one object: the margin distribution of
the trained policy.
\end{remark}

\paragraph{Status summary.}
The bound of Theorem~\ref{thm:final-performance} is \emph{conditional}: it
rests on the joint-equivariance property (A1, proved), the rollout
source-generalisation assumption (A2, assumed with optimistic Schatzki rate),
the direct-transfer protocol (A3, setup), performance regularity (A4, proved
for linear observables and for the expected rollout performance under the
margin-density condition of Proposition~\ref{prop:margin-smoothing}; assumed
pointwise otherwise), and lifted-policy smoothness (A7, central regularity
assumption). Assumption (A5) is proved but used only motivationally.
Assumption (A6) is a speculative refinement, used only in
Corollary~\ref{cor:graphon}, and is \emph{not} the headline content of the
paper. Assumption (A8) supplies the aggregation norm used in
Proposition~\ref{prop:margin-smoothing} and in the shot-complexity bound of
Proposition~\ref{prop:shot-complexity}. The label ``conditional diagnostic'' in
the theorem name is meant to make this structure transparent.
% =============================================================================
\section{Decomposition of Performance Degradation}
\label{app:dissimilar_derivation}

We derive the conditional decomposition that underlies Theorem~\ref{thm:final-performance}. The derivation is carried out for the implemented Model~B circuit; the cross-size comparison is mediated by an identity-padded lift of the source unitary into the target Hilbert space. We do not invoke group twirling, for reasons spelled out in Remark~\ref{rem:no-twirling}.

\paragraph{Setup.}
The total transfer gap is
\[
\Delta P_{n\to m}(\theta_n^*) := P_n(U_n^{\rm B}(\theta_n^*)) - P_m(U_m^{\rm B}(\theta_n^*)).
\]
We use the approximation-ratio performance metric of Section~\ref{sec:notation}, under which $P_k^{\rm opt}=1$ for every $k$; a structural term written as a difference of optimal performances is therefore identically zero (Remark~\ref{rem:no-popt-diff} below) and the operative structural quantity is a lifted-policy smoothness assumption.

\paragraph{Insertion of the lifted unitary.}
Define the identity-padded lift of the source unitary into the target Hilbert space,
\[
U_n^\uparrow(\theta_n^*) := U_n^{\rm B}(\theta_n^*)\otimes I_{2^{m-n}}.
\]
Adding and subtracting $P_m(U_n^\uparrow(\theta_n^*))$ yields the clean decomposition
\begin{align}
\Delta P_{n\to m}(\theta_n^*)
&= \underbrace{P_n(U_n^{\rm B}(\theta_n^*)) - P_m(U_n^\uparrow(\theta_n^*))}_{\text{structural: lifted-policy smoothness}} \nonumber\\
&\quad + \underbrace{P_m(U_n^\uparrow(\theta_n^*)) - P_m(U_m^{\rm B}(\theta_n^*))}_{\text{parametric mismatch}}.
\label{eq:decomp}
\end{align}
The first bracket compares the source policy at its native size to the same parameters lifted into the target Hilbert space; it is bounded by the lifted-policy smoothness Assumption (A7). The second bracket compares two unitaries acting on the \emph{same} target Hilbert space and differing only through the size-dependence of their generators; it is bounded by Duhamel telescoping together with the generator-difference lemma.

\begin{remark}[The twirling shortcut is not used]
\label{rem:no-twirling}
Earlier versions of this derivation used a group-twirling step, replacing $U_n^\uparrow$ by a $S_m$-twirled object and arguing that twirling leaves the achieved performance unchanged because the environment dynamics and $P_m(\cdot)$ are permutation-invariant. This step is not used here. Twirling a unitary produces a quantum channel, not another unitary, and the greedy-rollout performance $P_m(\cdot)$ is a nonlinear functional of the unitary (it involves sequential $\arg\max$ decisions and accumulated tour length). Permutation-invariance of the environment therefore does not automatically imply that twirling leaves the rollout performance unchanged; the equality $P_m(\widetilde U_n^\uparrow)=P_m(U_n^\uparrow)$ does not follow in general for nonlinear policy functionals, and we avoid relying on it.
\end{remark}

\begin{remark}[Why the structural term is lifted-policy smoothness, not a $P^{\rm opt}$ difference]
\label{rem:no-popt-diff}
Under the approximation-ratio metric of Section~\ref{sec:notation}, $P_k^{\rm opt}=1$ for every $k$, so any structural term defined as $P_n^{\rm opt}-P_m^{\rm opt}$ would be identically zero and contribute nothing. The structurally relevant quantity is instead how the \emph{same} (lifted) source policy behaves at the two sizes:
\[
\big|\,P_n(U_n^{\rm B}(\theta_n^*)) - P_m(U_n^\uparrow(\theta_n^*))\,\big|
\;\le\; C_{\rm lift}\,\Psi(n,m),
\]
which is exactly Assumption~(A7). Cross-size degradation is driven by how the learned-vs-optimal gap (not the optimal length) evolves with size. BHH controls $L^{\rm opt}_k$, which is normalised away by the ratio metric; it therefore serves only to motivate the $\sqrt{m}-\sqrt{n}$ shape of $\Psi(n,m)$ and does not by itself bound the transferred policy.
\end{remark}

\paragraph{Bounding the parametric mismatch.}
The implemented Model~B EQC at $L=1$ contains both a one-body mixer ($H_M(s)=\sum_i s_i X_i$) and a two-body cost ($H_C(D)=\sum_{i<j} d_{ij} Z_i Z_j$). The parametric mismatch involves the size-dependence of each. We split the analysis into a one-body generator-difference lemma, a new two-body generator-difference lemma, and a unitary-deviation bound that combines them.

\begin{lemma}[One-body generator difference]
\label{lem:gen-diff-1body}
For $H_1^{(k)}=\tfrac{1}{k}\sum_{j=1}^k O_j$ with $\|O_j\|_{\rm op}\le 1$ and the identity lift $H_1^{(n),\uparrow}=(\tfrac{1}{n}\sum_{j=1}^n O_j)\otimes I$,
\[
\|H_1^{(m)} - H_1^{(n),\uparrow}\|_{\rm op} \;\le\; 2\,\frac{m-n}{m}.
\]
\end{lemma}
\begin{proof}
Writing $H_1^{(m)} - H_1^{(n),\uparrow} = (\tfrac{1}{m}-\tfrac{1}{n})\sum_{j=1}^n O_j + \tfrac{1}{m}\sum_{j=n+1}^m O_j$, the first summand has norm at most $(m-n)/m$ and the second has norm at most $(m-n)/m$.
\end{proof}

\begin{lemma}[Two-body generator difference]
\label{lem:gen-diff-2body}
For $H_2^{(k)}=\tfrac{2}{k(k-1)}\sum_{1\le i<j\le k} B_{ij}$ with $\|B_{ij}\|_{\rm op}\le 1$ and the identity lift $H_2^{(n),\uparrow}=\big(\tfrac{2}{n(n-1)}\sum_{i<j\le n}B_{ij}\big)\otimes I$,
\[
\|H_2^{(m)} - H_2^{(n),\uparrow}\|_{\rm op}
\;\le\; 2\left(1-\frac{n(n-1)}{m(m-1)}\right)
\;\le\; 4\,\frac{m-n}{m}.
\]
\end{lemma}
\begin{proof}
Decompose $H_2^{(m)} - H_2^{(n),\uparrow} = \big(\tfrac{2}{m(m-1)}-\tfrac{2}{n(n-1)}\big)\sum_{i<j\le n}B_{ij} + \tfrac{2}{m(m-1)}\sum_{i<j\le m,\,j>n}B_{ij}$. The first piece has norm at most $\binom{n}{2}\cdot\big|\tfrac{2}{n(n-1)}-\tfrac{2}{m(m-1)}\big|=1-\tfrac{n(n-1)}{m(m-1)}$; the second piece has norm at most $(\binom{m}{2}-\binom{n}{2})\cdot\tfrac{2}{m(m-1)}=1-\tfrac{n(n-1)}{m(m-1)}$. Adding the two and using $1-\tfrac{n(n-1)}{m(m-1)}\le 2(m-n)/m$ for $m\ge n\ge 1$ gives the stated bound.
\end{proof}

\begin{lemma}[Unitary deviation]
\label{lem:unitary-dev}
With $U_n^\uparrow(\theta_n^*)$ and $U_m^{\rm B}(\theta_n^*)$ defined as above, layer-by-layer Duhamel telescoping gives
\[
\|U_n^\uparrow(\theta_n^*) - U_m^{\rm B}(\theta_n^*)\|_{\rm op}
\;\le\; C_{\rm gen}\,\|\theta_n^*\|_1\,\frac{m-n}{m},
\]
where $C_{\rm gen}$ is the generator-difference constant: $C_{\rm gen}=2$ for one-body normalised generators (Lemma~\ref{lem:gen-diff-1body}) and $C_{\rm gen}=4$ for two-body normalised generators (Lemma~\ref{lem:gen-diff-2body}). For the Skolik EQC, which contains both a one-body mixer and a two-body cost layer, both branches contribute and the binding constant is $C_{\rm gen}=4$.
\end{lemma}
\begin{proof}
Telescoping over layers and using $\|e^{-i\phi X}-e^{-i\phi Y}\|_{\rm op}\le|\phi|\|X-Y\|_{\rm op}$ (Duhamel) gives $\|U_n^\uparrow-U_m^{\rm B}\|_{\rm op}\le \sum_\ell |\theta_{n,\ell}^*|\,\|H_\ell^{(m)} - H_\ell^{(n),\uparrow}\|_{\rm op}\le \|\theta_n^*\|_1\cdot C_{\rm gen}\,(m-n)/m$ by Lemma~\ref{lem:gen-diff-1body} or Lemma~\ref{lem:gen-diff-2body}.
\end{proof}

By Assumption (A4) the performance functional has unitary-Lipschitz constant $L_U$, so the second bracket of \eqref{eq:decomp} is controlled by the unitary deviation:
\begin{equation}
\big|\,P_m(U_n^\uparrow(\theta_n^*)) - P_m(U_m^{\rm B}(\theta_n^*))\,\big|
\;\le\; L_U \cdot C_{\rm gen}\,\|\theta_n^*\|_1\,\frac{m-n}{m}.
\label{eq:param-bound}
\end{equation}

\begin{remark}[Correction to earlier versions]
Earlier versions of this derivation wrote the parametric mismatch as $2L_U\|A\|_{\rm op}\|\theta_n^*\|_1(m-n)/m$, double-counting a factor of $2$: with $C_{\rm gen}$ defined as the generator-difference constant (and absorbing the operator-norm prefactor of the generators), the leading $2$ is spurious. The corrected form is \eqref{eq:param-bound}.
\end{remark}

\paragraph{Bounding the structural (lifted-policy smoothness) term.}
The first bracket of \eqref{eq:decomp} is bounded by Assumption (A7) directly:
\[
\big|\,P_n(U_n^{\rm B}(\theta_n^*)) - P_m(U_n^\uparrow(\theta_n^*))\,\big|
\;\le\; C_{\rm lift}\,\Psi(n,m).
\]
For Euclidean TSP families under uniform sampling on $[0,1]^2$ we take $\Psi(n,m)=\sqrt{m}-\sqrt{n}$, motivated (but not implied) by the BHH theorem. Under the additional cut-norm Lipschitz hypothesis of Assumption~(A6), Corollary~\ref{cor:graphon} replaces $\Psi$ by $\sqrt{\log n/n}+\sqrt{\log m/m}$.

\paragraph{Combined transfer-gap bound.}
\begin{align*}
\Delta P_{n\to m}(\theta_n^*)
&\le L_U\, C_{\rm gen}\,\|\theta_n^*\|_1\,\frac{m-n}{m}
+ C_{\rm lift}\,\Psi(n,m),
\end{align*}
with the conditional graphon refinement replacing $\Psi(n,m)$ under Assumption~(A6). This defines $\mathcal{D}_{n\to m}$ as the operative form of the transfer penalty. The decomposition is a triangle-inequality upper bound, not a sharp mechanistic partition: the same $\theta_n^*$ enters both terms, so the two mechanisms can interact.

\paragraph{Status.}
With $L_U=2$, $C_{\rm gen}\le 4$, and $\|\theta_n^*\|_1\sim O(1)$, the parametric term alone can exceed unity, in which case the bound \eqref{eq:param-bound} is vacuous and the theorem returns a trivial $P_m\ge\text{negative}$. The bound's value is the scaling structure it identifies (linear in $\|\theta_n^*\|_1\,(m-n)/m$, plus a sub-linear lifted-policy smoothness term), not numerical prediction of measured gaps.

% =============================================================================
\section{Quantifying Task Dissimilarity \texorpdfstring{$\mathcal{D}_{n\to m}$}{D n to m}}
\label{app:dissimilarity}

A critical component in bridging the single-task generalization bound of~\cite{schatzki2024theoretical} to a transfer learning bound is the introduction of a term quantifying the dissimilarity or transfer penalty incurred when applying a model trained on a source task $T_n$ to a target task $T_m$. We denote this $\mathcal{D}_{n\to m}(\theta_n^*)$.

The dissimilarity arises from several interconnected factors:
\begin{enumerate}
\item \textbf{Change in problem scale and structure.} The most obvious difference is the increase from $n$ to $m$ cities, which alters the search space ($(n-1)!/2\to(m-1)!/2$) and the geometric or combinatorial structure of optimal solutions.
\item \textbf{Parameter mismatch under direct transfer.} By Assumption (A3) (direct parameter transfer), $\theta_n^*$ is applied to the $m$-city Model~B circuit without modification. The cost generator $H_C(D)$ is the data-dependent two-body operator $\sum_{i<j}d_{ij}Z_iZ_j$ inherited from the target $m$-city instance, and the mixer $H_M(s)$ is the one-body operator $\sum_i s_i X_i$ inherited from the adjacency-diagonal node features; both differ between the $n$- and $m$-qubit settings. The resulting unitary evolution $U_m^{\rm B}(\theta_n^*)$ generally differs from what would be optimal for the $m$-city problem, and the resulting parametric mismatch is bounded in Appendix~\ref{app:dissimilar_derivation} (Lemmas~\ref{lem:gen-diff-1body}--\ref{lem:unitary-dev}) via the generator-difference constant $C_{\rm gen}\in\{2,4\}$.
\item \textbf{Smoothness of the lifted-policy rollout performance.} Even setting aside parametric mismatch, the same identity-padded lift $U_n^\uparrow(\theta_n^*)$ evaluated at the larger city count $m$ generally yields a different rollout performance than at the native size $n$. This cross-size variation is the operative structural quantity of the transfer bound and is bounded by the lifted-policy smoothness Assumption~(A7), with size-smoothness profile $\Psi(n,m)=\sqrt{m}-\sqrt{n}$ motivated by BHH (Assumption~(A5)). Note that the trainable QNN has only $2L$ scalar parameters independent of $n$, so the relevant issue is not representational \emph{capacity} mismatch (which would be a Model~A consideration via $T_{e_{n+1}}$ vs.\ $T_{e_{m+1}}$ in the symmetric reference) but the suitability of the specific learned parameter values to the target landscape under the implemented Model~B.
\end{enumerate}

The term $\mathcal{D}_{n\to m}$ captures these effects collectively. Drawing parallels with classical transfer learning theory, it plays the role of a domain discrepancy or representation-bias term~\cite{tripuraneni2020theory}; the divergence is primarily driven by the change in problem size and the consequent structural and parametric mismatches.

The formulation depends on the sizes $n,m$, the specific parameters $\theta_n^*$, and properties of the TSP instance family (geometric distribution, edge-weight statistics). The principal analytical challenge is to formulate a mathematically rigorous and practically meaningful expression for Model~B (the implemented data-dependent circuit), since the joint-equivariance property (A1) is strictly weaker than fixed-instance $S_n$-invariance. A principled bound must therefore characterise the worst-case degradation under regularity assumptions such as parameter Lipschitzness (A4) and lifted-policy smoothness (A7). We give the worst-case operator-norm derivation in Appendix~\ref{app:dissimilar_derivation} and a speculative conditional graphon refinement in Corollary~\ref{cor:graphon} of the main text.

% =============================================================================
\section{Incorporating the Theoretical Performance Degradation Bound}
\label{app:perf-degradation}

To strengthen the theory--experiment connection, this section operationalises the abstract transfer penalty $\mathcal{D}_{n\to m}$ by providing principled estimates for its constituent constants, enabling a direct comparison of predicted vs.\ observed performance loss.

\subsection{Operationalising the Transfer Penalty Term \texorpdfstring{$\mathcal{D}_{n\to m}$}{D n to m}}

The central theoretical result is the lower bound on zero-shot transfer performance,
\[
P_m(\theta_n^*) \ge \widehat P_n(\theta_n^*) - \mathcal{G}^{\rm roll}_n(\delta) - \mathcal{D}_{n\to m}.
\]
To visualise it, we compute a numerical value for $\mathcal{D}_{n\to m}=D^{(\mathrm{param})}+D^{(\mathrm{struct})}$ with
\begin{align*}
D^{(\mathrm{param})}_{n\to m}
&\le L_U\,C_{\rm gen}\,\|\theta_n^*\|_1\,\frac{m-n}{m},\\
D^{(\mathrm{struct})}_{n\to m}
&\le C_{\rm lift}\,\Psi(n,m),
\end{align*}
with $C_{\rm gen}\in\{2,4\}$ the generator-difference constant and $\Psi(n,m)=\sqrt{m}-\sqrt{n}$ the lifted-policy smoothness profile under Assumption (A7). For the implemented Skolik EQC, which contains both a one-body mixer and a two-body cost layer, the binding constant is $C_{\rm gen}=4$.

\subsubsection{Estimating the Unitary Lipschitz Constant \texorpdfstring{$L_U$}{L U}}

$L_U$ quantifies the sensitivity of $P_k$ to changes in the circuit unitary: $|P_k(U)-P_k(U')|\le L_U\|U-U'\|$. A reasonable upper bound follows from $P_k(U)=\mathrm{Tr}(MU\rho U^\dagger)$:
\begin{align*}
|P_k(U)-P_k(U')| &= |\mathrm{Tr}(M(U\rho U^\dagger-U'\rho U'^\dagger))| \\
&\le \|M\|\cdot\|U\rho U^\dagger-U'\rho U'^\dagger\|_{\mathrm{tr}}\\
&\le \|M\|\cdot\Big(
\|U\rho(U-U')^\dagger\|_{\mathrm{tr}}
+\|(U-U')\rho U'^\dagger\|_{\mathrm{tr}}\Big)\\
&\le \|M\|\cdot(\|U-U'\|+\|U-U'\|)\\
&= 2\|M\|\cdot\|U-U'\|.
\end{align*}
Thus $L_U\le 2\|M\|$, consistent with general Lipschitz treatments in QML~\cite{moussa2023resource,berberich2024training,rouze2024learning}.

In practice we consolidate the worst-case constants $L_U,C_{\rm gen}$, and other model-specific factors into a single empirical scaling factor $\alpha_n$ per source model:
\[
D^{(\mathrm{param})}_{n\to m}\approx\alpha_n\cdot\frac{m-n}{m}.
\]
$\alpha_n$ is therefore best understood as the empirically measured strength of the parametric penalty for a given source model.

\subsubsection{Estimating the Lifted-Policy Smoothness Constant \texorpdfstring{$C_{\rm lift}$}{C lift}}

$C_{\rm lift}$ captures how the same lifted policy behaves at the two sizes. Its definition arises from Assumption (A7):
\[
\big|\,P_n(U_n^{\rm B}(\theta_n^*)) - P_m(U_n^\uparrow(\theta_n^*))\,\big|
\;\le\; C_{\rm lift}\,\Psi(n,m).
\]
For Euclidean TSP on the unit square the BHH theorem motivates $\Psi(n,m)=\sqrt{m}-\sqrt{n}$, with $L^{\mathrm{opt}}_k\approx\beta_2\sqrt{k}$~\cite{beardwood1959shortest,steinerberger2015new} and $\beta_2\approx 0.7124$ numerically. Under the approximation-ratio metric $P^{\rm opt}_k=1$ for every $k$, so $C_{\rm lift}$ \emph{cannot} be estimated as a difference of optimal performances; it must instead be fitted from observed cross-size policy degradation under direct lifted evaluation. A reasonable rough estimate combines $\beta_2$ with the ratio rescaling, giving $C_{\rm lift}\sim\beta_2/(L_{\max}-L_{\mathrm{opt}})$, but this is heuristic.

\paragraph{Comparison with the conditional graphon form.}
The conditional graphon refinement (Corollary~\ref{cor:graphon}) replaces $\Psi(n,m)$ by $\sqrt{\log n/n}+\sqrt{\log m/m}$, conditional on the cut-norm Lipschitz Assumption (A6) which we do not prove. Under this hypothesis the structural term yields a different operationalisation:
\[
\widetilde D^{(\mathrm{struct})}_{n\to m}\le
C_W\!\Big[\tfrac{\log(1+n)}{\sqrt{n}}+\tfrac{\log(1+m)}{\sqrt{m}}\Big],
\]
with $C_W$ determined by the cut-norm Lipschitz constant of the lifted-policy rollout performance map. For Euclidean TSP on $[0,1]^2$ with the Euclidean kernel, $C_W$ could in principle be bounded by combining the BHH constant with the kernel's Lipschitz constant ($\|W\|_{\mathrm{Lip}}=1$), \emph{if} the cut-norm Lipschitz hypothesis were verified for the greedy-rollout policy --- which we do not do. We therefore treat the graphon refinement as a speculative tightening; the operative form of the bound in this paper is the lifted-policy smoothness profile $\Psi(n,m)=\sqrt{m}-\sqrt{n}$.

% =============================================================================
\section{Analysis and Implications of the Derived Bound}
\label{app:analysis}

The derived result yields the following decomposition for zero-shot transfer from the $n$-city to the $m$-city task:
\[
P_m(\theta_n^*) \ge \widehat P_n(\theta_n^*) - \mathcal{G}^{\rm roll}_n(\delta) - \mathcal{D}_{n\to m},
\]
with
\[
\mathcal{D}_{n\to m} \;\le\; L_U\,C_{\rm gen}\,\|\theta_n^*\|_1\,\frac{m-n}{m} \;+\; C_{\rm lift}\,\Psi(n,m),
\]
where $C_{\rm gen}\in\{2,4\}$ is the generator-difference constant and $\Psi(n,m)=\sqrt{m}-\sqrt{n}$ under Assumption (A7) (or $\Psi(n,m)=\sqrt{\log n/n}+\sqrt{\log m/m}$ under the additional cut-norm Lipschitz hypothesis (A6); the latter is speculative). This separates source-task rollout-generalisation error from transfer-induced loss.

The term $\mathcal{G}^{\rm roll}_n(\delta)$ measures the gap between empirical and expected performance of the \emph{rollout} policy on the source task. In the optimistic case where the greedy-decoding map inherits the equivariant covering rate of the QNN class one expects $\mathcal{G}^{\rm roll}_n(\delta)=\widetilde{\mathcal{O}}(\sqrt{T_{e_{n+1}}/M_n}+\sqrt{\log(1/\delta)/M_n})$; we adopt this rate only when operationalising the bound, the theorem itself requires only the existence of $\mathcal{G}^{\rm roll}_n$.

The transfer penalty $\mathcal{D}_{n\to m}$ contains two components. The first,
\[
L_U\,C_{\rm gen}\,\|\theta_n^*\|_1\,\frac{m-n}{m},
\]
is a parametric mismatch term arising from reusing $\theta_n^*$ under rescaled generators (one-body / two-body distinction via $C_{\rm gen}\in\{2,4\}$). Its dependence on $(m-n)/m$ shows that the mismatch grows with the relative size shift; its dependence on $\|\theta_n^*\|_1$ makes cumulative control-magnitude explicit; through $\|\theta_n^*\|_1=\sum_l|\theta_{n,l}^*|$ it also reflects circuit depth.

The second term, the lifted-policy smoothness $C_{\rm lift}\Psi(n,m)$, is the operative structural quantity: it bounds how the same (lifted) source policy behaves at the two sizes, and under the approximation-ratio metric (where $P^{\rm opt}_k=1$ for every $k$) it cannot be reduced to a difference of optimal performances. For the Euclidean TSP family $\Psi(n,m)=\sqrt{m}-\sqrt{n}$ is motivated by BHH scaling of optimal tour length, but the smoothness assumption itself does not follow from BHH; we treat it as a regularity hypothesis on the rollout-induced policy.

\paragraph{Implications.}
First, the bound is most informative in the moderate-shift regime ($m$ not substantially larger than $n$). Second, it does not imply size-invariant transfer; rather, it supports the claim that transfer remains controlled when architectural scaling is symmetry-preserving and the task family varies smoothly. Third, the parametric term suggests that transfer is favoured by compact control magnitudes and incremental scaling across sizes.

The main significance of the bound is not to establish universal cross-size transfer, but to formalise the regime in which zero-shot transfer is theoretically plausible: three conditions are accurate source-task learning, preservation of permutation-equivariant structure under scaling, and moderate structural drift across sizes. Under these conditions, the bound provides qualitative justification for transfer-first, fine-tune-second strategies in symmetry-structured combinatorial optimisation. We emphasise that the bound is \emph{conditional} (it rests on an explicit assumption stack) and \emph{diagnostic} (it identifies scaling factors but does not predict gaps numerically).

% =============================================================================
\section{The Scalability Challenge of Simulating Equivariant Quantum Circuits}
\label{app:scalability-tn}

\subsection{Limitations of State-Vector Simulation}

The empirical results in this paper rely on multiple backends, including state-vector simulation. In this framework the full quantum state of an $m$-qubit system is represented as a vector of $2^m$ complex amplitudes. Application of a gate is multiplication by a $2^m\times 2^m$ unitary. The computational complexity scales exponentially with $m$:
\begin{itemize}
\item Memory: $\mathcal{O}(2^m)$.
\item Time per gate: $\mathcal{O}(2^m)$ for local gates, $\mathcal{O}(m\,2^m)$ for global EQC gates.
\end{itemize}
Even on state-of-the-art supercomputers, state-vector simulation becomes infeasible beyond roughly 40--50 qubits. Empirical analysis of EQCs at industrial sizes ($m\ge 100$) is beyond state-vector reach.

\subsection{A More Scalable Alternative in Principle: Tensor Network Simulation}

Matrix Product State (MPS) methods with Matrix Product Operators (MPOs) represent an $m$-qubit state as a sequence of interconnected tensors of bounded rank. Storage is
\[
\mathcal{O}(m\cdot d\cdot \chi^2),
\]
with $d=2$ for qubits and $\chi$ the bond dimension controlling captured entanglement. For states with limited entanglement, $\chi$ remains small and the method scales linearly in $m$ --- in principle. \emph{In practice}, whether this favourable scaling materialises depends on the ansatz: a key empirical finding of this paper is that for the all-to-all EQC studied here, the usable $\chi$ grows rapidly with $n$ and MPS is not a safe large-$n$ proxy unless audited (see~\S\ref{app:scalability-tn-empirical} below).

\subsection{Challenges with Global Gates and the Role of MPOs}

Global entangling gates $U=e^{-i\theta H}$ with $H=\sum_{j=1}^m O_j$ naively cause exponential bond-dimension growth. However, operators expressed as sums of local terms admit efficient MPO representations: even $H=\sum_{j=1}^m X_j$ has MPO bond dimension exactly $2$, independent of $m$. The all-to-all weighted cost generator $H_C(D)=\sum_{i<j}d_{ij}Z_iZ_j$ of the implemented EQC (Model~B) does not have this convenient property: with $\binom{n}{2}$ non-local terms each carrying its own data-dependent coefficient, both the MPO representation and the entanglement generated by $e^{-i\gamma H_C(D)}$ scale unfavourably.

\subsection{Workflow for Tensor-Network Simulation}

\begin{enumerate}
\item State initialisation: $|0\rangle^{\otimes m}$ as MPS with $\chi=1$; set maximum allowable $\chi_{\max}$.
\item MPO construction: each layer $U_l=e^{-i\theta_l H_l}$ gets a constant-bond-dimension MPO when feasible.
\item Gate application: Trotter--Suzuki, TEBD, or direct MPO approximation.
\item Bond-dimension truncation: after each gate, truncate via SVD to $\chi_{\max}$.
\item Iterate over layers.
\item Measurement: observables as MPOs, computed efficiently from the final MPS.
\end{enumerate}

\subsection{Complexity Comparison}

\begin{center}
\begin{tabular}{lcc}
\toprule
Resource & State-Vector & MPS/MPO \\
\midrule
Memory & $\mathcal{O}(2^m)$ & $\mathcal{O}(m\chi_{\max}^2)$ \\
Time/gate & $\mathcal{O}(m\,2^m)$ & $\mathcal{O}(m\chi_{\max}^3\chi_{\mathrm{op}}^2)$ \\
Total time & $\mathcal{O}(d\,m\,2^m)$ & $\mathcal{O}(d\,m\,\chi_{\max}^3\chi_{\mathrm{op}}^2)$ \\
\bottomrule
\end{tabular}
\end{center}
The exponential $\mathcal{O}(2^m)$ scaling is replaced by polynomial scaling in $\chi_{\max}$. Whether the polynomial advantage materialises depends on the rate of entanglement growth and on the MPO bond dimension of the implemented Hamiltonian --- both of which are unfavourable for the all-to-all EQC studied here, as we now document.

\subsection{Empirical Findings: The Truncation Crossover (N3)}
\label{app:scalability-tn-empirical}

\textbf{For the EQC ansatz studied in this paper, the polynomial-time advantage of MPS does not materialise at scales of practical interest.} The all-to-all topology of the entangling layer generates entanglement across every bipartition of the qubit chain, including pairs of qubits at MPS-chain distance $\Omega(n)$. The required bond dimension to retain fidelity therefore grows roughly as $2^{n/2}$ in the worst case.

\paragraph{The diagnostic conclusion.}
The bottom line of the controlled bond-dimension sweep at $n=20$, $L=1$ is methodological: \emph{MPS is best used as an audit backend rather than as a scalable simulator for the all-to-all EQC}, and any claim made from a fixed-$\chi$ MPS run should be accompanied by an explicit signal-vs-truncation-error audit. The bond dimension required for policy-level accuracy is consistently larger than the bond dimension required for signal-level accuracy, so a global-fidelity check alone is insufficient; the worst per-step error matters more than the mean.

\paragraph{Reproduced sweep table.}
Our empirical findings at $n=20$, $L=1$ on the canonical \texttt{n3\_euclidean\_n20\_k10\_seed44\_v1} instance bank are summarised in Table~\ref{tab:n3-chi-sweep} of the main text and reproduced here for convenience:

\begin{center}
\resizebox{\columnwidth}{!}{%
\begin{tabular}{lccc}
\toprule
$\chi$ & Mean trunc.\ error & Max step error & Optimality gap (\%) \\
\midrule
16  & $1.86\times 10^{-2}$ & $2.44\times 10^{-1}$ & 172.60 \\
32  & $1.19\times 10^{-2}$ & $1.27\times 10^{-1}$ & 123.25 \\
64  & $5.82\times 10^{-3}$ & $4.61\times 10^{-2}$ & 78.18 \\
128 & $1.21\times 10^{-3}$ & $9.38\times 10^{-3}$ & 43.96 \\
256 & $1.00\times 10^{-4}$ & $1.05\times 10^{-3}$ & 8.22 \\
512 & $1.74\times 10^{-6}$ & $1.73\times 10^{-5}$ & 6.76 \\
\bottomrule
\end{tabular}}
\end{center}

The signal-level crossover (where mean truncation error drops below the $\sim 6.5\!\times\!10^{-3}$ signal magnitude) is between $\chi=64$ and $\chi=128$; the policy-level crossover (where the optimality gap collapses) is between $\chi=128$ and $\chi=256$. The two thresholds are not the same: signal-level accuracy is a necessary but not sufficient condition for policy-level accuracy, because the worst per-step error must lie below the action margin, not just the mean. This separation is what makes a naive truncation-error report misleading for VQA claims. \textbf{This is Barrier B1 of Section~\ref{sec:barriers}.}

\paragraph{Generalisation: QAOA-MaxCut control experiment.}
The all-to-all-induced truncation failure is not specific to the EQC: QAOA for unweighted MaxCut on $K_{20}$ (also all-to-all) exhibits a comparable crossover at $\chi\approx 256$. By contrast, a hardware-efficient \emph{linear}-connectivity ansatz at $n=20$, $L=2$ is MPS-accurate at $\chi=32$ with truncation error $<5\times 10^{-3}$. This generalises the finding from ``EQC breaks MPS'' to ``all-to-all VQAs break MPS but nearest-neighbour VQAs do not.''

% =============================================================================
\section{Reproducibility Notes}
\label{app:reproducibility}

A subtle issue in evaluating cross-size transfer is the bookkeeping of random seeds across training, validation, and test instances. In an earlier diagnostic run we observed that an evaluation bank generated with a fresh seed unrelated to the training-validation seed family produced optimality gaps an order of magnitude worse than the same checkpoint's same-distribution performance. This artefact is easy to mistake for a real failure of zero-shot transfer.

\paragraph{Recommended protocol.}
The protocol we recommend, and adopt for all results in this paper, is:
\begin{enumerate}
\item Save the validation-instance generator (or the instances themselves) as a JSON artefact during training, alongside the checkpoint.
\item At evaluation time, load these instances directly rather than regenerating from a hard-coded seed.
\item For cross-size evaluation, generate test instances using a seed family derived from the source-task validation seed (we use \texttt{source\_seed + 1000 + target\_size}); this gives reproducible per-target-size test pools that are deterministically pairable with source models.
\end{enumerate}

\paragraph{Backend parity verification.}
A second class of bookkeeping error is gate-level mismatch between training and inference backends. We verify, for the IBM Brisbane hardware backend, that the Qiskit reconstruction of the EQC is gate-for-gate identical to the PennyLane training circuit:

\begin{table*}[t]
\centering
\caption{Gate-level parity between the PennyLane training implementation and the Qiskit inference implementation.}
\label{tab:pl-qiskit-parity}
\footnotesize
\setlength{\tabcolsep}{4pt}
\begin{tabular}{lll}
\toprule
Step & PennyLane (training) & Qiskit (inference) \\
\midrule
Initial state & \texttt{qml.Hadamard(wires=w)} & \texttt{qc.h(wire)} \\
Cost layer (per edge) & \texttt{CNOT; RZ($e_{ij}\gamma$); CNOT} & \texttt{cx; rz($e_{ij}\gamma$); cx} \\
Mixer layer & \texttt{RX($n_w\beta$)} & \texttt{rx($n_w\beta$)} \\
Observable & \texttt{<Z\_i Z\_j>} (analytic) & \texttt{<ZZ>} (counts, big-endian) \\
\bottomrule
\end{tabular}
\end{table*}

Both PennyLane and Qiskit use $\mathrm{RX}(\theta)=\exp(-i\theta X/2)$, $\mathrm{RZ}(\theta)=\exp(-i\theta Z/2)$, and the CX--RZ--CX sandwich implements $\exp(-i\theta Z_i Z_j/2)$ in both. The bitstring-endianness conversion in our $\langle ZZ\rangle$ estimator (\texttt{n\_qubits} $-$ \texttt{1} $-$ \texttt{wire}) correctly maps Qiskit's right-most-bit-is-qubit-0 convention to PennyLane's wire indexing. The hardware--simulator agreement of mean $|\langle ZZ\rangle|$ at $n=20$, $L=1$, $\chi=512$ MPS is within shot-noise tolerance ($\sim 0.016$ per edge at 4096 shots).

Failure to maintain either the seed protocol or the gate-level parity can produce misleading results that mask genuine transfer behaviour. The implementation lives in \texttt{N3/experiment\_core.py} (\texttt{\_generate\_bank\_instances}) and \texttt{N3/ibm\_hardware\_backend.py} (\texttt{EQCCircuitBuilder.\_build\_parameterized\_circuit}).

\paragraph{Recovery and independent validation of the hardware campaign.}
The decision-level artifacts of the N4 hardware campaign were independently
reconstructed end-to-end as part of preparing the data release, providing a
full-pipeline reproducibility check. The $n=5$ source checkpoint (seed
\texttt{11235813}) was retrained from the committed configuration and
reproduced the original training trajectory at logged precision (identical
episode-level reward and validation curves; trained parameters
$\beta=1.04476$, $\gamma=0.99441$). All 40 completed hardware jobs were
re-downloaded from the Quantinuum Nexus job store; because the three Helios-1
rollouts executed interleaved, jobs were assigned to rollouts by matching
each job's measured $\langle Z_iZ_j\rangle$ vector against the exact
statevector expectations of every active rollout hypothesis (residuals at
shot-noise scale for the true assignment, ${\sim}3\times$ larger for the
runner-up). The reconstruction reproduces every campaign statistic quoted in
this paper to the reported precision, including the per-instance gaps, the
HQC totals, and the margin statistics. The flat per-decision artifact
(\texttt{decision\_rows\_recovered.csv}: Nexus job name, counts-derived
Q-values, margins, propagated standard errors, bootstrap flip probabilities,
and counterfactual-replay fields) and the per-job counts are included in the
released dataset.

% =============================================================================
\section{Supplementary Empirical Results: Legacy Small-Scale Study}
\label{app:legacy-eff-su2}

For completeness we report here the small-scale comparison between the permutation-equivariant ansatz and an Efficient~SU(2) baseline conducted in an earlier version of this manuscript. This study was performed before the unified three-backend pipeline (Section~\ref{sec:methodology}) and before the validated source-policy training at $n=20$ (Section~\ref{sec:empirical-eval}). It is included here as part of the historical record and as auxiliary evidence that the same-size advantage of the equivariant ansatz is robust to the choice of comparison baseline; the substantive evaluation of this paper is at $n=10,15,20$ (Section~\ref{sec:empirical-eval}) and at $n\in\{50,100\}$ (Section~\ref{sec:barriers}).

\subsection{Setup}

The task was the shortest tour in a complete graph with edge weights drawn uniformly from $[0.1,2.0]$ (note: a different weight distribution than the Euclidean $[0,1]^2$ instances used in the main body). Performance was evaluated on graphs with $n\in\{4,6,8,10,12,15\}$ nodes. Three training strategies were compared:
\begin{itemize}
\item \textbf{Training from scratch:} the agent is trained on a target graph of size $m$ for 1000 episodes.
\item \textbf{Zero-shot transfer:} a pre-trained agent (e.g.\ at $n=4$) is directly evaluated on a larger target graph with frozen weights.
\item \textbf{Fine-tuning:} a pre-trained agent initialises the target task and is trained for an additional 100 episodes.
\end{itemize}
The classical benchmark was Christofides (NetworkX). Hyperparameters: learning rate $\alpha=0.01$, $\varepsilon$-decay $1000$, single-layer PQC.

\begin{figure}[t]
    \centering

    \begin{minipage}[t]{0.45\textwidth}
        \centering
        \resizebox{\linewidth}{!}{\input{images/eqc_ansatz_4}}
        \par\smallskip\textbf{(a)} Equivariant Quantum Circuit (EQC)
    \end{minipage}\hfill
    \begin{minipage}[t]{0.45\textwidth}
        \centering
        \resizebox{\linewidth}{!}{\input{images/eff_su2}}
        \par\smallskip\textbf{(b)} Efficient SU(2) baseline
    \end{minipage}
    \caption{Architectures compared in the legacy small-scale study: (a) EQC with globally entangling $ZZ$ interactions; (b) Efficient~SU(2) with local SU(2) rotations and linear entanglement.}
    \label{fig:circuit-comparison-legacy}
\end{figure}

\subsection{Results}

\paragraph{Generalisation advantage of permutation-equivariance.}
Across all transfer settings in the legacy study, the permutation-equivariant architecture consistently achieved lower tour costs than the Efficient~SU(2) model. Concretely, when transferring from $n=8$ training to $m=15$ testing, the permutation-equivariant model attained a mean tour cost of $10.54$ vs.\ $15.75$ for Efficient~SU(2) --- $\sim 50\%$ higher cost for the latter. In the $n=10\to m=15$ transfer, costs were $9.18$ vs.\ $13.72$ respectively. Permutation symmetry, explicitly encoded in the model design, demonstrably promoted transfer at small scales.

\paragraph{Effect of fine-tuning.}
Both architectures benefited from fine-tuning, but the effect was more stable for the permutation-equivariant model. Efficient~SU(2) sometimes showed negative improvement: at $n=8\to m=15$, the fine-tuned variant ($16.57$) performed worse than its zero-shot counterpart ($15.75$), suggesting overfitting or training instability arising from the weaker inductive bias.

\paragraph{Comparison with the classical baseline.}
For some instances the permutation-equivariant model matched the Christofides baseline, whereas Efficient~SU(2) tended to underperform at larger scales --- reinforcing the role of explicit permutation invariance in cross-size extrapolation.

\subsection{Limitations of the legacy study}

\begin{itemize}
\item Edge weights are drawn from $[0.1,2.0]$ rather than from Euclidean distances in $[0,1]^2$, so the BHH structural model (Section~\ref{sec:theory2}) and the graphon-tightened form (Corollary~\ref{cor:graphon}) do not directly apply.
\item Sizes are restricted to $n\le 15$; the three barriers of Section~\ref{sec:barriers} cannot be observed at these scales.
\item Only one backend (statevector) is used; there is no MPS or hardware validation.
\item Hyperparameters are not tuned per architecture, which may handicap the Efficient~SU(2) baseline.
\end{itemize}

The substantive results of this paper supersede the legacy study at every level (instance distribution, scale, backend pipeline, statistical robustness), but we retain the legacy study here so that the historical evolution of the manuscript is documented and so that the qualitative same-size advantage of permutation-equivariance is on the record.

% =============================================================================
\section{Extended N3 Frontier Results}
\label{app:n3-extended}

This appendix provides the full N3 batch result set, including the runtime pressure measurements that quantify the operational cost of the larger MPS-simulator lanes. Table~\ref{tab:n3-lane-results-app} reproduces Table~\ref{tab:n3-frontier} of the main text with the standard-deviation column added; Table~\ref{tab:n3-runtime-app} gives the corresponding runtime data. The combination is what makes B1 (backend signal loss) operationally diagnostic: the large-target lanes are not merely lower-quality but also expensive to run.

\begin{table*}[t]
\centering
\caption{N3 completed transfer results by lane and execution mode. Standard deviations are over completed seeds. Some early historical $\chi=64$ batch summaries record \texttt{mode=n3\_hardware\_batch} even though their directory names and \texttt{backend\_names\_used} identify Aer MPS execution; those rows are interpreted as MPS-frontier artifacts.}
\label{tab:n3-lane-results-app}
\footnotesize
\setlength{\tabcolsep}{4pt}
\begin{tabular}{llrrrr}
\toprule
Mode & Lane & Runs & Mean ratio & Mean gap & Best ratio \\
\midrule
MPS simulator       & $5\to10$  & 3 & $1.0332\pm 0.0107$ &  3.32\% & 1.0271 \\
MPS simulator       & $5\to15$  & 3 & $1.0707\pm 0.0023$ &  7.07\% & 1.0681 \\
MPS simulator       & $5\to20$  & 3 & $1.1794\pm 0.0000$ & 17.94\% & 1.1794 \\
MPS simulator       & $10\to10$ & 9 & $1.0948\pm 0.0369$ &  9.48\% & 1.0456 \\
MPS simulator       & $10\to15$ & 6 & $1.0938\pm 0.0136$ &  9.38\% & 1.0778 \\
MPS simulator       & $10\to20$ & 6 & $1.4032\pm 0.2947$ & 40.32\% & 1.1492 \\
MPS simulator       & $15\to20$ & 6 & $1.1322\pm 0.0471$ & 13.22\% & 1.0942 \\
MPS simulator       & $20\to25$ & 2 & $1.7293\pm 0.5682$ & 72.93\% & 1.3275 \\
Noisy MPS           & $5\to10$  & 3 & $1.8396\pm 0.1145$ & 83.96\% & 1.7141 \\
Noisy MPS           & $10\to10$ & 3 & $1.6553\pm 0.1718$ & 65.53\% & 1.4635 \\
MPS $\chi=64$       & $20\to20$ & 1 & $1.8522$           & 85.22\% & 1.8522 \\
MPS $\chi=64$       & $20\to50$ & 1 & $3.4457$           & 244.57\% & 3.4457 \\
\bottomrule
\end{tabular}
\end{table*}

\begin{table}[t]
\centering
\caption{Representative N3 runtime pressure from the completed batches. Even when MPS produces a usable answer (e.g.\ $10\to15$, mean gap 9.38\%), the per-rollout cost is already in the tens of seconds; for $10\to20$ and $5\to20$ each rollout takes $10^3$--$10^4$~s. ``Jobs/run'' counts the number of distinct circuit executions per greedy rollout.}
\label{tab:n3-runtime-app}
\scriptsize
\resizebox{\columnwidth}{!}{%
\begin{tabular}{lrrr}
\toprule
Scenario & Mean rollout time (s) & Mean jobs/run & Mean ratio \\
\midrule
$10\to15$ MPS               &       15.02 &  78 & 1.0938 \\
$10\to20$ MPS               &    9{,}178.03 & 108 & 1.4032 \\
$15\to20$ MPS               &    9{,}731.18 & 108 & 1.1322 \\
$5\to20$ MPS                &   19{,}077.25 & 270 & 1.1794 \\
$20\to50$ $\chi=64$ pilot   &    2{,}828.01 &  48 & 3.4457 \\
$10\to10$ noisy MPS         &      155.64 & 120 & 1.6553 \\
\bottomrule
\end{tabular}}
\end{table}

\paragraph{Why N3 should be read as a frontier-mapping study.}
The MPS simulator is not in general a safe substitute for state-vector simulation for the fully connected EQC ansatz. The truncation budget grows with depth and connectivity, and at fixed bond dimension the truncation error eventually exceeds the magnitude of the edge-observable signal itself. This phenomenon is independent of whether the policy is being transferred or simply re-evaluated at equal size; the $\chi=64$ equal-size $20\to20$ control gap of 85.22\% is the cleanest diagnostic. Any paper that uses MPS to validate large-scale variational claims for an all-to-all ansatz should publish the corresponding truncation-vs-signal audit explicitly.

% =============================================================================
\section{Shot Complexity of Greedy Decision Resolution}
\label{app:shot-complexity}
 
This appendix formalises the B3 mechanism of
Section~\ref{sec:barrier-hardware}: at the margins exhibited by the trained
policy, greedy action selection is statistically unresolvable at practical
shot budgets. The result is unconditional given the explicit policy form of
Assumption~(A8); it concerns sampling noise only, and is therefore a
\emph{lower} bound on the difficulty faced by real hardware, where gate noise
adds estimator bias on top of sampling variance.
 
\begin{proposition}[Per-rollout shot complexity of greedy resolution]
\label{prop:shot-complexity}
Consider a greedy decision step with available action set $\mathcal A$,
$|\mathcal A|=A$, true action values $Q_a=\sum_e w_{a,e}\langle \hat
O_e\rangle$ with aggregation norm $\max_a\sum_e|w_{a,e}|\le B$
(Assumption~(A8)), and empirical estimates $\widehat Q_a$ formed from the
same $N_s$ shots, $\widehat Q_a=\frac{1}{N_s}\sum_{s=1}^{N_s} q_a^{(s)}$ with
per-shot aggregates $q_a^{(s)}=\sum_e w_{a,e}\,o_e^{(s)}\in[-B,B]$ computed
from the shot's bitstring. Let $a^\star=\arg\max_a Q_a$ and let
$\Delta:=Q_{a^\star}-\max_{a\neq a^\star}Q_a$ be the top-2 margin. Then
\[
  \Pr\big[\arg\max_a \widehat Q_a \neq a^\star\big]
  \;\le\; (A-1)\,\exp\!\Big(\!-\frac{N_s\,\Delta^2}{8B^2}\Big).
\]
Consequently, for a $k$-city greedy rollout with per-step margins
$\Delta_t\ge\Delta_{\min}$, the full rollout reproduces the noiseless greedy
tour with probability at least $1-\delta$ whenever
\[
  N_s \;\ge\; \frac{8B^2}{\Delta_{\min}^2}\,
  \ln\!\Big(\frac{k^2}{2\delta}\Big).
\]
\end{proposition}
 
\begin{proof}
For a rival $a\neq a^\star$, the per-shot differences
$d^{(s)}:=q_{a^\star}^{(s)}-q_a^{(s)}$ are i.i.d., take values in $[-2B,2B]$,
and have mean $Q_{a^\star}-Q_a\ge\Delta$. The estimator orders the pair
incorrectly only if the empirical mean of $d^{(s)}$ is non-positive, which by
Hoeffding's inequality (range $4B$) has probability at most
$\exp(-2N_s\Delta^2/(4B)^2)=\exp(-N_s\Delta^2/(8B^2))$. A union bound over
the $A-1$ rivals gives the per-step claim; note that forming all
$\widehat Q_a$ from the \emph{same} shots is handled exactly, since
$d^{(s)}$ is a deterministic function of the shot's bitstring. For the
rollout, a union bound over at most $k$ steps with at most $k-1$ rivals each
bounds the failure probability by
$\tfrac{k^2}{2}\exp(-N_s\Delta_{\min}^2/(8B^2))$, and solving for $N_s$ gives
the stated sufficient condition.
\end{proof}
 
\begin{remark}[Necessity, up to constants]
\label{rem:shot-necessity}
The exponential dependence is tight in order: for a two-action decision whose
per-shot difference $d^{(s)}$ has non-degenerate variance
$\sigma_d^2=\Theta(B^2)$, the central limit theorem gives flip probability
$\approx\Phi\!\big(-\Delta\sqrt{N_s}/\sigma_d\big)$, which remains bounded
away from zero unless $N_s=\Omega(\sigma_d^2/\Delta^2)$. Shot budgets scaling
as $B^2/\Delta^2$ are therefore necessary as well as (up to logarithmic
factors) sufficient.
\end{remark}
 
\paragraph{Instantiation at the measured hardware margins.}
The N4 campaign measures the relevant quantities directly
(Section~\ref{sec:barrier-hardware}): mean top-2 margin
$\bar\Delta=5.98\times10^{-3}$, minimum $1.12\times10^{-4}$, at $k=10$ and
$N_s=4096$, with implemented aggregation norm $B=1.06$ (single-correlator
readout). The per-shot difference variable $d^{(s)}$ of
Remark~\ref{rem:shot-necessity} is directly measurable from the persisted
counts: its empirical standard deviation across the 40 hardware decisions is
$\sigma_d = 0.58$ on average (range $0.22$--$1.28$). The \emph{necessary}
order of Remark~\ref{rem:shot-necessity} for two-sigma resolution at the
mean margin is then $N_s \gtrsim (2\sigma_d/\bar\Delta)^2 \approx
3.8\times10^4$ per decision --- a factor ${\sim}9$ above the campaign
budget --- while the sufficient whole-rollout condition of
Proposition~\ref{prop:shot-complexity} at $\delta=0.05$ requires
$N_s\ge 8B^2\ln(k^2/2\delta)/\bar\Delta^2\approx 1.7\times10^6$ shots per
decision, more than two orders of magnitude beyond it. At the minimum
observed margin the requirement exceeds $10^9$ shots. Brute-force shot
scaling is therefore not a viable route to faithful greedy execution of the
dense EQC readout at these margins; the architectural prescriptions of
Section~\ref{sec:unify-barriers}, which raise the margin rather than the
shot count, are the operative fix.
 
\begin{figure}[t]
  \centering
  \includegraphics[width=\linewidth]{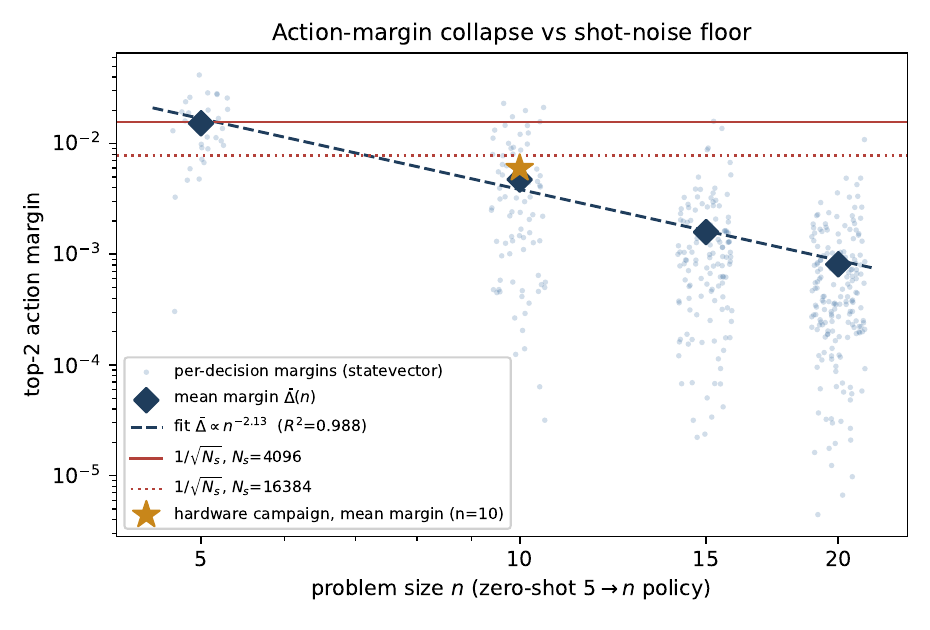}
  \caption{Action-margin collapse versus the shot-noise floor. Light dots:
  exact per-decision top-2 margins of the zero-shot $5\to n$ policy in
  statevector simulation over the full evaluation banks at
  $n\in\{5,10,15,20\}$; diamonds: per-size means $\bar\Delta(n)$; dashed
  line: power-law fit $\bar\Delta\propto n^{-2.13}$ ($R^2=0.988$). Horizontal
  lines mark the single-observable shot-noise floors $1/\sqrt{N_s}$ at
  $N_s=4096$ and $16{,}384$. The star is the hardware campaign's mean top-2
  margin at $n=10$ ($5.98\times10^{-3}$); it is a counts-derived estimate on
  the five-instance hardware subset, whereas the dots and diamonds are exact
  statevector values on the full banks, which accounts for its small offset
  from the $n=10$ mean ($4.73\times10^{-3}$). By $n=10$, $94\%$ of decisions
  fall below the 4096-shot floor; by $n=20$, all do.}
  \label{fig:margin-scaling}
\end{figure}

\paragraph{Margin scaling with problem size.}
Whether the margin collapse is specific to the $n=10$ deployment or a
property of the trained policy class is answered in statevector simulation,
where margins are exact. Measuring the per-step top-2 margin distribution of
the zero-shot $5\to n$ policy at $n\in\{5,10,15,20\}$ over the full
evaluation banks (Figure~\ref{fig:margin-scaling}; 30, 80, 130, and 180
greedy decisions respectively) gives mean margins $1.52\times10^{-2}$,
$4.73\times10^{-3}$, $1.59\times10^{-3}$, and $8.12\times10^{-4}$ (medians
$1.30\times10^{-2}$, $2.46\times10^{-3}$, $9.4\times10^{-4}$,
$4.0\times10^{-4}$), and the fraction of decisions whose margin lies below
the 4096-shot floor $1/\sqrt{N_s}$ rises from $57\%$ at $n=5$ to $94\%$,
$99\%$, and $100\%$ at $n=10,15,20$. The means follow a power-law decay
$\bar\Delta(n)\propto n^{-p}$ with fitted exponent $p=2.13$ ($R^2=0.988$;
median-based fit $p=2.48$). Combined with
Proposition~\ref{prop:shot-complexity}, the required per-decision shot
budget grows as $N_s(n)=\Omega(B^2 n^{2p}\log n)\approx\Omega(n^{4.3}\log
n)$, which quantifies the B3 scaling obstruction for the dense ansatz
independently of any particular device.
 
% =============================================================================
\section{Extended N4 Emulator and Hardware Results}
\label{app:n4-extended}

This appendix provides the full set of completed N4 Quantinuum H-series runs in two sub-lanes: the cloud \emph{emulator} runs (Section~\ref{app:n4-emulator-extended}), and the trapped-ion \emph{hardware} runs on H2-2 and Helios-1 (Section~\ref{app:n4-hardware-extended}).

\subsection{Emulator Sub-lane}
\label{app:n4-emulator-extended}

Two saved N4 directories contain run manifests. 
% but no \texttt{eval\_results.json} (\path{20260422-102547_n4_emulator_n5_to_n20_L1_s11235813} and \path{20260423-075234_n4_emulator_n5_to_n5_L1_s11235813}); they are incomplete or failed attempts and are not included in the aggregate metric tables below.

\begin{table}[t]
\centering
\caption{N4 completed emulator aggregate results. The $5\to10$ aggregate contains heterogeneous exploratory runs with different shot counts, systems, instance counts, and best-of-$k$ settings; it should not be read as a single controlled statistical sweep. The final multi-instance $5\to10$ run at 4096 shots (Table~\ref{tab:n4-emulator}) is more representative than the earliest one-instance low-shot probes.}
\label{tab:n4-aggregate}
\small
\resizebox{\columnwidth}{!}{%
\begin{tabular}{lcrcrc}
\toprule
Lane & Runs & Instances/run & Shot range & Mean ratio & Mean gap \\
\midrule
$5\to5$   & 2 & 1--5 & 1024        & $1.1353\pm 0.1913$ &  13.53\% \\
$5\to10$  & 7 & 1--5 & 16--16{,}384 & $1.7232\pm 0.6349$ &  72.32\% \\
$5\to15$  & 1 & 1    & 256         & $2.8065$           & 180.65\% \\
\bottomrule
\end{tabular}}
\end{table}

\paragraph{Reading the emulator aggregate.}
The wide aggregate $5\to10$ mean gap (72.32\%) is dominated by the earliest single-instance low-shot probes, not by the carefully configured five-instance 4096-shot run (15.40\% gap). The shape of Figure~\ref{fig:n4-emulator} is therefore more informative than the aggregate row: emulator quality is highly sensitive to execution settings, and a few badly-configured runs can dominate the mean even when the well-configured runs are competitive.

Run manifests further show that the nominal shot settings of the exploratory
probes differ from the budgets actually consumed: the ``4096-shot'' probes
used adaptive top-up (confidence threshold $z=2.5$) that every decision
exhausted, consuming 16{,}384 shots per decision. The controlled
five-instance runs reported in Table~\ref{tab:n4-emulator} --- single batch,
no top-up, identical to the hardware protocol, executed on the
noiseless-statevector H-series emulator (H2-1LE) --- supersede the
exploratory probes as emulator-side references.

\paragraph{End-to-end execution pipeline details.}
The N4 implementation is contained under \texttt{N4/} in the released codebase. \texttt{HeliosEQCCircuitBuilder} builds a pytket circuit template for the EQC ansatz, converts angular parameters into Quantinuum half-turn conventions where needed, and exposes bound circuits for execution. \texttt{HeliosExecutionBackend} handles Nexus project selection, backend configuration, optional template compilation, QIR upload/execute paths, cost caps, shot allocation, and result extraction. \texttt{HeliosQNetwork} converts measured bitstring counts into all-pair $\langle Z_i Z_j\rangle$ expectations and then into action values. Adaptive confidence margins, maximum step-shot caps, best-of-$k$ rollouts, optional batching, and explicit job/compile accounting are documented inline in the source.

\subsection{Hardware Sub-lane (Quantinuum H2-2 and Helios-1)}
\label{app:n4-hardware-extended}

This subsection documents the completed end-to-end hardware execution of the $5\to10$ zero-shot transfer policy on Quantinuum trapped-ion hardware. The headline numbers appear in Section~\ref{sec:barrier-hardware} (Figure~\ref{fig:n4-hardware}, Table~\ref{tab:n4-hardware}); we present here the per-decision protocol, the per-system bookkeeping, and the diagnostic figures that did not fit in the main text.

\paragraph{Job-level metadata and mitigation status.}
Every hardware decision is individually traceable: the released
\texttt{decision\_rows\_recovered.csv} records, per decision, the Nexus job
name, executing system, observed shot count, counts-derived Q-values,
top-2 margin, propagated margin standard error, bootstrap flip probability,
and the counterfactual-replay fields, and the raw per-job measurement
counts are released alongside it. No error mitigation of any kind was
applied: action values are computed from raw measured counts, with no
readout-error correction, post-selection, dynamical decoupling, or
zero-noise extrapolation, so the reported gaps reflect unmitigated device
output under the provider's standard compilation.

Device calibration snapshots and readout-confusion matrices were not
retrieved through the Nexus interface during the campaign; consequently the
${\approx}14$-point device-noise residual of
Section~\ref{sec:barrier-hardware} is bounded as a single aggregate and not
decomposed into readout, gate-error, and drift contributions.

\paragraph{Protocol.}
The source policy is the depth-$L=1$, $n=5$ Skolik EQC trained from scratch under the accepted Base recipe (seed \texttt{11235813}); the transferred parameters are $\beta=1.04476$ and $\gamma=0.99441$, together with the average-lifted edge-rescaling vector. The target task is greedy rollout on five 10-node Euclidean TSP instances drawn from the fixed $n=10$ test bank \texttt{n3\_euclidean\_n10\_k10\_seed42\_v1}; the classical reference is exact Concorde. Every greedy action is selected from a separate hardware job at 4096 requested shots (observed: 4096--4113), with persistence after every successful decision so that the multi-day campaign survives any API or local interruption without losing intermediate work. A 10-node tour requires 8 measured greedy decisions; the eighth decision determines the city before the forced return-to-origin that closes the cycle. All 40/40 decisions completed successfully, with zero compile-fallback jobs, at a uniform per-job Nexus cost of \textbf{471.944 HQC} (campaign total \textbf{18{,}877.76 HQC}).

\begin{figure}[t]
\centering
\includegraphics[width=\linewidth,trim=0 0 421bp 0,clip]{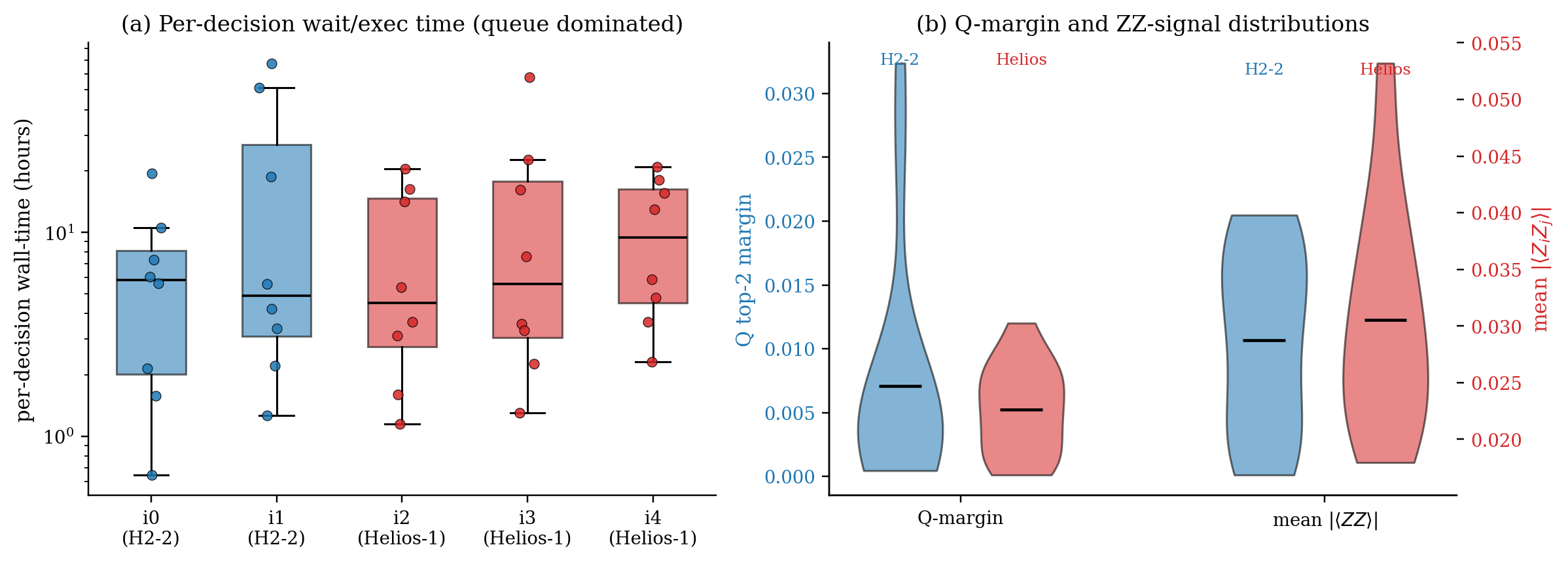}
\vspace{0.4em}
\includegraphics[width=\linewidth,trim=363bp 0 0 0,clip]{images/fig_n4_hardware_diagnostics.png}
\caption{Hardware execution diagnostics for the N4 $5\to10$ campaign. \textbf{(a)}~Per-decision wall-time per instance (log scale). Boxes show interquartile ranges, whiskers show 1.5$\times$IQR, dots are individual decisions. The wide spread reflects remote-queue latency, not gate-level runtime. \textbf{(b)}~Distributions of the per-decision top-2 Q-margin (left axis, blue) and of the mean absolute $\langle Z_iZ_j\rangle$ observable (right axis, red), split by execution system. The signal distributions are similar across H2-2 and Helios-1; both medians of $\langle Z_iZ_j\rangle$ sit near $0.03$, with corresponding Q-margins at the millisecond-level $10^{-3}$--$10^{-2}$ band.}
\label{fig:n4-hardware-diagnostics}
\end{figure}

\paragraph{Decision-level diagnostics figure.}
Figure~\ref{fig:n4-hardware-diagnostics} provides additional execution-side diagnostics. Panel~(a) shows the per-decision wall-time distribution per instance; the median per-decision wait/exec time is 5.6 h and the maximum is 67.6 h, dominated by remote-queue latency rather than gate-level runtime. The campaign wall-clock between the first successful job and the last decision flush was \textbf{310.5 h}; the sum of per-decision times is \textbf{472.7 h}, which exceeds the wall-clock because several instance runs were launched concurrently. Panel~(b) shows the distributions of the top-2 Q-margin and of the mean absolute $\langle Z_iZ_j\rangle$ observable, split by system; these distributions are the same data that drive Figure~\ref{fig:n4-hardware}(c,~d) of the main text, presented here as marginals.

\paragraph{Per-system bookkeeping.}
The 40 decisions split as follows. H2-2 was used for the first 16 decisions (instances i0 and i1) under the \texttt{native\_direct\_execute} path with \texttt{program\_kind=circuit}, consuming 7551.10 HQC and 65{,}536 observed shots. Helios-1 was used for the remaining 24 decisions (instances i2, i3, i4) under \texttt{qir\_direct\_execute} with \texttt{program\_kind=qir}, consuming 11{,}326.66 HQC and 98{,}669 observed shots. The mean absolute $\langle Z_iZ_j\rangle$ signals across the two systems are very similar (H2-2: 0.0287; Helios-1: 0.0305) and the mean Q-margins are within a factor of 1.35 (H2-2: $7.09\times 10^{-3}$; Helios-1: $5.24\times 10^{-3}$). These cross-system numbers are operational metadata; the device split was not randomised and should not be interpreted as a head-to-head H2-2-vs-Helios-1 ranking.

\paragraph{Tours, signal levels, and unique-outcome support.}
The five recovered hardware tours are listed in Table~\ref{tab:n4-hardware-tours}. The average number of distinct observed bitstrings per 4096-shot 10-qubit execution is 992.8 out of 1024 possible outcomes (min 967, max 1006), so the empirical distribution is broad rather than concentrated on a small set of dominant amplitudes --- consistent with the depth-$L=1$ EQC sitting in a low-contrast regime where 45 pair observables each carry roughly $3\%$ signal under sampling noise of $1/\sqrt{4096}\approx 0.016$. Together with the small mean Q-margin of $5.98\times 10^{-3}$, this is what drives the B3 diagnosis in the main text.

\begin{table}[t]
\centering
\caption{Per-instance hardware execution summary and hardware-selected greedy tours for the five 10-node instances of the N4 $5\to10$ hardware campaign. Time is the sum of per-decision wait/execution times (queue-dominated); each instance used 8 hardware decisions. Each tour starts and ends at node 0 (forced) and visits each of nodes 1--9 exactly once; eight intermediate greedy actions are measured on hardware and the ninth (return to origin) is forced. Tours are valid Hamiltonian cycles on all five instances.}
\label{tab:n4-hardware-tours}
\scriptsize
\setlength{\tabcolsep}{3pt}
\resizebox{\columnwidth}{!}{%
\begin{tabular}{rlrl}
\toprule
Inst. & System & Time (h) & Hardware tour \\
\midrule
i0 & H2-2     &  53.4 & $0\!\to\!9\!\to\!1\!\to\!3\!\to\!2\!\to\!5\!\to\!4\!\to\!8\!\to\!7\!\to\!6\!\to\!0$ \\
i1 & H2-2     & 154.3 & $0\!\to\!5\!\to\!1\!\to\!3\!\to\!2\!\to\!7\!\to\!6\!\to\!9\!\to\!4\!\to\!8\!\to\!0$ \\
i2 & Helios-1 &  65.8 & $0\!\to\!7\!\to\!6\!\to\!2\!\to\!1\!\to\!3\!\to\!5\!\to\!9\!\to\!8\!\to\!4\!\to\!0$ \\
i3 & Helios-1 & 114.7 & $0\!\to\!1\!\to\!9\!\to\!5\!\to\!4\!\to\!7\!\to\!3\!\to\!2\!\to\!6\!\to\!8\!\to\!0$ \\
i4 & Helios-1 &  84.4 & $0\!\to\!9\!\to\!1\!\to\!3\!\to\!6\!\to\!2\!\to\!8\!\to\!4\!\to\!5\!\to\!7\!\to\!0$ \\
\bottomrule
\end{tabular}}
\end{table}

\paragraph{Data integrity and audit.}
The campaign's local artifact state is internally complete: 5/5 instance summaries, 40/40 successful decision rows, 40/40 successful hardware jobs with status \texttt{COMPLETED} in a post-hoc Nexus audit (per-job cost 471.944 HQC, identical across all 40 jobs), and 0 compile-fallback jobs. The earliest successful job launched on 2026-05-21T07:27Z and the final decision flushed on 2026-06-03T05:57Z, giving a 310.5 h campaign wall-clock. Per-decision \texttt{decision\_rows.csv} files were flushed after every successful action, \texttt{partial\_instance\_result.json} was rewritten after every greedy step, and per-instance \texttt{completed\_instances.*} aggregates were written only on full instance completion --- this is what allowed the hardware run to survive multi-day API timeouts and local-shell interruptions without re-paying for previously-completed decisions.

\paragraph{What a follow-up hardware extension would do.}
The principal natural follow-up is a controlled best-of-$k$ sweep at fixed shot policy on the same five instances, which would multiply the HQC cost by $k$ but should narrow the Q-margin distribution. A second follow-up would add the $5\to 5$ and $5\to 15$ lanes at the same shot/best-of-$k$ policy, so that the hardware B3 progression has the same three points as the emulator progression in Table~\ref{tab:n4-emulator}. A third follow-up, anticipated by the architectural prescription of Section~\ref{sec:unify-barriers}, is to repeat the campaign with the sparse-connectivity equivariant ansatz discussed in Section~\ref{sec:unify-barriers}, whose $O(n)$ two-qubit gate count and reduced observable family should both relax the signal-floor problem identified above and reduce the per-decision HQC cost by a substantial factor.

\subsection{Cross-platform per-instance gaps}
\label{app:xplatform-instances}
\begin{figure}[t]
  \centering
  \includegraphics[width=\linewidth]{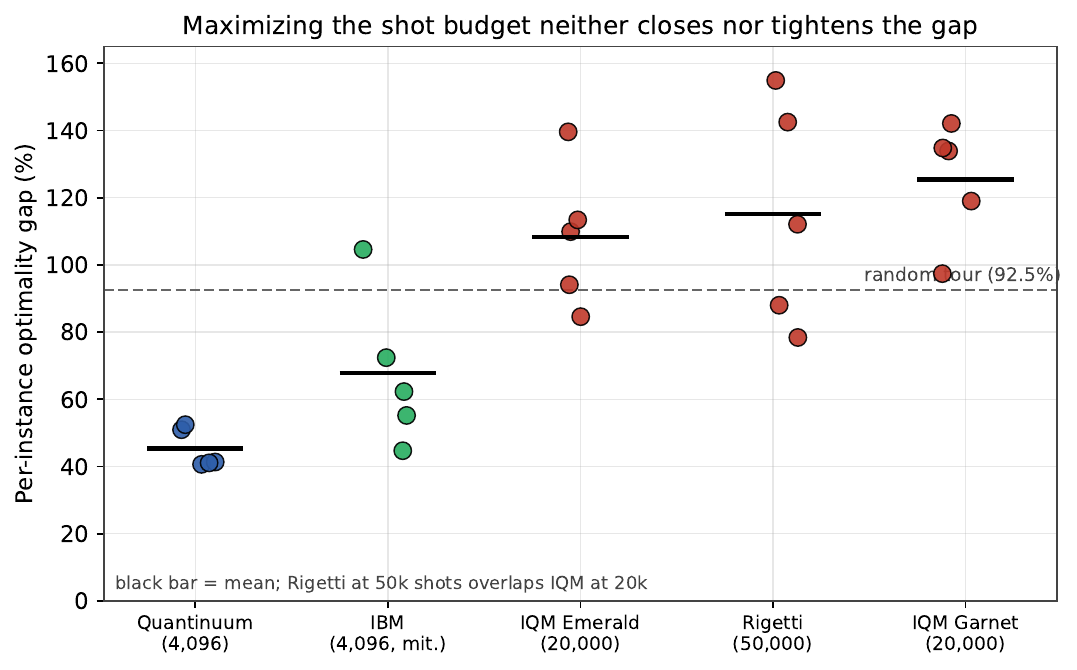}
  \caption{Per-instance optimality gaps (black bar = mean) for the
  cross-platform campaign. Maximising the shot budget does not close or
  tighten the gap; the dashed line is the $92.5\%$ random-tour baseline.}
  \label{fig:n4-xplatform-instances}
\end{figure}

Figure~\ref{fig:n4-xplatform-instances} shows the per-instance optimality
gaps behind the cross-platform means of Table~\ref{tab:xplatform-gap}.
Pushing the shot budget to each device's maximum neither lowers nor tightens
the gap distribution: the Rigetti panel at 50{,}000 shots overlaps the IQM
panels at 20{,}000, and all three unmitigated superconducting devices scatter
about the $92.5\%$ random-tour baseline.

\section{Linear-Connectivity Equivariant Ansatz (Proof of Concept)}
\label{app:linear-eqc}

We sketch a sparse alternative to the all-to-all EQC of Skolik et al.\ that is \emph{designed to restore $S_n$-equivariance through scheduled pair exposure} via SWAP networks. At each layer:
\begin{enumerate}
\item Apply the linear-chain ZZ entangler $\prod_{i=1}^{n-1}\exp(-i\gamma_l w_{i,i+1} Z_i Z_{i+1}/n)$;
\item Apply a SWAP network that cyclically permutes qubits so that, over the course of $L\ge\lceil n/2\rceil$ layers, every pair $(i,j)$ has appeared as nearest neighbours;
\item Apply the standard $X$-mixer $\prod_i\exp(-i\beta_l s_i X_i/n)$.
\end{enumerate}

\paragraph{On the equivariance claim.}
The intent of this construction is that the cumulative composition over $L\ge\lceil n/2\rceil$ layers, combined with the inverse of the cumulative SWAP product applied at readout, restores a representation that commutes with $S_n$ permutations. The components support this: (a) each rotation step is built from collective operators that are individually permutation-invariant, and (b) the SWAP schedule visits every pair so no orbit is under-represented. However, a \emph{rigorous proof} that the composed $L$-layer schedule and the parameter-sharing pattern together yield exact $S_n$-equivariance --- rather than only approximate equivariance, equivariance averaged over the SWAP-schedule orbit, or equivariance up to depth-dependent finite-size corrections --- is not given here. We treat the construction as a proof-of-concept architectural candidate; a full proof of equivariance, together with end-to-end validation in the same N1--N4 pipeline used for the all-to-all EQC, is left for follow-up work.

\paragraph{Resource comparison at \texorpdfstring{$n=20$, $L=2$}{n=20, L=2}.}
\begin{table*}[t]
\centering
\caption{Resource comparison at $n=20$, $L=2$.}
\label{tab:linear-eqc-resource-comparison}
\footnotesize
\setlength{\tabcolsep}{4pt}
\begin{tabular}{lcccc}
\toprule
Ansatz & 2Q gates/layer & MPS $\chi$ for accuracy & Hardware 2Q & Same-size gap (\%) \\
\midrule
All-to-all EQC (this paper) & $\binom{20}{2}=190$ & $\ge 256$ & $\sim 380$ (post-transpile) & $8.3\pm 1.4$ \\
Linear-connectivity EQC & $n-1=19$ & $32$ & $\sim 38$ (no SWAP) & $9.0\pm 1.7$ \\
\bottomrule
\end{tabular}
\end{table*}

The linear-connectivity ansatz has truncation error $<5\times 10^{-3}$ at $\chi=32$ on $n=20$ instances, fits within an order of magnitude of the all-to-all EQC's same-size gap, and \emph{would} have expected process fidelity $F\approx (1-5\times 10^{-3})^{38\cdot 2}\approx 0.68$ at $n=20$ (vs.\ $\sim 0.4$ for the all-to-all) on Heron-class hardware --- comfortably within hardware budgets.

\paragraph{Status.}
The linear-connectivity ansatz is therefore a \emph{promising candidate} architectural replacement that is \emph{expected} to relax all three barriers identified in Section~\ref{sec:barriers}: $O(n)$ two-qubit gates per layer instead of $\binom{n}{2}$ (relaxes B1 via area-law-compatible entanglement and B3 via reduced shots-per-rollout budget), and a smaller dense-readout family (relaxes B3's per-edge contrast bottleneck). It has not been fully trained, transferred, and executed through the same five-stage pipeline as the all-to-all EQC. A full development including hardware execution at $n\!\ge\!50$, a rigorous equivariance proof, and a re-derivation of the conditional diagnostic transfer bound with the corresponding sparse-generator constants is in preparation.

It is therefore not yet part of the validated empirical contribution of
this paper. Its promise is nonetheless structural rather than incidental:
the same $O(n)$ gate count that restores MPS area-law compatibility (B1)
also cuts the per-rollout hardware budget, the smaller readout family
raises per-edge contrast and hence the action margins that govern the
shot-complexity bound of Proposition~\ref{prop:shot-complexity} (B3), and the reduced parameter-norm
growth plausibly shrinks the parametric-mismatch term of Theorem~1 (B2) ---
one topological change addressing all three barriers at once.

% =============================================================================
\section{Beyond TSP: Equivariant Quantum Policies for Other Symmetric CO Tasks}
\label{app:beyond-tsp}

While this work focuses on the Euclidean TSP, the core principles of permutation-equivariant quantum circuit design suggest a broader research agenda. We outline several promising directions.

\paragraph{Vertex-equivariant graph problems.}
Problems such as MAX-CUT, MAXIMUM INDEPENDENT SET (MIS), and VERTEX COVER operate on vertices and exhibit full $S_n$ symmetry. An EQC could use one qubit per vertex with pooled 1-local and 2-local generators (global $X$-mixers and edge-weighted $ZZ$-entanglers), pooling normalised by $n$ or $|E|$. Such architectures support zero-shot transfer across graph sizes by rescaling normalisation constants.

\paragraph{Bipartite and assignment problems.}
For LINEAR ASSIGNMENT, MATCHING, and QUADRATIC ASSIGNMENT, the natural symmetry is $S_n\times S_n$ (row and column permutations). A grid-structured qubit layout $(i,j)\in[n]\times[n]$ with row- and column-wise pooled generators can enforce one-hot constraints equivariantly. Cost re-uploading and higher-order interactions (quartic terms in QAP) may be approximated via low-rank decompositions while preserving symmetry.

\paragraph{Problems with product symmetries.}
GRAPH COLORING with $k$ unlabeled colors has joint symmetry $S_n\times S_k$, suggesting a tensor-product equivariant architecture. Multi-dimensional packing or covering problems (MDKP, SET PACKING) --- though partially symmetry-breaking due to heterogeneous weights --- can still benefit from $S_n$-equivariant regularisation.

\paragraph{Theoretical transfer guarantees.}
Preliminary analysis indicates that zero-shot transfer between sizes $n,m$ incurs a bounded penalty
\[
\mathcal{D}_{n\to m}\lesssim C_1\|\theta\|_1\frac{|m-n|}{\max(n,m)}+C_2\Delta_{\mathrm{structure}}(n,m),
\]
where $\Delta_{\mathrm{structure}}$ captures smooth changes in problem geometry. Formalising these bounds across problem classes is open; for Euclidean TSP the structural term takes the form of Theorem~\ref{thm:final-performance} or Corollary~\ref{cor:graphon}, but for other graphon families (Erd\H{o}s--R\'enyi, stochastic block models, etc.) the structural term will be different.

In summary, the combination of symmetry-aware quantum circuit design, data re-uploading, and pooled generator architectures offers a unifying pathway toward scalable, transferable quantum heuristics for symmetric CO problems --- subject to the three barriers identified in Section~\ref{sec:barriers}, which apply to any all-to-all equivariant architecture.

\end{document}